\documentclass[useAMS,usenatbib]{mn2e}
\usepackage{graphicx}
\usepackage{color}
\usepackage{amsmath}
\usepackage{lscape}  

\usepackage{subfigure}  %% permette di impostare piu' figure 

\usepackage{natbib}
\usepackage{txfonts}
\usepackage{lineno}
\usepackage[english]{babel}
\usepackage{graphicx}
\usepackage{color}
\def\Ha{H${\alpha}$}
\def\Hb{H${\beta}$}
\def\NII{[N\,\textsc{ii}]}
\def\SII{[S\,\textsc{ii}]}
\def\OIIIb{[O\,\textsc{iii}]}
\def\OII{[O\,\textsc{ii}]}
\def\NeV{[Ne\,\textsc{v}]}
\newcommand{\ub}{\ensuremath{U\!-\!B}}%                         % U-B

\newcommand{\apj}{ApJ}%                                         % Journal abbreviations
\newcommand{\apjs}{ApJS}
\newcommand{\apjl}{ApJL}
\newcommand{\aap}{A{\&}A}

\newcommand{\mnras}{MNRAS}
\newcommand{\aj}{AJ}
\newcommand{\araa}{ARAA}
\newcommand{\pasp}{PASP}
\newcommand{\nat}{Nature}
\newcommand{\nar}{NewAR}
\title[AGN in COSMOS: host galaxy properties]{Accreting SMBHs in the COSMOS field and the connection to their host galaxies}

\author[A. Bongiorno et al.]
{\parbox{\textwidth}{A. Bongiorno,$^{1,2}$\thanks{E-mail: \texttt{angela.bongiorno@oa-roma.inaf.it (OAR)}}
A. Merloni,$^{1}$
M. Brusa,$^{1}$ 
B. Magnelli,$^{1}$ 
M. Salvato$^{1}$
M. Mignoli,$^{3}$ 
G. Zamorani,$^{3}$ 
F. Fiore,$^{2}$ 
D. Rosario,$^{1}$
V. Mainieri,$^{4}$ 
A. Comastri,$^{3}$
C. Vignali,$^{3,5}$ 
I. Balestra,$^{1}$
S. Bardelli,$^{3}$
S. Berta,$^{1}$
F. Civano,$^{6}$
P. Kampczyk,$^{7}$
E. Le Floc'h,$^{8}$
E. Lusso,$^{9}$
D. Lutz,$^{1}$
L. Pozzetti,$^{5}$
F. Pozzi,$^{3}$
L. Riguccini,$^{8}$
F. Shankar,$^{10}$ and
J. Silverman,$^{11}$
}\vspace{0.4cm}\\
\parbox{\textwidth}{$^{1}$Max-Planck-Institut f\"ur extraterrestrische Physik (MPE), Giessenbachstrasse 1, D-85748, Garching bei M\"unchen, Germany.\\
$^{2}$INAF-Osservatorio Astronomico di Roma, Via di Frascati 33, 00040, Monteporzio Catone, Rome, Italy.\\
$^{3}$INAF-Osservatorio Astronomico di Bologna, Via Ranzani 1, 40127, Bologna, Italy.\\
$^{4}$ESO, Karl-Schwarschild-Strasse 2, D–85748 Garching bei M\"unchen, Germany.\\
$^{5}$ Dipartimento di Astronomia, Universit\`a di Bologna, via Ranzani 1, I-40127 Bologna, Italy.\\
$^{6}$Harvard Smithsonian Center for Astrophysics, 60 Garden Street, Cambridge, MA 02138, USA.\\
$^{7}$ Institute of Astronomy, ETH Zürich, CH-8093, Zürich, Switzerland.\\
$^{8}$Laboratoire AIM, CEA/DSM-CNRS-Universit\'e Paris Diderot, IRFU/Service d’Astrophysique, B\^at.709, CEA-Saclay, 91191 Gif-sur-Yvette Cedex, France.\\
$^{9}$ Max-Planck-Institut f\"ur Astronomie, K\"onigstuhl 17, D-69117, Heidelberg, Germany.\\
$^{10}$ GEPI, Observatoire de Paris, CNRS, Univ. Paris Diderot, 5 Place Jules Janssen, F-92195 Meudon, France.\\
$^{11}$ Institute for the Physics and Mathematics of the Universe (IPMU), University of Tokyo, Kashiwanoha 5-1-5, Kashiwa-shi, Chiba 277-8568, Japan.\\
}}
\begin{document}

\date{Accepted Year Month Day. Received  Year Month Day; in original
  form  Year Month Day} 

\pagerange{\pageref{firstpage}--\pageref{lastpage}} \pubyear{2002}

\maketitle

\label{firstpage}

\begin{abstract}
Using the wide multi-band photometry available in the COSMOS field we
explore the host galaxy properties of a large sample of Active
Galactic Nuclei (AGN)
($\sim$1700 objects) with $L_{\rm
  bol}$ ranging from 10$^{43}$ to 10$^{47}$ erg s$^{-1}$ obtained by combining X-ray and optical spectroscopic
selections. Based on a careful study of their Spectral Energy 
Distribution (SED), which has been parametrized using a 2-component
(AGN+galaxy) model fit, we derived dust-corrected rest-frame magnitudes, colors and 
stellar masses of the obscured and unobscured AGN  hosts up to high
redshift (z$\la$3). 
Moreover, for the obscured AGN sample we also derived reliable star formation rates (SFRs).
We find that AGN hosts span a large range of
stellar masses and SFRs. No color-bimodality is seen at any redshift
in the AGN hosts, which are found to be mainly massive, red
  galaxies. Once accounting for the color-mass degeneracy in 
well defined mass-matched samples, we find a residual
(marginal) enhancement of
AGN incidence in redder galaxies with lower specific star formation
rates, and we argue that this result might emerge because of our
ability to properly account for AGN light contamination and dust
extinction, compared to surveys with a more limited multi-wavelength coverage.  
However, since these color shifts are relatively small, systematic effects could still be held responsible for some of the observed trend.
Interestingly, we find that the probability for a galaxy to host a
black hole growing 
at any given ``specific accretion rate'' (i.e. the ratio of X-ray
luminosity to the host stellar mass) is almost
{\it independent} of the host galaxy mass, while it decreases
as a power-law with $L_{\rm X}/M_*$. By analyzing the normalization
of such probability distribution, we show how the incidence of AGN
increases with redshift as rapidly as (1+z)$^{4}$, in close
resemblance with the overall evolution of the specific star formation rate of
the entire galaxy population. We do provide the analytic fitting formulae
that describe the probability of a galaxy of any mass (above the
completeness limit of the COSMOS survey) to host an AGN of any given
specific accretion rate as a function of redshift. These can be useful
tools for theoretical 
studies of the growing black holes population within galaxy evolution models. 
Although AGN activity and star
formation in galaxies do appear to have a common triggering mechanism,
at least in a 
statistical sense, within the COSMOS sample we do not find
strong evidence of any 'smoking gun' signaling powerful AGN
influence on the star-forming properties of their hosts galaxies.    
\end{abstract}

\begin{keywords}
Surveys, Catalogues, Galaxies:active, Galaxies: fundamental
parameters, Galaxies: evolution 
\end{keywords}

\section{Introduction}

Most of the times since their discovery, more than 40 years ago, Active
Galactic Nuclei (AGN) powered by accretion onto Supermassive Black
Holes (SMBH) have been considered rare and exotic 
objects with peculiar (and extreme) astrophysical properties.  
In the past decade, studies in the local Universe 
established the presence of supermassive Black Holes in the
nuclei of virtually all galaxies with a bulge/spheroidal component
 \citep[e.g.][]{Kormendy1995,Kormendy2011};
changing dramatically the perception of this class of objects. Since
cosmological SMBH growth is mostly due to accretion of matter during 
their active phases \citep{Soltan1982}, we are led to conclude that 
most bulges went through a phase of strong nuclear activity \citep{Marconi2004,
  Merloni2008,Shankar2009R}. Moreover, as the energy released in the process of accretion
on the SMBH can be higher than the total binding energy of a massive
galaxy, AGN may represent  a key ingredient of the formation
and evolution of all galaxies. If, and how, the nuclear black holes influence their host 
galaxies, and vice-versa, became a major focus 
in studies of structure formation and evolution. 

Indeed, many observational discoveries do support the notion of 
a close link between SMBH growth
and the assembly of the host galaxy. The most crucial ones
are the following: (i) The tight
correlation between the central SMBH mass and various properties of
their host galaxies i.e. luminosity: \citep{Kormendy1995,
  McLure2001, Marconi2003}, stellar mass \citep{Magorrian1998}, and
velocity dispersion \citep{Ferrarese2000, Gebhardt2000,
  Tremaine2002, Kormendy2011}.  
(ii) As populations, 
SMBHs and galaxies seem to evolve with redshift in a similar way. 
In fact, the AGN evolution is luminosity-dependent with low-luminosity
AGN reaching the peak in their space density later in the history of
the Universe  (i.e. at lower redshift) than higher luminosity AGN
\citep{Miyaji2000, LaFranca2005, Ueda2003, Fiore2003,Cirasuolo2005,
  Hasinger2005, Bongiorno2007}. This pattern 
is similar to the so-called ``cosmic downsizing''
of star forming galaxies \citep{Cowie1996,Menci2008} seen also in the spheroid
galaxies \citep{Cimatti2006,Thomas2010}.  
(iii)  The shape of the integrated AGN activity over cosmic time is
similar to the one of the 
global star formation rate (SFR) with a peak at z$\sim$2 and a rapid
decline at both lower \citep[e.g.][]{Dickinson2003,
  Merloni2004,Hopkins2007,Shankar2009}  and
higher \citep[e.g.][]{Franceschini1999,Wilkins2008, Brusa2009hzlf} redshift.
(iv) Finally, luminous AGN show a
correlation between the AGN bolometric luminosity and the star formation
(SF) of the host galaxy over more than five orders of magnitude in
luminosity \citep{Netzer2009}, albeit with a substantial scatter; possible
deviation from such a behavior have however been reported for lower
luminosity AGN \citep{Shao2010, Mullaney2012,Santini2012}.

While the evidences of a tight link between the formation of galaxies
and the growth of the SMBHs at their centers are many and clear, the
physical processes behind this interplay remain unclear. In
particular, the fundamental question of whether AGN-driven feedback
processes are ultimately responsible for determining global properties
of the galaxy population (e.g. luminosity functions, color-magnitude
distributions and the evolution thereof) or, on the contrary, they
happen to be triggered and fueled as a by-product of star-formation
activity and morphological evolution of their hosts still remains
unanswered. This is of
fundamental importance for understanding both galaxy assembly and the
accretion density of the Universe. The work we present here is indeed
motivated by the desire to obtain robust, and statistically
significant, constraints on the possible scenarios for the
co-evolution of galaxies and AGN.  

A slew of models \citep[e.g.][]{Somerville2001, Granato2004,
  Monaco2005, Croton2006, Hopkins2006a, Cen2011, Springel2005,Schawinski2006} have
been developed in recent years to explain this co-evolution and to
describe the main mechanism that fuel the central SMBH and build the
galaxies' bulge. 
Some of these semi-analytical models and hydrodynamical simulations
\citep[e.g.][]{Springel2005,Hopkins2006a,Menci2008} invoke major mergers of gas-rich galaxies
as the main fueling mechanism.   
Alternative mechanisms have been discussed in the literature,
including minor-mergers \citep[e.g.][]{Johansson2009}, disk
instabilities \citep[e.g.][]{Genzel2008}, and recycled gas from dying
stars \citep[e.g.][]{Ciotti2010}.  
Different fueling mechanisms are usually associated to different
luminosity ranges e.g. major mergers are invoked to trigger bright
quasars \citep[e.g.][]{Hopkins2008}, while the role of major mergers
in triggering more typical ($\sim L_{*}$) AGN is called into question 
\citep[see e.g.][]{Cisternas2011b,Schawinski2011,Kocevski2012}. 
On the contrary,  secular processes seem to be enough
to fuel low luminosity AGN i.e. Seyfert-like objects
\citep{Hopkins2008, Allevato2011}.
 Due to the large mismatch in physical scales between AGN and
 galaxies, all these  models have to rely on specific assumptions
 regarding the mechanism
responsible for the link between the nuclear activity, which releases most
of its energy on the scale of few Schwarzschild radii $R_{\rm S}$ ($\sim$
10$^{-5}$ pc), and the stellar population on much larger scales
($\sim$ a few kpc, some $\sim 10^8$ times $R_{\rm S}$). This mechanism is
in most cases an energetic feedback from the central engine that
deposits the energy liberated by the accretion process within the
host galaxy, or its dark matter halo \citep{Silk1998} The
physical description of the feedback itself and its effect on the AGN host
galaxy is, however, very different among different models.   

In particular, due to the complex interplay between gas inflow,
outflow and nuclear obscuration during active AGN phases, 
a correct and complete identification of unobscured, obscured, and
highly obscured AGN at all redshifts (and especially in the $z=1-3$
interval, where most of the feedback is expected to happen) is
therefore crucial for a comprehensive understanding of the still
little explored phase of the common growth of SMBHs and their host
galaxies. 

Generally speaking,  two different approaches can be 
followed to test the predictions of different theoretical models of AGN
feedback and to constrain the physical mechanisms at play: 
a detailed study of the physical properties of relatively
small but well defined  
samples of selected sources or a statistical investigation
of a large sample of AGN which includes different sub-classes. 
Within the COSMOS survey, the first
approach has been followed in our previous papers
\citep[i.e.][]{Merloni2010,Mainieri2011,Cisternas2011b,Lusso2011},
while the latter approach is adopted here.  

To study growing SMBHs and their influence on the host galaxies over
different periods of their cosmic evolution, one needs to compile a 
sample of AGN (both unobscured and obscured) as complete as possible, spanning a
wide range in luminosity and redshift. X-ray surveys have been proved
to be the most efficient way to compile nearly unbiased samples of
Compton Thin AGN 
\citep{Brandt2005}, but they may miss Compton thick AGN
\citep[e.g.][]{Comastri2008,LaMassa2009}.  
On the other hand, the optical spectroscopic selection, even though allowing
a straightforward selection of unobscured AGN and being less biased
against Compton 
Thick AGN, can fail to select those AGN for which the nuclear
optical/IR light is diluted
by their host 
galaxy. To overcome these limitations, we decided to consider and
combine X-ray and 
optically selected AGN samples compiling a highly homogeneous and
representative  
sample of obscured and unobscured AGN selected in the COSMOS field over a
wide redshift baseline (0$<$z$<$4). 

Other selections criteria (e.g. from IR color-color diagrams,
\citealt{Stern2005,Donley2012},  
or from radio surveys, e.g. \citealt{Smolcic2008}) may provide
additional AGN and  
are needed in order to get a full picture on the AGN-galaxy
co-evolution \citep[e.g.][]{Hickox2009}.  
However, since  X-ray and optical selections make up more than 50\%
of the full AGN population in COSMOS (Brusa et al. in prep) for the
purpose of this paper we will focus  
on only these two methods quantifying, whenever relevant, the
incompleteness due to the not inclusion of the other samples.

The major observational challenge in any comprehensive study
of the AGN-galaxy co-evolution is
the accurate separation of the AGN and galaxy emission
components, at all optical-IR wavelengths. This is a crucial step for
a number of reasons: depending on the intrinsic spectral energy
distribution of the nuclear (AGN) and of the stellar light, and of their
respective level of extinction, inaccurate de-blending might not only
hamper any precise determination of the galaxies' physical properties,
but also mask AGN signatures and bias our view of the SMBH growth.  

To this end, we used a two-component Spectral Energy
Distribution (SED) fitting procedure. The observed SED is
fitted with a large grid of models based on a combination of the
\citet{Richards2006SED} AGN template and several host-galaxy models
\citep[i.e. synthetic spectra created from the stellar population
synthesis models of][]{Bruzual2003}. Given the wide multi-wavelength
coverage available in the COSMOS field, this fitting technique allows
to decompose, typically with a high level of confidence,
 the entire SED into nuclear AGN and host galaxy
components and to derive robust measurements of the host galaxy physical
properties, e.g. rest-frame colors, stellar mass, K-band luminosity
and star formation rate.  
The strength of the SED fitting method is that, given sufficiently wide
photometric coverage, it is applicable to all AGN, obscured and
unobscured, independent of their luminosity. In particular, once a
comprehensive set of templates for the SED components is chosen, the
method can be applied (almost) blindly to any detected object in a
multi-wavelength survey, irrespective of the specific selection
criteria, reaching an accuracy that depends 
crucially on the number of bands and depth of the available
photometric catalogs. In this respect, COSMOS is a uniquely
suited field for our investigation.

The structure of the paper is as follows: In Section
\ref{sec:sample} we will introduce our sample, before proceeding to a
discussion of the 2-component SED fitting method in Section
\ref{sec:sed}. In Section \ref{sec:GALprop} we will discuss the physical
properties of the AGN host galaxies (i.e. colors, masses and SFRs)
compared to normal galaxies focusing in Section \ref{sec:AGNfr_mass} 
 and \ref{sec:AGNfr_sfr} on the incidence of the AGN as a function of
 their host properties i.e. mass and star-formation rate, respectively. 
A final discussion and a summary of our main results are presented in
Section \ref{sec:discussion} and \ref{sec:summary}. 
Additionally, in
Appendix \ref{sec:1component}  we discuss the effect on the  derived
host galaxy properties if the AGN component is not properly subtracted, 
while Appendix \ref{sec:sfr} shows a detailed comparison between
two SFR indicators i.e. UV-optical SED fitting and the FIR bands based method. In Appendix \ref{sec:table} two tables
with the AGN catalog and the parameters derived from the SED fitting
(e.g. rest-frame magnitudes,  masses and SFRs) can be found together
with a short explanation of the columns. These tables show only 16
lines as an example of the ones published online.  
Throughout this paper, we use the standard cosmology 
($\Omega_{\rm m}$=0.3, $\Omega_{\Lambda}$=0.7, with H$_0$=70 km s$^{-1}$
Mpc$^{-1}$). Magnitudes, if not differently specified, are AB magnitudes.

\section{AGN samples from the COSMOS survey}
\label{sec:sample}

In order to construct a sample of AGN as complete as possible, including both
obscured and unobscured objects, we first selected  XMM-COSMOS point-like
sources \citep{Hasinger2007,Cappelluti2009}, to which we
added a smaller sample of purely optically selected AGN from the
zCOSMOS bright spectroscopic survey
\citep{Lilly2009}, spanning a lower AGN luminosity range
(\citealt{Bongiorno2010}, Mignoli et al. in prep). In this section,
we give the details of our sample selection.  

\begin{figure*}
\includegraphics[width=0.49\textwidth]{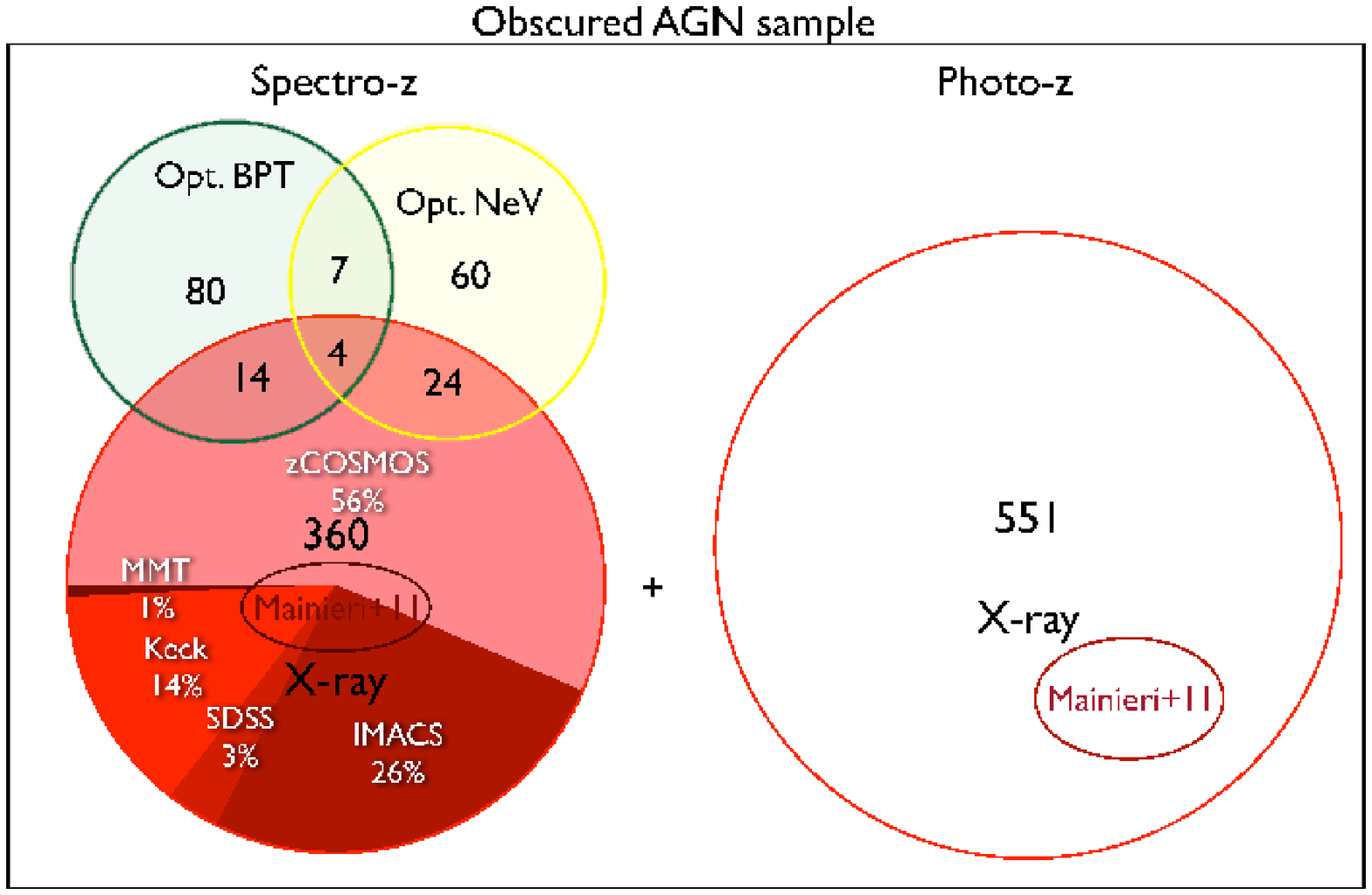}
\includegraphics[width=0.49\textwidth]{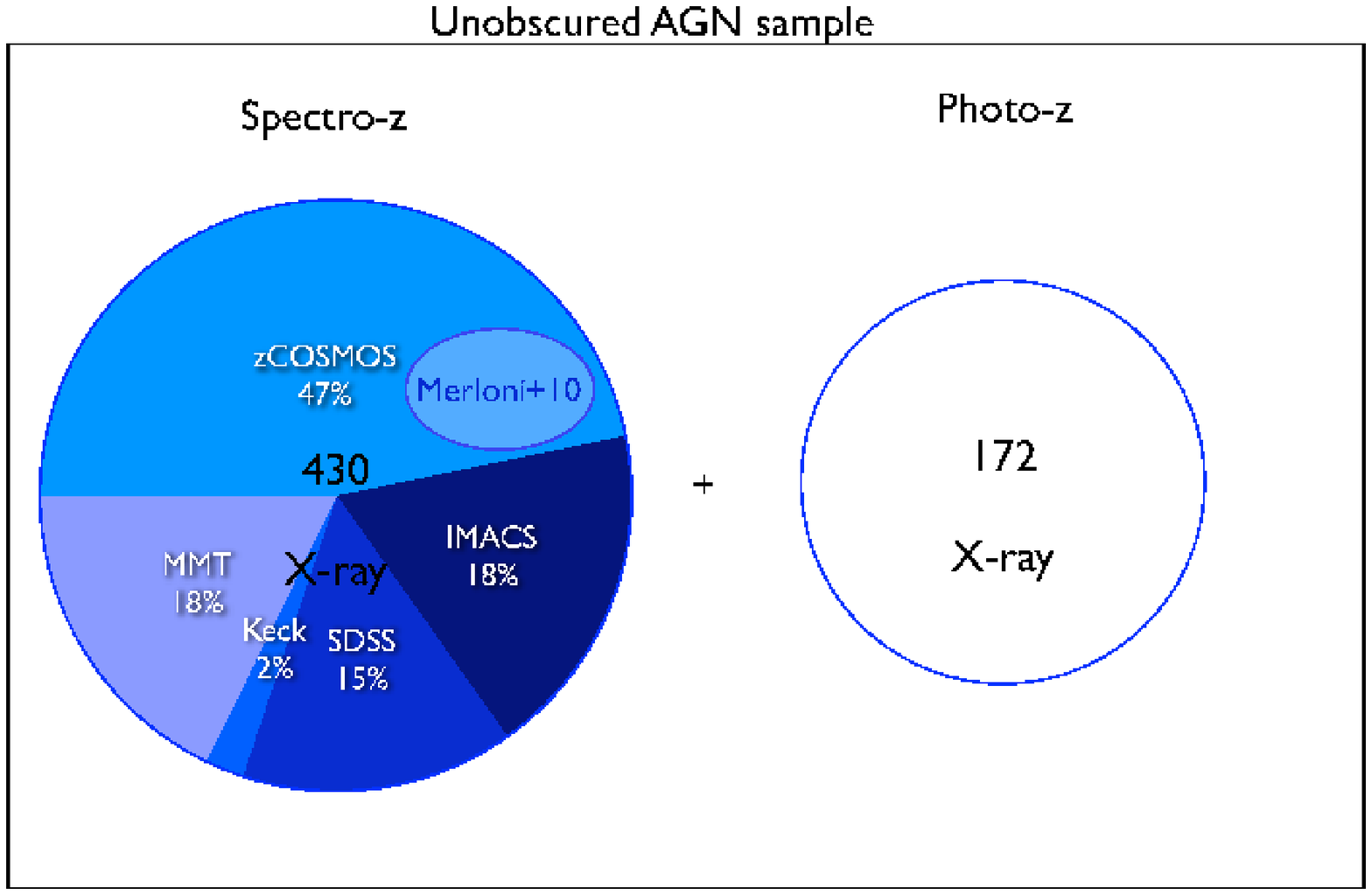}
\caption{
Schematic view of the analyzed sub-samples for obscured (left) and
unobscured (right) AGN. Different circles represent the different
methods used to select them. Blue and red circles correspond to X-ray
unobscured and obscured AGN, green and yellow to optically selected
AGN using the diagnostic diagrams and the \NeV\ line. Open and filled
symbols are for photometric and spectroscopic redshift whose origin is
highlighted with the slices of different color gradations.
The combined analyzed sample consists of 1702 AGN: 602 type--1 AGN (of
which 430 with spectra) and 1100 type--2 AGN (of which 549 with
spectra).}
\label{fig:sample}
\end{figure*}

\subsection{The X-ray selected sample} 
The XMM-COSMOS catalog  includes $\sim$1800 point-like X-ray
sources. The X-ray catalog has been presented in
\citet{Cappelluti2009}, while the optical identifications and
multi-wavelength properties have been discussed in
\citet{Brusa2010}\footnote{In this paper we use an improved version of
  the XMM-COSMOS identifications catalog with respect to the one
  published in \citet{Brusa2010}, the main difference being the
  availability of new photometric redshifts from
  \citet{Salvato2011}. The updated version of the XMM-COSMOS
  identification catalog can be downloaded here:
  http://www.mpe.mpg.de/XMMCosmos/xmm53\_release/}. The sources of the XMM-COSMOS
catalog of \citet{Brusa2010} have
been classified as normal galaxies, obscured and unobscured AGN
according to their properties, i.e. X-ray luminosities, spectra and
multi-wavelength SEDs \citep{Salvato2011}; here we consider only those 
sources classified as AGN, both unobscured and obscured, as we
describe below.
 
Optical spectra are available for
more than half of the sample from different instruments: VIMOS/VLT
\citep[zCOSMOS project,][]{Lilly2007,Lilly2009}, IMACS/Magellan
\citep{Trump2007,Trump2009}, MMT \citep{Prescott2006}, and
DEIMOS/KeckII (P.I. Scoville, Capak, Salvato, Kartaltepe, Mobasher). 

From the XMM-COSMOS subsample for which spectra are available, we
identified 430 sources which show broad (FWHM$>$2000 km s$^{-1}$)
emission lines in their optical spectra, which we classify as type--1
(unobscured). Further 402 spectroscopically confirmed AGN were instead
classified as type--2 (obscured), by making use of 
 both their spectral
properties and X-ray luminosities. In particular, the type--2 AGN class includes:
(a) all the sources that do not show broad emission lines and for
which the ratio between their high-ionization narrow lines indicates
AGN activity \citep{Baldwin1981} and (b) in case  the high ionization
lines are not detected in the observed wavelength range, all the
objects without broad lines and with rest-frame (2-10 keV) X-ray luminosity
greater than 2$\times 10^{42}$ erg s$^{-1}$.  

For about half of the XMM-COSMOS sources (723 objects), only photometric redshifts
are available (the ``photo-z'' AGN sample), 
which are however very accurate with $\sigma_{\Delta
  z/(1+z)}\sim$0.015 
and a fraction of outliers of just 5.8\% \citep{Salvato2009,Salvato2011}. 	 
Thus, the XMM-COSMOS sources for which no spectra are available have been
classified as unobscured type--1 in 172 cases, and as obscured type--2
in 551 cases. In practice, the source classification for ``photo-z'' AGN 
was performed  as in \citet{Salvato2011} on the basis of the template\footnote{chosen from a library of hybrid template obtained combining the observed galaxy templates from \citet{Polletta2007} with different levels of AGN contamination \citep{Salvato2011}} that best describes their
SED, and applying the same threshold in X-ray luminosity ($L_{\rm [2-10] keV}>2 \times 10^{42}$ erg s$^{-1}$).
  
Summarizing, the XMM-COSMOS sample analyzed in this work comprises 602
unobscured type--1 AGN (430 with spectroscopic redshift and broad
emission lines; 172 with photometric redshift) and 953 obscured type--2 AGN (402 with
spectroscopic redshift and 551 with photometric redshift), for a total
of 1555 X--ray selected AGN. 

\begin{table}
\begin{center}
\begin{tabular}{|cccc|c|}
\hline\hline
\multicolumn{5}{|c|}{Obscured AGN sample}\\
\hline
\multicolumn{4}{|c|}{Spec-z}&\multicolumn{1}{|c|}{Phot-z}\\
\hline
&X-ray&OptBPT&OptNeV&Xray\\
X-ray&\textbf{360}&18&28&\textbf{551}\\
OptBPT&18&\textbf{105}&11&\\
OptNeV&28&11&\textbf{95}&\\
\hline
\multicolumn{5}{c}{}\\
  \hline\hline
\multicolumn{5}{c}{Unobscured AGN sample}\\
\hline
&\multicolumn{3}{|c|}{Spec-z}&\multicolumn{1}{|c|}{Phot-z}\\
\hline
X-ray&\multicolumn{3}{|c|}{\textbf{430}}&\multicolumn{1}{|c|}{\textbf{172}}\\
 \hline\hline

\end{tabular}
\caption{Analyzed sub-samples of obscured and unobscured AGN. Bold numbers indicate the total number of sources while the rest highlight the sources in common between two or more classification methods. The combined analyzed sample consists of 1702 AGN: 602 unobscured AGN (of
which 430 with spectra) and 1100 obscured AGN (of which 549 with
spectra).}  
\label{tab:sample}
\end{center}
\end{table}

\subsection{Optically selected Seyfert--2 galaxies: diagnostic diagrams}
For type--1 AGN the X-ray and optical selection match very well,
i.e. in addition to the 430 spectroscopic broad line AGN with XMM
detection, there are only 25 ($\sim$6\%) objects purely optically selected
from the zCOSMOS bright survey which are not detected in the X-ray band by
XMM (the number goes down to just 14, $\sim$3\%, if we consider also
Chandra detections).   
This is not the case for obscured type--2 AGN, for which a more substantial fraction 
($>$25\%) can be selected from optical surveys showing no X-ray
counterparts. Moreover, while the redshift and luminosity distribution
of X-ray and optically selected type--1 AGN is very similar, for
type--2 objects this is not the case and different bands can allow
us to retrieve different sub-classes of sources.  
For this reasons, we
include in this study also obscured type--2 AGN selected purely on
optical properties. Since optically selected AGN (Seyfert--2 galaxies)
tend to have much lower 
luminosities (Fig. \ref{fig:AGNlum}) this allows us to enlarge the 
analyzed luminosity range. 

The sample of optical type--2 AGN was selected in the following way:
from the final 20\,000  VIMOS/VLT zCOSMOS bright, and in analogy with the 
procedure adopted in \citet{Bongiorno2010} for the first 10\,000 spectra
\citep[10k,][]{Lilly2009}, we have selected 392 sources using
 the standard diagnostic diagrams
\citep[\OIIIb/\Hb\ versus \NII/\Ha\ and \OIIIb/\Hb\ versus
\SII/\Ha,][BPT diagram]{Baldwin1981} to isolate AGN in the redshift
range 0.15$<z<$0.45 and the diagnostic diagram  
\OIIIb/\Hb\ versus \OII/\Hb\ \citep[][``blue''
diagram]{Rola1997,Lamareille2004} to extend the selection to higher
redshift (0.5$<z<$0.92). 
These diagnostic diagrams allow the classification of emission line objects as
Seyfert--2 galaxies, candidate Seyfert--2 galaxies, and LINERs. Here
we include only the 105 objects classified as secure Seyfert--2 galaxies
\citep[the LINER sample is studied separately in][]{Tommasin2012}.  
Of these 105 AGN, 18 have X-ray counterparts and therefore are already
included in the X-ray selected sample described above. Hence, from the
optical selection based on the spectral diagnostic diagrams,
we isolated 87 pure optically selected Seyfert--2 which
do not show any X-ray emission at the depth of our XMM observations.

\subsection{Optically selected obscured AGN: the \NeV\ emission
  line sample}
A relatively new method to select obscured AGN in the optical surveys uses the detection of 
\NeV\ line whose presence is a distinctive feature of AGN activity
\citep[see e.g.][]{Gilli2010}. 
Using this criterion on the same 20\,000
zCOSMOS bright spectra, Mignoli et al. (in prep) identified 95 obscured
type--2 AGN, 11 of which were already identified through the emission
line diagnostic diagrams.  
%(4 of the 11 being also X-ray sources).  
Of the remaining 84 sources, 24 have X-ray counterparts and are hence
already included in the  X-ray selected sample described above. From
the \NeV-based optical selection, we thus isolated additional 60 pure
optically selected Seyfert--2 which do not show any X-ray detection.

In total, the optically selected obscured AGN amount to 189 sources, of which
147 are purely optically selected with no XMM detection\footnote{We
  note here that 7 of the 147 purely optically selected AGN are
  actually detected in the deeper Chandra observation covering the
  central part of the COSMOS field (\citealt{Elvis2009}, Civano et
  al. in prep). In order to keep selection effects under control and
  exploit the full 2 deg$^{2}$ area of COSMOS, in this paper we limit
  the X-ray analysis to the XMM-COSMOS sample.}. Since both these two criteria are based on spectral features, the number of spectra available for obscured AGN is artificially higher than for the unobscured ones. 

\begin{figure*}
\includegraphics[width=0.48\textwidth]{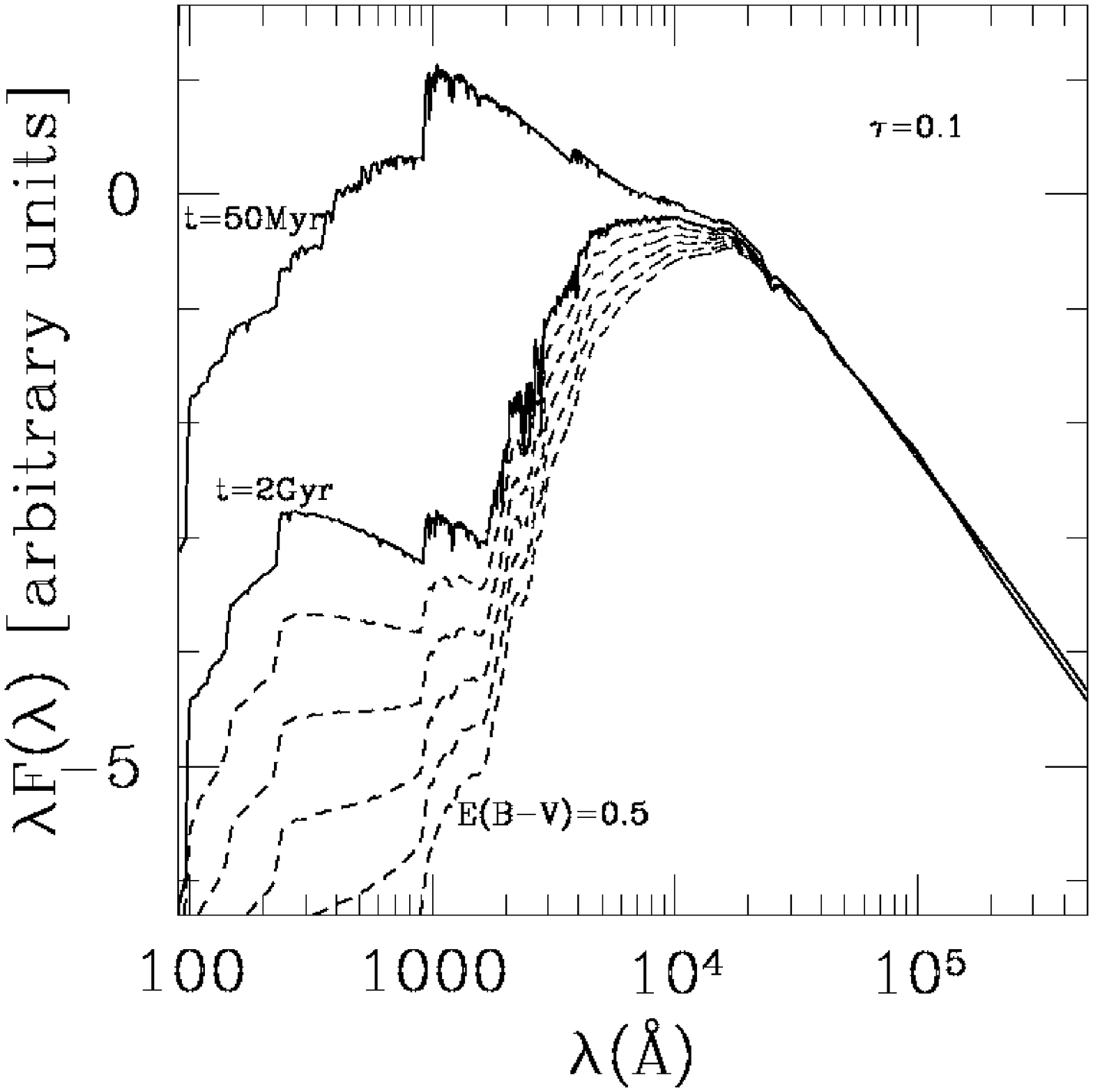}
\includegraphics[width=0.48\textwidth]{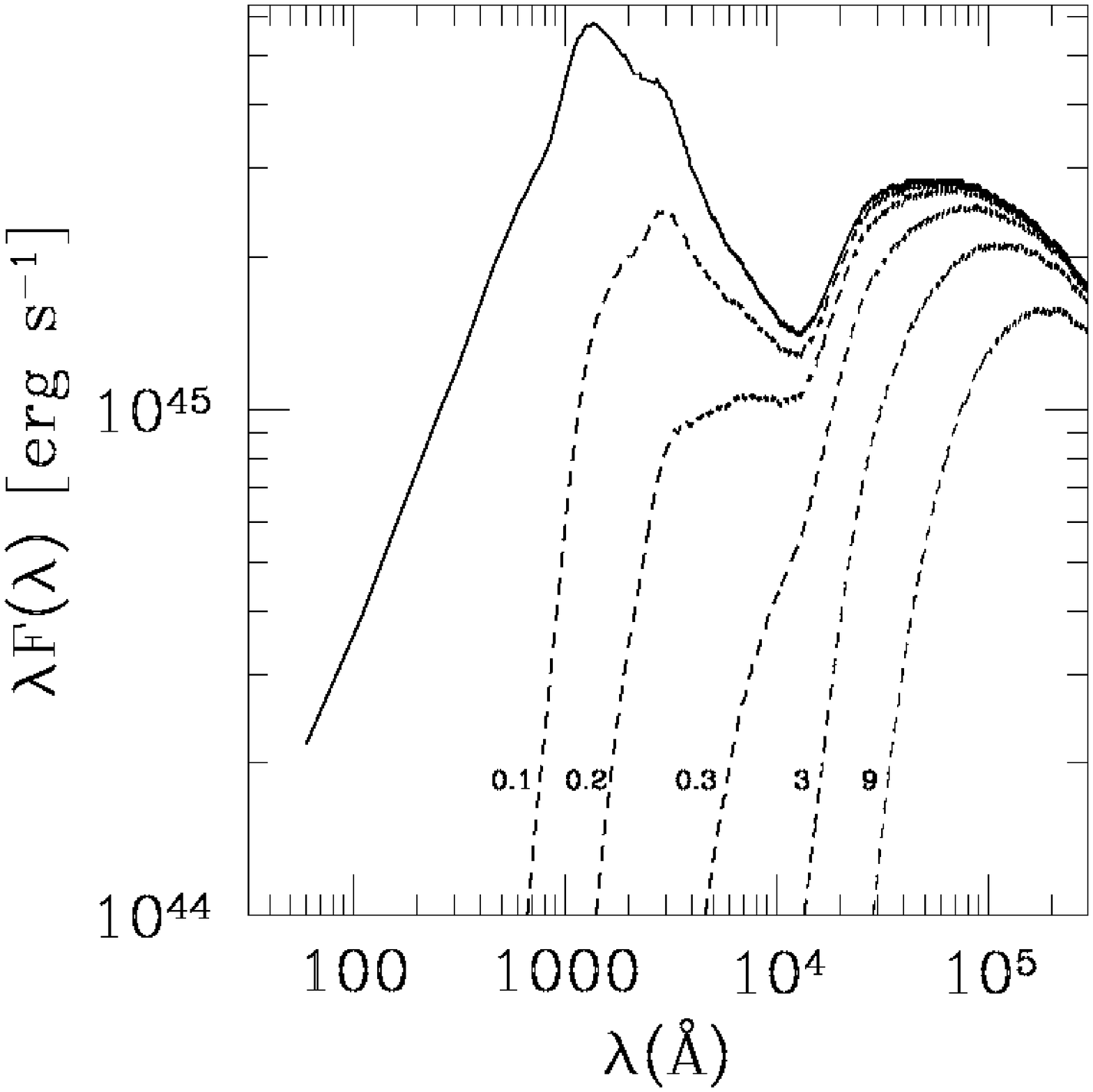}
\caption{ \textit{Left:} Examples of galaxy templates derived from
  \citet{Bruzual2003} models. The two solid curves correspond to a
  model with $\tau$=0.1 and with ages of 50Myr and 2Gyr,
  respectively. The dashed curves correspond to the latter SED
  dust-reddened using Equation \ref{eq:GALext} with
  $E_{g}(B-V)$=$[0.1, 0.2, 0.3, 0.4, 0.5]$. \textit{Right:} AGN
  template by \citet{Richards2006SED} (solid line). The dashed curves
  correspond to the same SED dust-reddened using Equation
  \ref{eq:AGNext} with different $E_{a}(B-V)$ values $[0.1, 0.2, 0.3,
  3, 9]$.} 
\label{fig:temp_ex}
\end{figure*} 

\subsection{The total AGN sample and the parent galaxy sample}
\label{sec:parent}

The final combined sample analyzed in this paper consists of 1702 AGN: 
602 type--1 AGN (of which 430 with optical spectra) and 1100 type--2 AGN (of
which 549 have spectra).  A graphical representation of the sample
and its basic substructure, highlighting classification, the method
used, and overlaps among the various sub-samples, is
 shown in Figure~\ref{fig:sample} and in table \ref{tab:sample}.
The color code adopted in Fig.~\ref{fig:sample} will be
 used throughout the entire paper: blue and red colors will denote
X-ray unobscured and obscured AGN, respectively; green those Seyfert--2 galaxies
selected optically using the diagnostic diagrams, and yellow the
obscured AGN selected via \NeV\
line emission. Open and filled symbols are used instead for photometric and
spectroscopic redshifts, respectively. 
The full sample can be found in the online material. An example of the
published table is given in  
Table \ref{tab:tab1}.

Since the aim of this work is to study the AGN host galaxy properties
and compare them to normal galaxies,  throughout this paper we will use
the IRAC selected galaxy sample in the COSMOS area as comparison ``parent''
galaxy  sample. For the whole galaxy sample photometric redshifts,
masses and star-formation rates are available \citep{Ilbert2010}. 
Moreover, in order to disentangle multiple dependencies and thus isolate the
real differences in the properties of galaxies hosting an AGN from
those that do not,
we have also created two \textit{matched} parent samples: a ``mass-matched
parent sample'' and an ``i-band-apparent-magnitude-matched parent
sample''.   
In both cases we matched each AGN with 3 non-AGN chosen from the whole IRAC sample. 
For the mass-matched sample we considered the $(z, M)$ plane and for
each AGN with a given redshift and host galaxy stellar mass (see
section~\ref{sec:masses} below) we extracted the 3 non-AGN objects 
from the parent sample which lie closest in the ($z,M$) plane to the AGN allowing as maximum difference a factor of 10 in Mass.   
The same procedure was followed to build the I-band matched sample,
this time working on the $(z,i)$ plane, allowing as maximum difference a factor of 4 in luminosity.

\subsection{Multi-wavelength observations}

The Cosmic Evolution Survey (COSMOS) is a multi- wavelength
observational project over 1.4$\times$1.4 deg$^2$ of equatorial field
centered at (RA, DEC)$_{\rm J2000}$ = (150.1083, 2.210) \citep{Scoville2007}. 
The COSMOS field is characterized by an unprecedented deep
multi-wavelength coverage, and  
in order to compile the SED for the AGN sample described above, we
exploited such a large number of multi-wavelength data available in the field.
For most of the sources we have constrained the SED using 14 different
bands that encompass optical to MIR wavelengths.  
More specifically, for all our sources we used 6 SUBARU bands (B, V,
g, r, i, z; from \citealt{Taniguchi2007}), and  the CFHT/ U-, J- and
K-bands \citep{Capak2007,McCracken2010}. All but a
few ($\sim$97\%) of our sources are detected in all 4 Spitzer/IRAC bands (3.6$\mu$m,
4.5$\mu$m, 5.8$\mu$m and 8.0$\mu$m; \citealt{Sanders2007,Ilbert2010}) and
$\sim$80\%  of them are also  detected in the 24$\mu$m Spitzer/MIPS
band \citep{LeFloch2009}.  
For most of the sources we can thus constrain the SED in a very large
wavelength interval, ranging from $\sim$3800 \AA (U$_{\rm CFHT}$) to
24$\mu$m (Spitzer/MIPS).  
Moreover, for a few percent of the sample we can also rely on longer
wavelength detections. In particular, for the X-ray selected sample,
109 sources have detection at 70 $\mu$m
\citep[Spitzer/MIPS,][]{Sanders2007},  216 are visible at 100$\mu$m
and 201 at 160$\mu$m (77 of them are detected in both wavelengths) by the
\textit{Herschel} Space Observatory as part of the PACS Evolutionary
Probe
\citep[PEP\footnote{http://www.mpe.mpg.de/ir/Research/PEP};][]{Lutz2011}
guaranteed time key program.

\section{Spectral Energy Distributions of COSMOS AGN}
\label{sec:sed}

In this section, we describe the main tool in our analysis, namely the
study of the overall Spectral Energy Distribution of the entire AGN
sample. The approach we follow is to consider, with as little {\it a
  priori} assumptions as possible, each and every observed SED in the
sample as the result of the superposition of the nuclear AGN
emission and of the galactic stellar light entering the fixed
apertures of 3'' diameter used to derive all optical/IR photometry
\citep{Salvato2009}. 

The typical SED of a pure QSO, shown in the right panel of
Fig. \ref{fig:temp_ex}, is characterized by two bumps in the UV and IR
regimes \citep{Sanders1989,Elvis1994,Richards2006SED} which create a dip at around
1$\mu$m. The UV bump is interpreted as thermal emission from the
accretion disk \citep{Czerny1987}, while the IR bump is thought to be
due to absorption of intrinsic AGN UV, X-ray and optical radiation by
dusty clouds in the AGN `torus' on $\sim$pc-scales, which subsequently
re-radiate this energy at IR frequencies \citep{Barvainis1987}. 
Despite the apparent large scatter in the observed SED of large
QSO/AGN samples \citep{Richards2006SED}, there appear to be little
evidence for substantial systematic changes in the {\it shape} of such intrinsic
SED (apart from those introduced by dust extinction) 
across wide luminosity ranges, as recently demonstrated by the
analysis of broad H$\alpha$ line selected local AGN from the SDSS
(DR7) survey \citep{Stern2012}.

On the other hand, the optical SED of a galaxy (see e.g. left panel of
Fig. \ref{fig:temp_ex}) is usually modeled as due to the integrated light of the
stellar populations of the galaxy which are generated by different 
star-formation histories. Galaxy SEDs peak typically at around 1$\mu$m
for a very wide range of
SFR histories.  
%(which corresponds to the dip of the AGN SED, see Fig. \ref{fig:sed_decompose}). \\

In most AGN (with the exception of the most luminous ones) 
both components significantly contribute and the global SED is the
result of the combination of the central QSO's and the host galaxy's
SED \citep{Hao2009,Merloni2010,Lusso2011}. How much these two components
contribute to the global SED in detail 
depends on a number of factors, the most important ones being their
relative luminosity and the level of obscuration affecting each of them.

\begin{figure*}
\includegraphics[width=0.32\textwidth,clip]{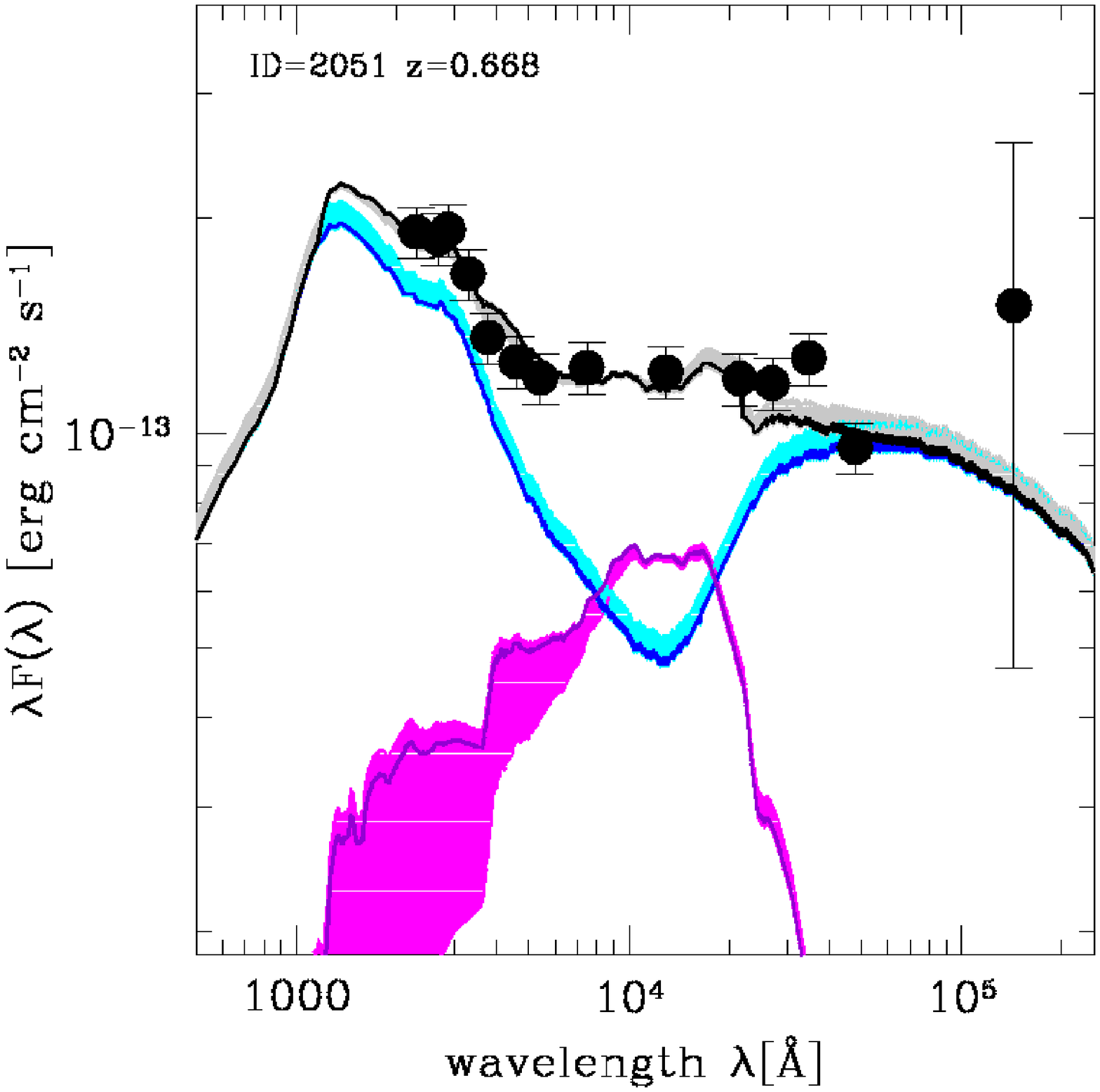}
\includegraphics[width=0.32\textwidth,clip]{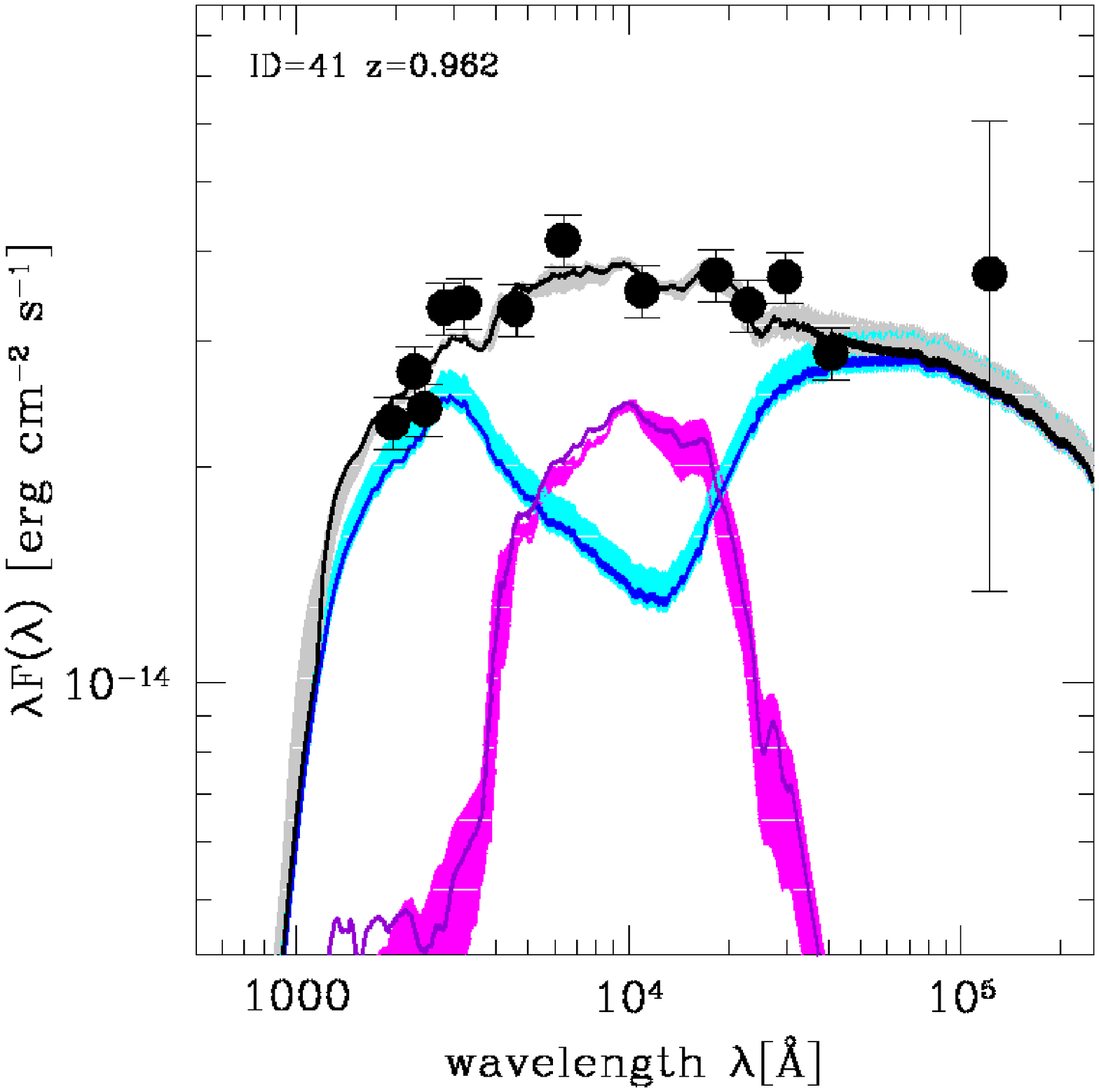}
\includegraphics[width=0.32\textwidth,clip]{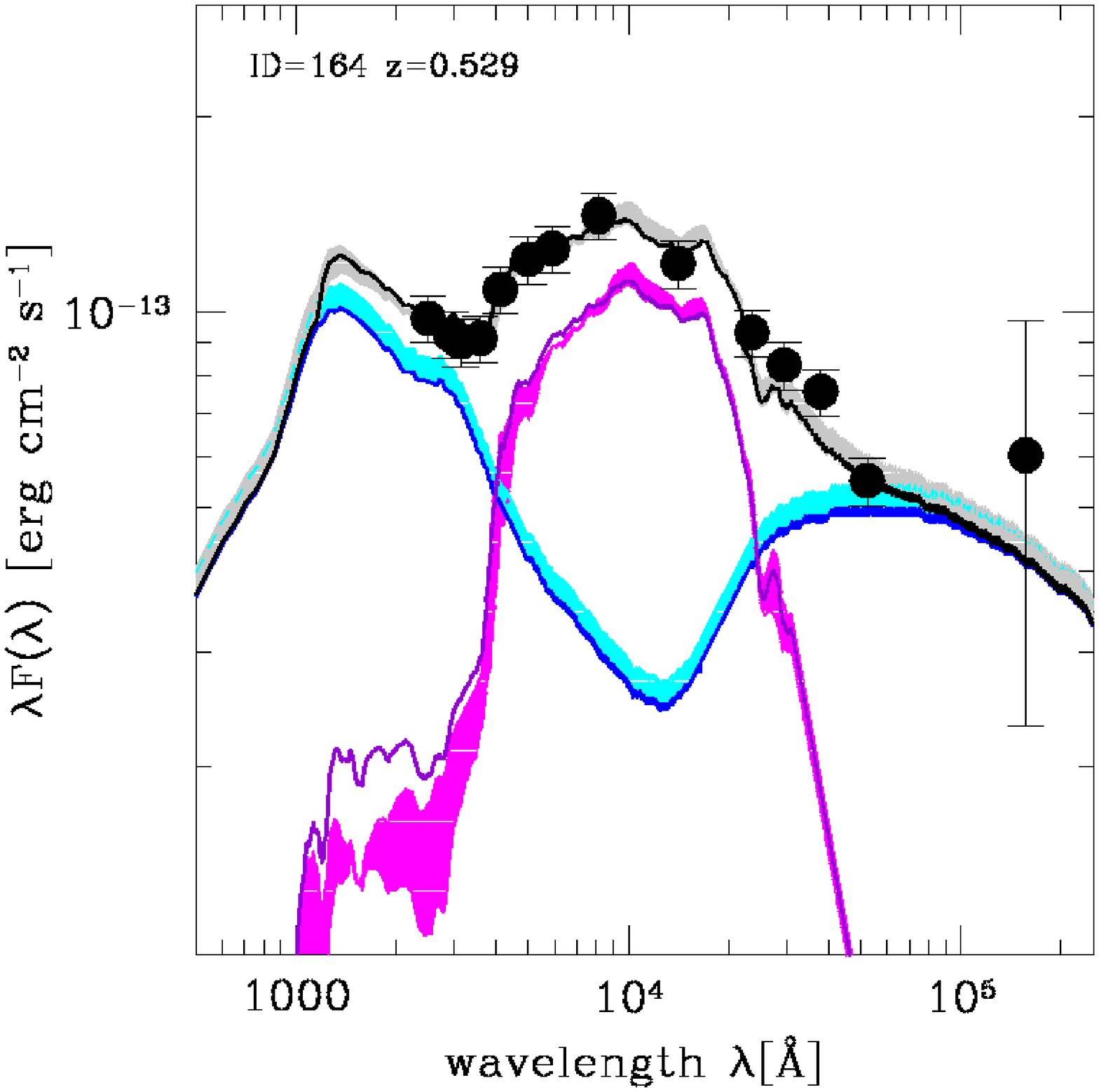}
\includegraphics[width=0.32\textwidth,clip]{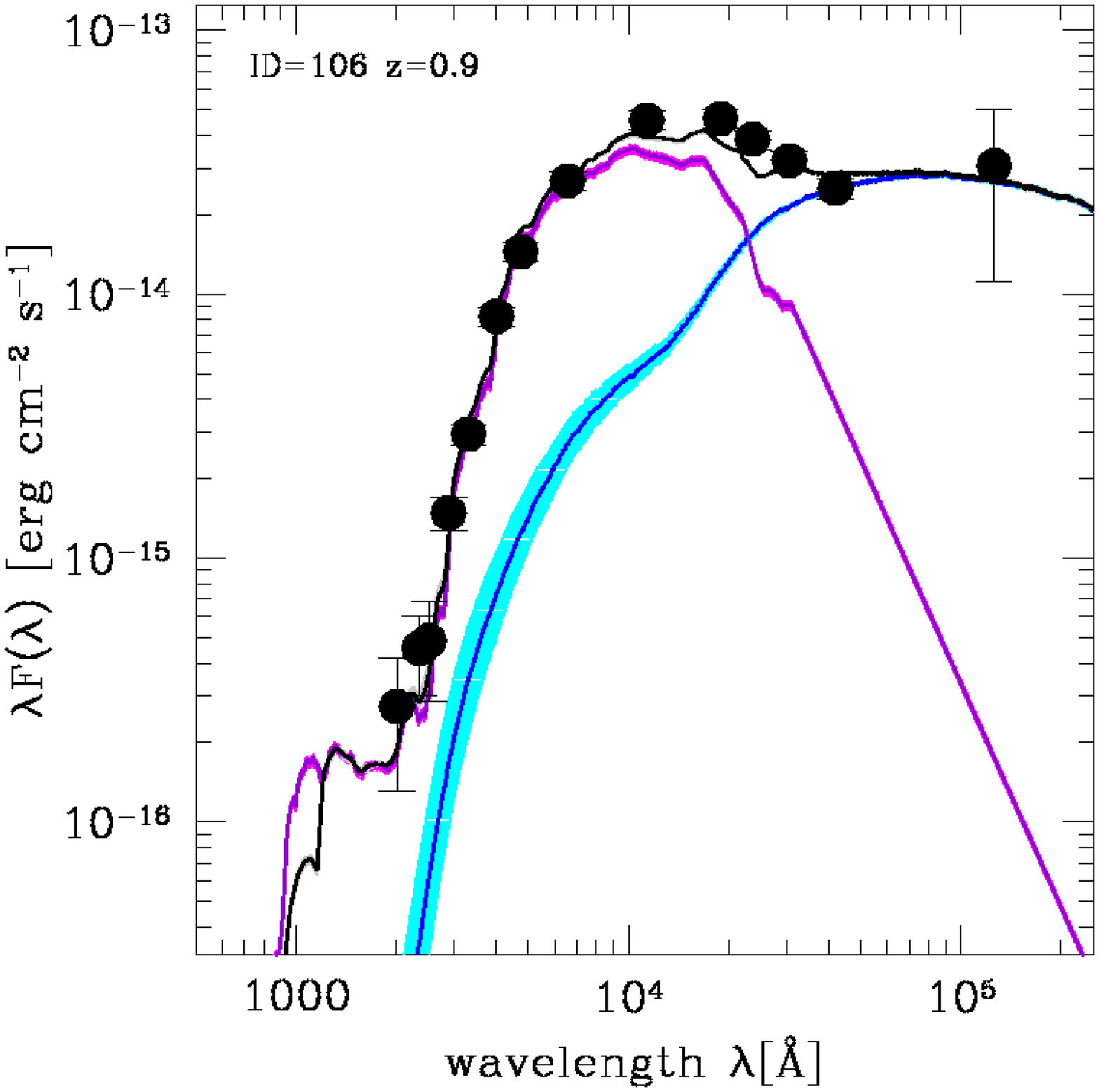}
\includegraphics[width=0.32\textwidth,clip]{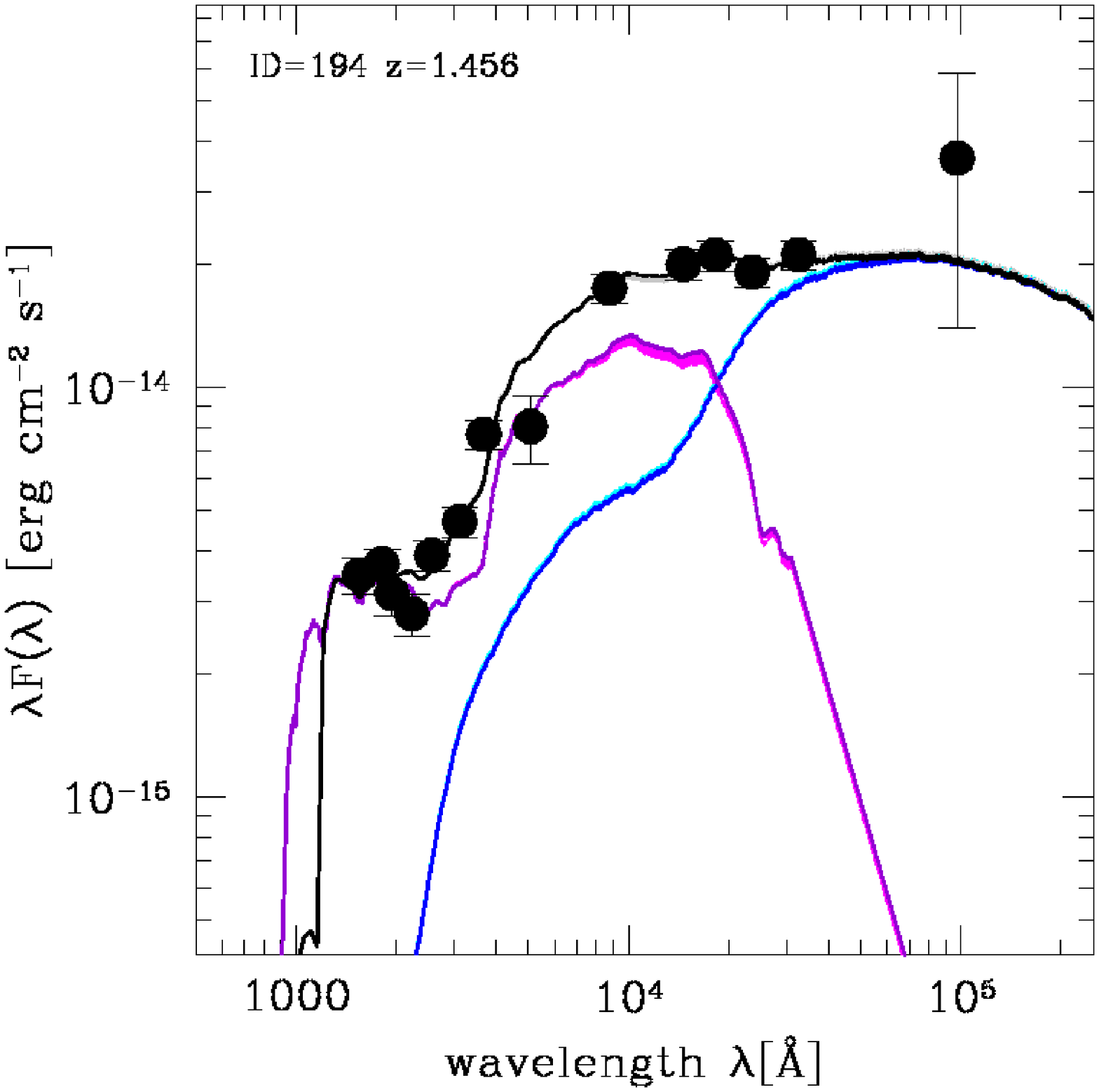}
\includegraphics[width=0.32\textwidth,clip]{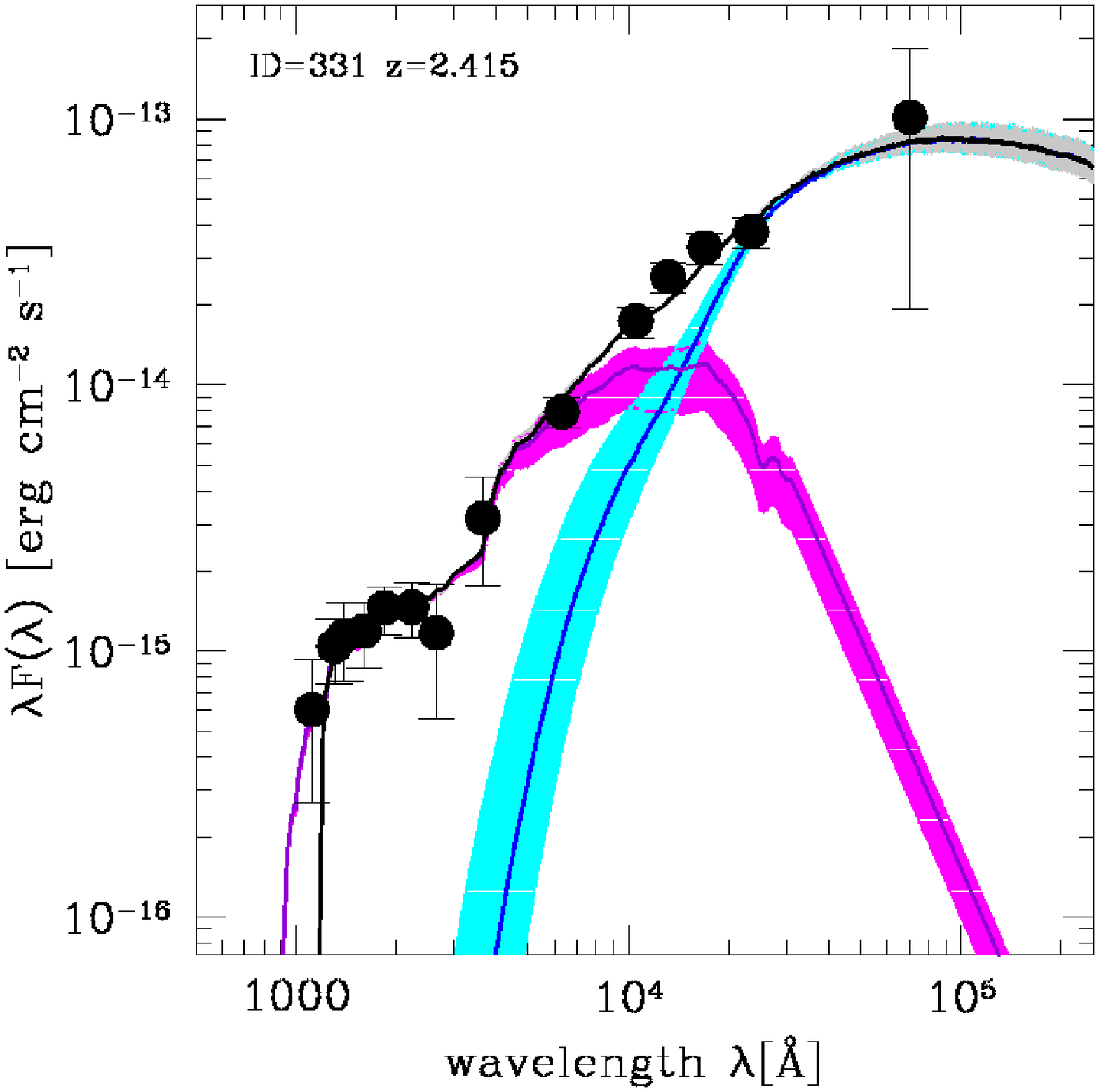}
\caption{Examples of SED decompositions for unobscured AGN (upper
  panels) and obscured ones (bottom panels). Black circles are
  rest-frame fluxes corresponding to the observed bands used to
  constrain the SED. Purple and blue lines correspond to the galaxy
  and the AGN template found as best-fit solution through the $\chi^2$
  minimization, while the black line shows their sum. Pink and cyan
  shaded areas show the range of the SED template library within
  1$\sigma$ of the best-fit template, and light gray their sum.} 
\label{fig:sed_decompose}
\end{figure*}

\subsection{Modeling the AGN and host galaxy emission components with SED fitting}
\label{sec:sed_fit}

Fixing the redshift of the source to the spectroscopic or photometric redshift, we fit the observed fluxes $f_{\rm OBS}$ with a two-component model using
a combination of AGN and host-galaxy templates
\citep[see e.g. similar approaches][]{Pozzi2010,Lusso2011} (but see Appendix 
\ref{sec:1component} for a comparison with the standard SED fitting
method without the AGN component). This fitting technique, initially presented in \citet{Bongiorno2007},
has been already successfully used in the analysis of smaller AGN
sub-samples in the COSMOS field \citep[i.e.][]{Merloni2010,Mainieri2011}. 
More complex analysis, including multiple dust emission components,
have been carried out by Lusso et al. (2012). As we mentioned before, 
the AGN and galaxy SEDs are shaped in a way that allows  a relatively
easy decoupling with the galaxy's emission peak falling in the
wavelength range of the AGN dip at $\sim 1\mu$m (rest frame).  
To be specific, we assume that: 

\begin{equation}
f_{\rm obs}=c_1f_{\rm AGN} + c_2f_{\rm GAL}
\label{eq:sed}
\end{equation}

Using a library of galaxy and AGN templates we find the 
best normalization parameters $c_1$ and $c_2$ that reproduce the
observed flux of each object, by minimizing the $\chi^2$. 
In type--1 unobscured AGN, we expect the AGN component to dominate in
the optical and IR bands with some degree of contribution from the
host galaxy, while the optical continuum of type--2 obscured AGN is
dominated by the host galaxy emission with the AGN component rising
mainly at IR wavelength.   
Notice that we are not implying that both components are always
detectable. Solutions in which the SED is a pure QSO (in case of
type--1 AGN) or a galaxy with a negligible contribution from the
central AGN (in case of type--2 AGN) are both  possible.

\subsection{Library templates}
\label{sec:library}

The data are fitted with a large grid of AGN and galaxy templates. 
For the AGN component we adopt the \citet{Richards2006SED} mean QSO
SED as derived from the study of 259 IR selected quasars with both
Sloan Digital Sky Survey and Spitzer photometry (see right panel of
Fig. \ref{fig:temp_ex}). The Richards SED is an extension and is consistent with the original \citet{Elvis1994} QSO SED. For the galaxy component we generated a library
of synthetic spectra using the models of stellar population synthesis
of \citet{Bruzual2003}.  
We assumed a universal initial mass function (IMF) from
\citet{Chabrier2003} and we built 10 exponentially declining star
formation histories (SFH) $SFR\propto e^{-t_{\rm age}/\tau}$ with
e-folding times, $\tau$, ranging from 0.1 to 30 Gyr and a model with
constant star formation.  
For each of the SFHs, the SED was generated for a grid of 13 ages
ranging from 50 Myr to 9 Gyr, subject only to the constraint that 
the age should be smaller than the age of the Universe at the redshift
of the source. Some examples of the used galaxy templates are given in
the left panel of Fig.~\ref{fig:temp_ex}. 

\subsubsection{Extinction}
\label{sec:extinction}

Both the AGN and the galaxy templates can be affected by dust extinction. 
The observed flux can be written as  

\begin{equation}
f_{\rm obs}(\lambda) = f_{\rm int}(\lambda)\times 10^{-0.4A_{\lambda}},
\label{eq:ext_gen}
\end{equation}
where $f_{\rm obs}$ and $f_{\rm int}$ are the observed and the
intrinsic fluxes, respectively, and A$_{\lambda}$ is the extinction at
a given wavelength $\lambda$ described by  
$A_{\lambda} = k(\lambda) E(B-V)$, with the color excess $E(B-V)$ and
the reddening curve $k(\lambda)$.

\subsubsection*{Galaxies:}
Recent studies on high redshift galaxies and star formation obscured
by dust have shown the importance of the reddening in the high-z
universe. Therefore, dust extinction, produced inside the galaxies
themselves, is an important effect to be taken into account. 

We chose as reddening curve for our galaxy templates the Calzetti's
law \citep{Calzetti2000}, which is the most used attenuation curve in
high-redshift studies. The Calzetti's law is an empirical relation
derived using a small sample of low-redshift starburst galaxies (SB)
and has the following form:  

\begin{equation}
k_g(\lambda) = \left\{
\begin{array}{ll} 
2.659\left(-2.156+\frac{1.509}{\lambda}-\frac{0.198}{\lambda^2}+\frac{0.011}{\lambda^3}
\right)+ R_V    \\ 
2.659\left(-1.857+\frac{1.040}{\lambda}\right)+ R_V  \\
\end{array}
\right. 
\label{eq:GALext}
\end{equation}
with $R_V=4.05$. The two equations are valid for $0.12\,\mu {\rm m}
\le \lambda \le 0.63\,\mu {\rm m}$ and  $0.63\,\mu {\rm m} \le \lambda
\le 2.20\,\mu {\rm m}$, respectively. Below and above the valid
wavelength range, the slope is computed extrapolating to lower and higher wavelengths the value of $k_g(\lambda)$ obtained by interpolating between 0.11 - 0.12$\mu$m and  2.19 - 2.20$\mu$m,
respectively.  

In the fitting procedure we considered for the galaxy 
component only $E_g(B-V)$ values in the range
$0\le E_g(B-V) \le 0.5$.  Following \citet{Fontana2006} and
\citet{Pozzetti2007}, we impose the prior $E_g(B-V) < 0.15$ if $t_{\rm
  age}/\tau > 4$ (i.e. we excluded the models implying large dust extinctions in the absence of a significant star formation rate, t$_{age}$/$\tau>$ 4).

In the left panel of Fig. \ref{fig:temp_ex} we show two examples of
galaxy templates with $\tau$=0.1 and $t_{\rm age}$=50Myr, 2Gyr  (solid
lines). For the latter template we also show the
corresponding dust-reddened templates obtained by applying the above
equations with  
different $E_g(B-V)$ values (dashed lines).

\subsubsection*{AGN:}
The extinction of the nuclear AGN light has been modeled with an 
SMC-like dust-reddening law from \citet{Prevot1984} for which:

\begin{figure*}
\includegraphics[width=0.45\textwidth,clip]{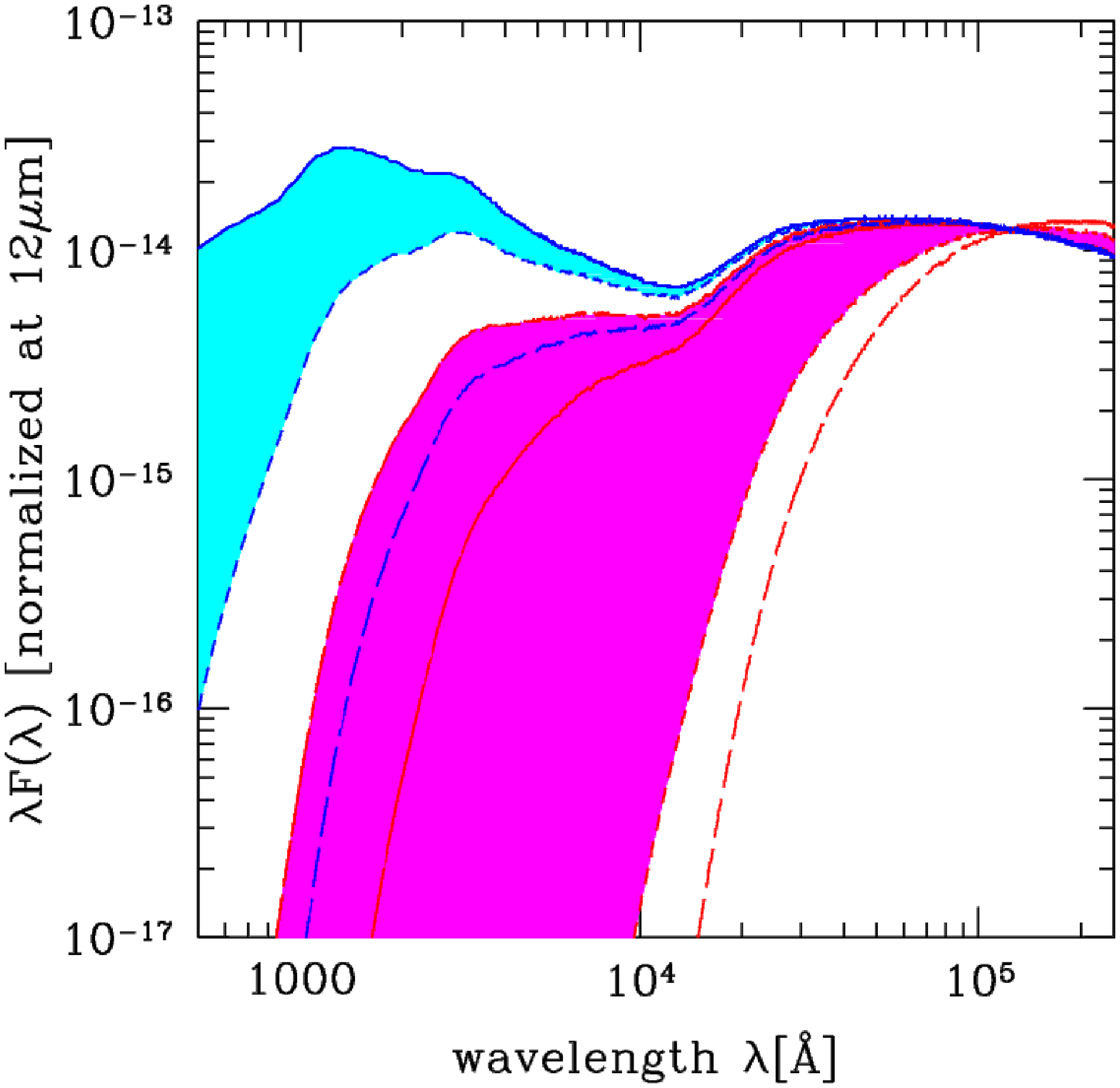}
\includegraphics[width=0.45\textwidth,clip]{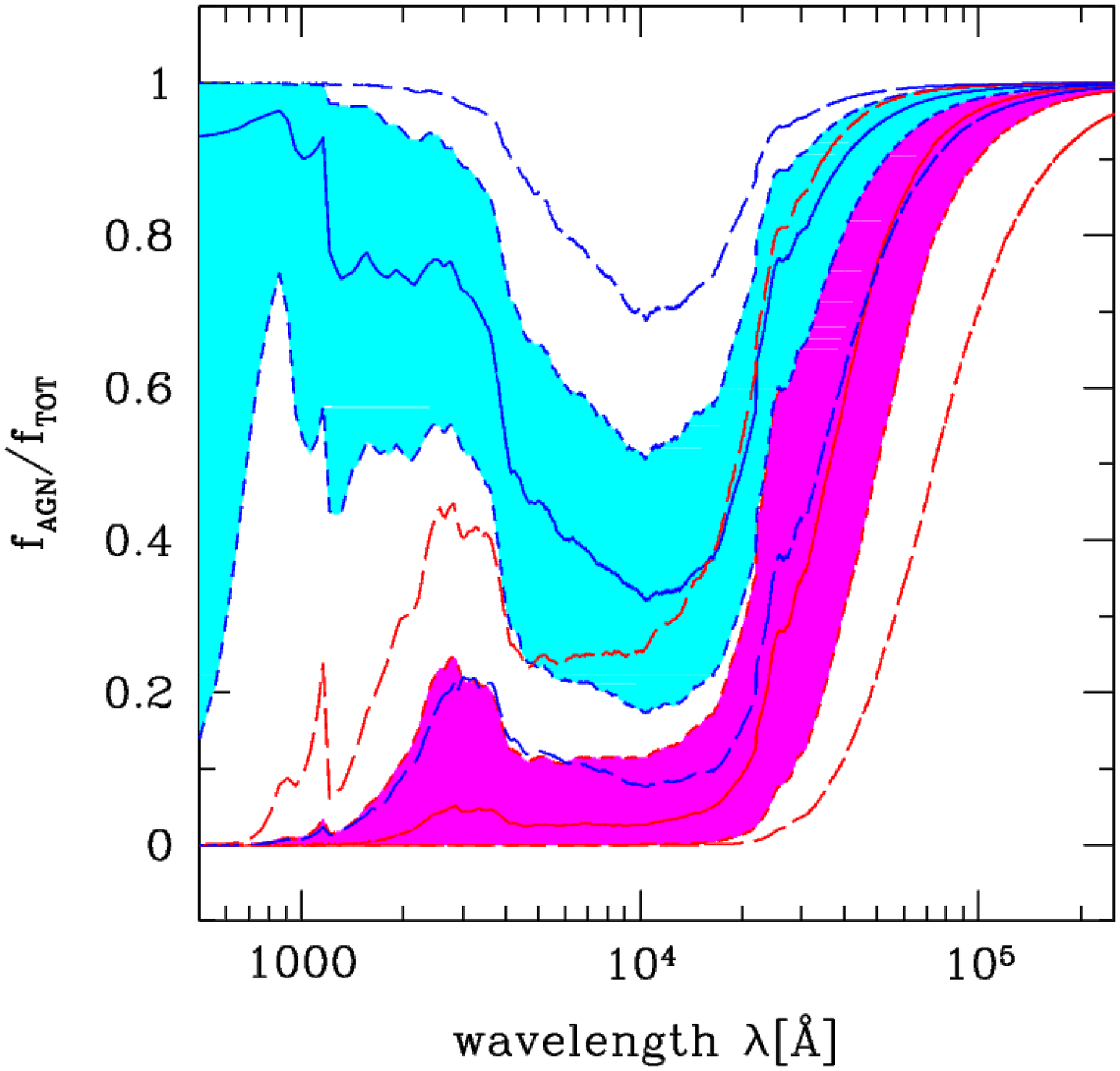}
\caption{\textit{Left Panel:} Median AGN SED chosen as best
    fit for our sample of unobscured (blue continuum line) and
    obscured (red continuum line) AGN, all normalized at rest-frame 12
    $\mu$m. Shaded areas correspond to the
    range within
    25th and 75th percentiles, while long dashed lines that within
    the 10th and 90th
    percentiles. \textit{Right:} Median AGN contribution to the total
    SED at different wavelengths for unobscured (blue continuum line)
    and obscured (red continuum line) AGN. Shaded areas correspond to
    the range within the 25th and 75th percentiles, while long dashed
    lines that within the 10th and 90th percentiles.}  
\label{fig:sed_fraction}
\end{figure*}

\begin{equation}
k_a(\lambda) =  1.39 \ \lambda_{\mu{\rm m}}^{-1.2}
\label{eq:AGNext}
\end{equation}

We considered $E_a(B-V)$ values in the range $0\le E_a(B-V) \le 1$ for
unobscured type--1 AGN and  $0.3\le E_a(B-V) \le 9$ for the obscured
ones, in steps of $\Delta E_a(B-V)=0.1$.
In the right panel of Fig. \ref{fig:temp_ex} we show the AGN template
(solid line) and the  
corresponding dust-reddened templates obtained for 
different $E_a(B-V)$ values. As visible in this figure, the upper
limit of $E_a(B-V)=9$ corresponds to a very extinct AGN SED, and an 
even higher upper limit would not change by much the SED shape. 

Taking into account the described extinction laws, we can re-write Eq. \ref{eq:sed} as

\begin{equation}
\begin{split}
f_{\rm obs}&=c_1 f_{\rm AGN}^{\rm int}(\lambda)\times 10^{-0.556 \ \lambda_{\mu m}^{-1.2} E_a(B-V)}\\ 
&+ c_2f_{\rm GAL}^{\rm int}(\lambda)\times 10^{-0.4k_{g}(\lambda) E_g(B-V)} 
\end{split}
\label{eq:sed2}
\end{equation}
with k$_{g} (\lambda)$ described by Eq. \ref{eq:GALext}

\begin{figure*}
\includegraphics[width=0.8\textwidth,clip]{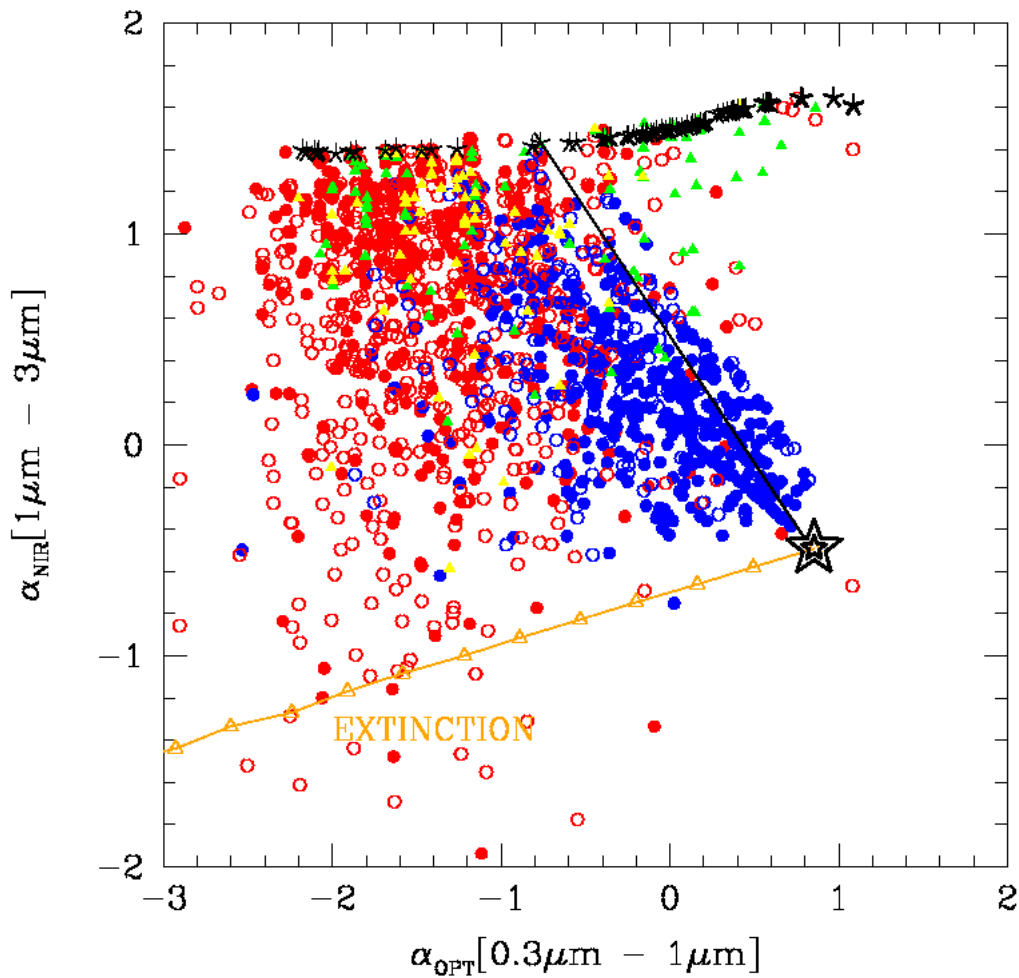}
\caption{\textit{Mixing Diagram from \citet{Hao2010}}: slope -  slope
  plot of the entire sample of analyzed AGN.  Circles correspond to
  XMM-COSMOS obscured (red) and unobscured (blue) AGN;  green and
  yellow triangles show the optically selected AGN (selected using the
  BPT diagram and the NeV line). Open and filled circles correspond to
  photometric and spectroscopic redshifts. The big star show the R06
  mean SED while the asterisks show different BC03 galaxy
  templates. The black line indicates the boundaries of the possible
  slopes obtained  by mixing the R06 template with different fractions
  (0\% - 100\%) of a given galaxy template. The orange line shows the
  reddening vector of R06 with the triangles corresponding to steps of
  0.1 in E(B-V).} 
\label{fig:HHplot}
\end{figure*}

\subsection{Results of the SED fitting}
\label{sec:results_sed_fit}

Given the wide multi-wavelength coverage, the fitting technique
described above
allows to decompose the entire spectral energy distribution into a
nuclear AGN and a host galaxy component and to derive robust
measurements of both the AGN and the host galaxy properties, for
almost all (98.6\%) of the objects in the original sample\footnote{In fact, 
due to an insufficient photometric coverage, 5 sources have been
excluded from the analysis, while for 18 sources no good solution
could be found due to problems in the photometry or in the SED: the
reduced $\chi^2$ in these cases is large ($>$20). For these sources no
physical parameters could be derived.}.  
Some examples of the SED fitting for unobscured (upper panels) and
obscured AGN (bottom panels) are shown in
Figure~\ref{fig:sed_decompose}, while the left panel of figure
  \ref{fig:sed_fraction} shows the median SED chosen as best fit in
  the sed fitting for unobscured (blue) and obscured (red) AGN
  together with the 10th, 25th, 75th and 90th percentiles of the
  samples. Finally,
  the right panel of the same figure highlights the median AGN
  contribution to the global SED at different wavelengths for
  unobscured (blue continuum line) and obscured (red continuum line)
  AGN. Again the shaded areas and the long dashed lines mark the 10th,
  25th, 75th and 90th percentiles of the samples, respectively. 
From this figure, we
see that, for obscured AGN, three quarters of the objects have AGN
contribution in the optical bands smaller than ~20\%. Thus, any residual AGN
contamination (or AGN over-subtraction), should not affect the
determination of the rest-frame optical colors of the host by more
than about 0.1 magnitude. On the other hand, at rest-frame wavelengths
of about 1$\mu$m, three quarters of type--1 AGN have AGN fractions
smaller than about 50\%, implying that, even for unobscured objects,
our method allow a robust determination of the host total stellar
masses (see also Merloni et al. 2010). 
   
Based on the best fit solution we derived the AGN
luminosity and the galaxies' rest-frame magnitudes, colors,
 stellar mass content and star formation rate.  
One sigma errors on the best fit parameters are computed considering
the range of values corresponding to the solutions for which
$\Delta\chi^2 = \chi^2(sol)-\chi^2(best)\le 1.0$, corresponding to
1$\sigma$ in the case of one parameter of interest \citep{Avni1976}. 
 
In some cases we are not able to derive a measurement for the galaxy
or the AGN components but we can still define a meaningful upper limit
on the unconstrained component.  
In unobscured type--1 AGN, if the rest-frame K-band luminosity of the
host galaxy is smaller than 10\% of the AGN luminosity in the same
rest-frame band, we assign an upper limit to the host galaxy SED
component. This is the case for 34 type--1 AGN for which the value of
the upper limit of the galaxy K-band luminosity is taken as the
highest possible value of the parameter within the
uncertainty. In these cases 
we assign an upper limit also to the object's total stellar mass
assuming the median mass-to-light ratio of all other objects in the
sample. No rest-frame magnitudes in U- and B-band are derived in
these cases.

On the other hand, for 279 obscured type--2 AGN which are not detected
in the 24$\mu$m Spitzer/MIPS band, we can provide only upper limits
on the AGN component (and thus on the AGN luminosity). This is because
in heavily obscured sources the AGN component rises almost exclusively at NIR
wavelengths, and without the 24$\mu$m detection it is not possible to
determine reliably the AGN contribution to the optical/IR SED. 
For these sources, we derived an upper
limit on the 12$\mu$m rest-frame AGN luminosity (see below) using the
24$\mu$m detection 
limit and applying the appropriate $k$-$correction$. We note here that
in these cases, we can still use the observed X-ray (or emission line)
luminosity to infer the level of AGN strength in the source.

Summarizing, from the initial sample of 1702 sources the SED fitting
was successfully applied to 1679 sources. Of these objects, 34 type--1
AGN have only an upper limit in the galaxy component parameters (mass
and rest-frame M$_K$) while no U- and B-band rest frame magnitudes were
assigned. For 279 type--2 AGN only an upper limit in the AGN component
parameters (L$_{12.3\mu{\rm m}}$ and thus L$_{\rm bol}$) could be derived. All
the derived parameters are available online.  
An example of the published table can be found in Table \ref{tab:tab2}.

\subsection{AGN-galaxy Mixing Diagram}

Based on our SED fitting procedure described in the previous section, 
Figure \ref{fig:HHplot} shows the rest-frame NIR and optical slopes of
the total best fit SED of the sources in our sample \citep[adapted from][]{Hao2010}.
For each source we plot the slopes on either side of the rest-frame
1$\mu$m which is approximately the inversion point in the typical QSO
SED in a $\nu$f($\nu$) versus
$\nu$ representation (i.e. where the exponent of a locally power-law
fit to the SED changes sign). In particular, we computed the IR slope
($\alpha_{\rm NIR}$) between 1$\mu$m and 3$\mu$m and the optical slope
($\alpha_{\rm OPT}$) between 0.3$\mu$m and 1$\mu$m \citep[for
more details see][]{Hao2010,Elvis2012}. The location of the sources in this
'color-color' space is then shown in Fig.~ \ref{fig:HHplot}, where 
circles represent XMM-COSMOS obscured (red)
and unobscured (blue) AGN, while triangles correspond to the optically
selected sources  without X-ray counterpart (green for AGN selected
with the diagnostic diagrams and yellow for NeV selected AGN). Open
and filled circles correspond to photometric and spectroscopic
redshift. Note that the same symbols and color-code is used throughout the
paper.

\begin{figure*}
\includegraphics[width=0.48\textwidth,clip]{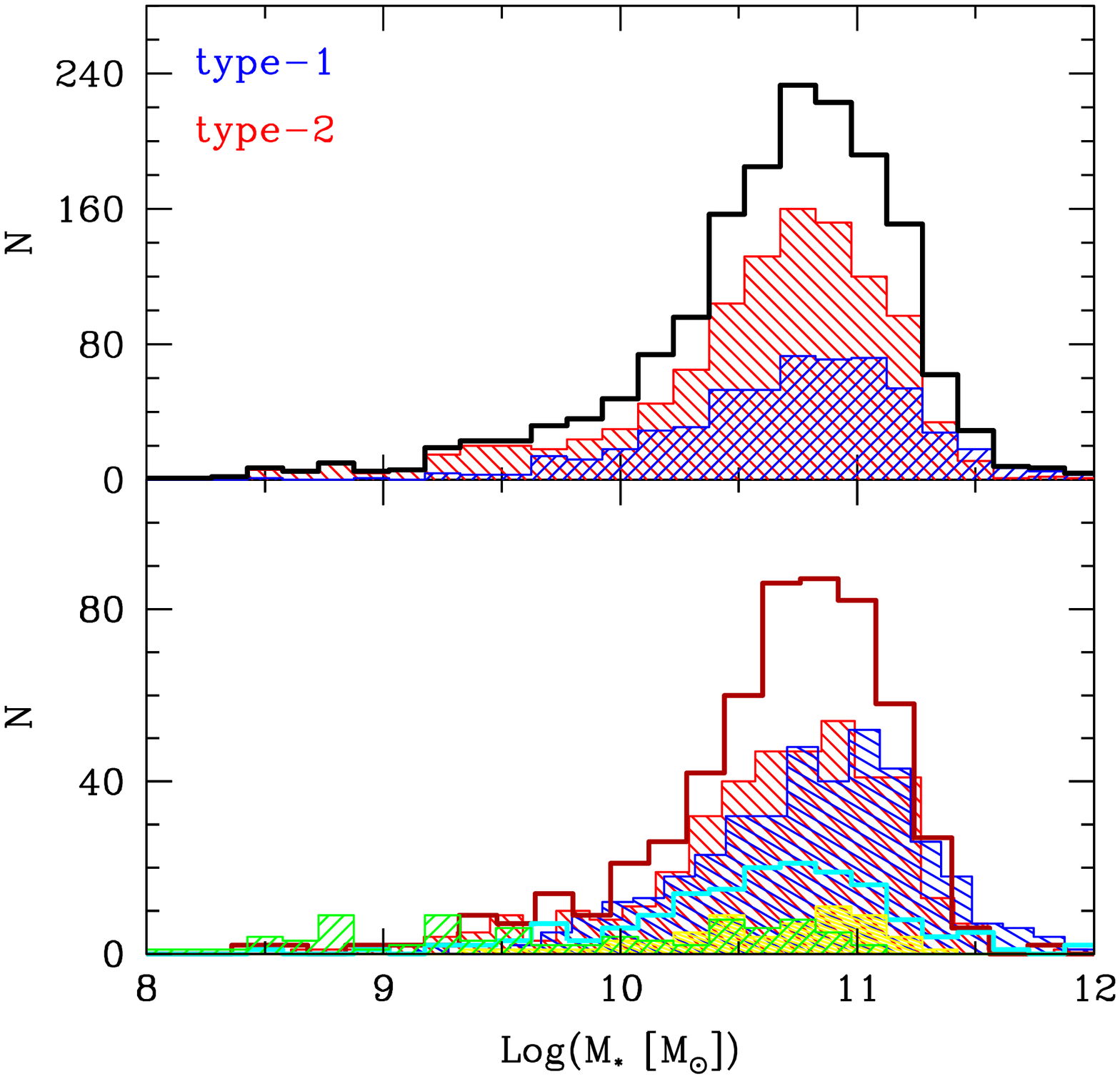}
\includegraphics[width=0.48\textwidth,clip]{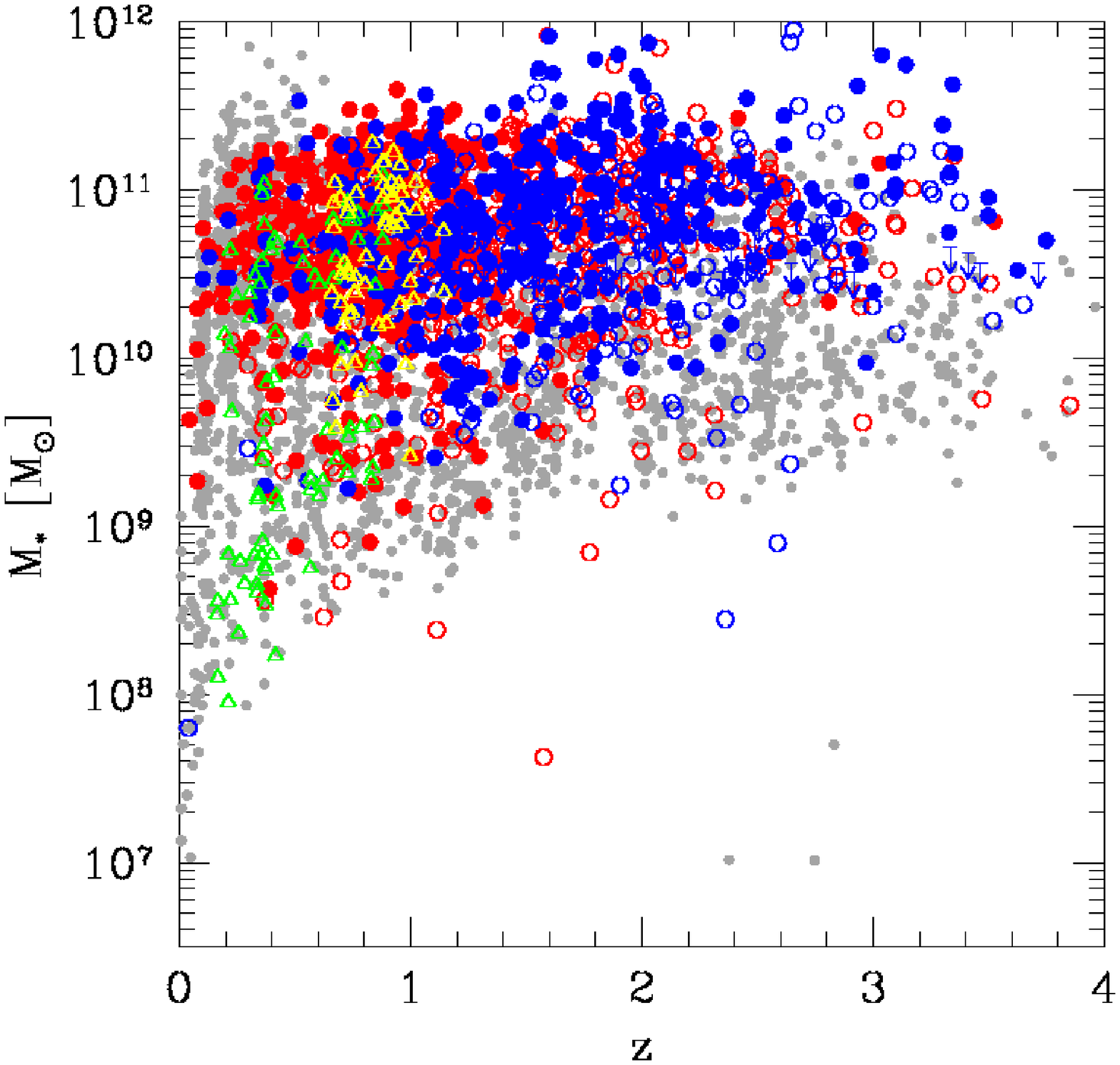}
\caption{\textit{Left:}  In the lower panel we show the stellar mass
  distribution of the AGN host galaxies for the different analyzed
  sub-samples: shaded blue and red  for spectroscopically confirmed
  X-ray type--1 and type--2 AGN, empty cyan and red for X-ray selected
  type--1 and type--2 AGN with photometric redshift, dashed green and
  yellow for optically selected through emission lines (green: BPT
  diagram and yellow: NeV line) without X-ray detection. The top
  panel shows instead the entire sample (thick open histogram) and
  divided in type--1 (shaded blue) and type--2 (shaded red) AGN. 
\textit{Right:} Host galaxy
  stellar masses versus redshift for the entire AGN sample with the
  usual color-coded sub-classes division. We also show in gray the
  distribution of masses for the I-band-apparent-magnitude-matched
  parent galaxy population.}
\label{fig:mass_histo}
\end{figure*}

In the plot, the big star indicates the SED of our pure type--1 AGN
template, for
which $\alpha_{\rm OPT}$=0.85 and $\alpha_{\rm NIR}$=-0.49 \citep[][hereafter
R06]{Richards2006SED}. Starting from this point, the orange line shows
the R06 template when reddening is applied (the orange triangles
correspond to steps in E(B-V) of 0.1). The asterisks show the 
position in this diagram of different galaxy templates derived from
the stellar population 
synthesis models of \citet[][hereafter BC03]{Bruzual2003} adopted:
from passive to extreme starbursts going from left to right. The black
line indicates the R06 template when mixed with different 
fractions (0\% - 100\%) of a given galaxy template.  This mixing curve
defines the boundaries of the possible slopes obtained by mixing the
R06 with this particular galaxy template. 
%All the type--1 AGN for which we cannot derive any reliable
%measurement because of that (98, see  Sec. \ref{sec:result}) are
%marked with an inner gray circle. 
This plot shows the three main components of the typical SED of the
analyzed sources (AGN, host galaxy and extinction) and their relative
contributions to the total SED and serves as a powerful illustration
of the level of ``contamination'' of the optical/IR SED of AGN, at
least at the luminosity levels probed by the COSMOS survey. 
%points out that most of the SED can be explained as a combination of
%a pure AGN possibly extinct and/or contaminated by the host galaxy
%light. 
%Driven and justified by this, we examined the global SED of each
%object trying to disentangle these three components.   
As visible in figure, for type--1 unobscured AGN the galaxy
contamination is less predominant than in type--2 obscured AGN, where
high value of extinction are associated to high contamination of the
host galaxy which can dominate in the 0.3$\mu$m - 3$\mu$m
interval. Moreover, there are no objects along the AGN extinct line
at E(B-V)$>$1 showing that when the AGN absorption is high, the galaxy
contribution becomes important.  

Most of the sources lie on the left-side region of the $\alpha_{\rm OPT} -
\alpha_{\rm NIR}$ plot, and only a small percentage (and only obscured AGN)
lie in the upper right corner. This is a consequence of the fact that
in nature galaxies with high star formation rates typically show high levels of
obscuration, too. This means that, while we do include non-extinct 
starburst templates in our library, these solutions are almost never chosen
because unrealistic (unless t$_{age}$ or $\tau$/t$_{age}$ are very low, as is
the case for the few points in that region of the plot).  

Unobscured AGN in this region of the
plot have been interpreted by \citet{Hao2010} as “hot-dust-poor” (HDP)
quasars i.e. normal quasar with a relatively weak IR bump. However,
for the obscured ones, another possible explanation is that these
sources are normal AGN hosted by starburst galaxies which steepen the
UV slope.  
Looking one by one at the SED of all these sources we found that all
but four of them are not detected at 24$\mu$m and their SED in the UV-optical
part is dominated by a starburst galaxy. Most of
these sources are Seyfert--2 galaxies optically selected using BPT
diagrams. Some of them, due to the errors associated to the line
measurements, can be misclassified normal star forming
galaxies. Alternatively they could be very faint AGN hosted in
starburst galaxies.  

\subsection{Measuring AGN host stellar masses and star formation rates}
\label{sec:masses_sfr}

\subsubsection{Stellar masses}
\label{sec:masses}
The SED fitting procedure allows us to estimate the total stellar mass     
and the star formation rate (SFR) of the AGN host galaxies. Knowing
the IMF, for a given star-formation history, each combination of
$\tau$ and $t_{\rm age}$ (see Sec.~\ref{sec:sed_fit}) is in fact
uniquely associated to a value of specific star formation rate
(sSFR=SFR/M, see Sec. \ref{sec:AGNfr_sfr}) 
and, given the normalization of the template, total stellar mass.

For our AGN COSMOS sample, we have thus derived 
a robust estimate of the host galaxy stellar masses for 1650 objects 
(out of a total number of 1702 attempted SED fits). 
As explained in Sec.~\ref{sec:sed_fit}, for 34
type--1 AGN we could derive only an upper limit on the stellar mass, while for
18 objects the SED fit failed due to problems in the photometry or in
the SED shape, so no mass estimation was possible.
 
\begin{figure*}
\includegraphics[width=0.9\textwidth,clip]{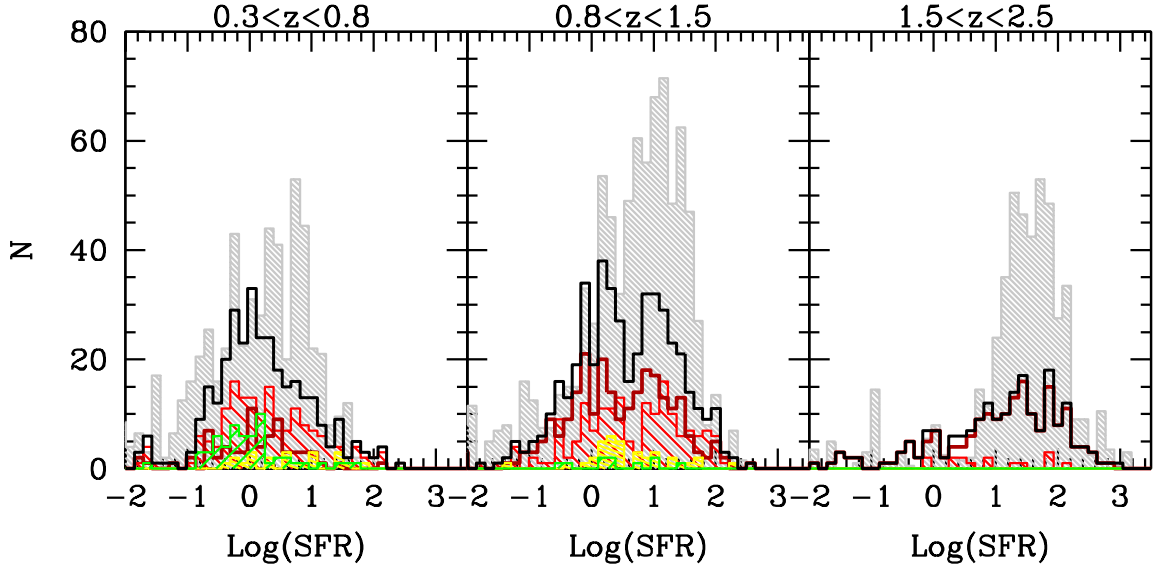}
\caption{Star formation rate distributions in three redshift bins of the AGN host
  galaxies for the entire type--2 AGN sample (thick open histogram)
  and divided in different sub-samples: shadow and empty red for X-ray
  spectroscopically confirmed and with only photometric redshift
  type--2 AGN,  dashed green and yellow for optically selected through
  emission lines (green: BPT diagram and yellow: NeV line) without
  X-ray detection. The gray histogram shows as comparison the distribution of the mass-matched parent sample. To make easier the comparison, the latter has been divided by a factor of two.} 
\label{fig:SFRhisto}
\end{figure*}

The top left panel of figure \ref{fig:mass_histo} shows the host
galaxies stellar mass
distribution for the entire AGN sample (open histogram) and divided in
type--1 (hatched blue) and type--2  (hatched red) objects. 
The bottom panel shows a detail of the $M_{*}$ distribution in the different
sub-samples, color-coded as usual. 
Masses range mainly from 10$^{10}$M$_{\sun}$ to 10$^{11.5}$M$_{\sun}$
with a peak at 10$^{10.9}$M$_{\sun}$. No significant difference is found between
the type--1 and type--2 AGN host population which show on
average the same host galaxy mass distribution.  
Interestingly, the hosts of optically selected Seyfert--2 galaxies without X-ray
counterpart 
show a tail in the distribution at lower masses
down to 10$^{8.5}$M$_{\sun}$.  

The right panel of Fig. \ref{fig:mass_histo} shows instead 
the stellar masses of the AGN hosts as a function of redshift
superimposed on the underlying distribution of stellar masses of
non-AGN in the COSMOS area drawn from \citet{Ilbert2010} and matched
to the AGN population in redshift and I-band apparent magnitude (see
Sec. \ref{sec:parent}).

\subsubsection{Star formation rate}
\label{sec:sfr_main}

The measure of the host galaxy SFR is subject to considerably larger
uncertainties than that of stellar mass, in particular for type--1 AGN. 
In general, the most reliable way to compute the SFR is summing up the
rate of unobscured star formation (emitted in the UV) with the
obscured one re-emitted in the (mid- and far-) IR by dust (SFR$_{\rm UV+IR}$). 
Our optical/IR SED fitting procedure relies only on the UV emission measurements, which
traces only the unobscured SF, which is however scaled up by the dust correction
factor computed using the full SED to account for the latter
contribution.

The issue of the reliability of UV-based indicators of star formation
rates in galaxies is a long-standing one. With the advent of the {\it
  Herschel} telescope, sensitive Far-IR measurements of cold dust
heated by star-forming processes has become available for deep and
medium-deep surveys of the extragalactic sky, providing a first direct
handle of SFR$_{\rm UV+IR}$ in large samples of galaxies \citep[see
e.g.][]{Rodighiero2010}. 
For complete samples of AGN selected in both X-rays and optical
spectroscopic surveys, however, the final answer to the question of
what are the intrinsic star-forming properties of the AGN hosts
requires SFR indicators that are sensitive to both star-forming
galaxies and passive/quiescent ones. This is not possible even
for {\it Herschel}, whose sensitivity to individual galaxies does not
reach into the quiescent population at the redshifts ($z>1$) of deep
extragalactic surveys such as COSMOS \citep[see
e.g.][]{Shao2010,Mullaney2012}. Indeed, only $\sim$10\% of the AGN in
our sample are detected above the $3\sigma$ limits of the PEP
catalogues in the COSMOS field.
In Appendix \ref{sec:sfr} we compare the SFR estimated from the SED
fitting with the one derived with other methods (i.e. FIR observation
and optical emission lines). We find that above $\sim$20 M$_{\sun}$
yr$^{-1}$ the agreement between the SFR derived from the SED and the
one derived from the FIR bands is quite good, while at lower SFR
($<20 M_{\sun}$ yr$^{-1}$) the disagreement becomes evident, with SFR(FIR) being systematically higher than SFR(SED).
We interpreted this discrepancy as a combination of a possible
contamination from SF in the NIR not properly taken into account in
the SED fitting together with a not-negligible AGN contamination to
the FIR bands not taken into account in the FIR SFR estimates\citep[see e.g.][]{Rosario2012}. This would imply that, while our SFR estimates can be in some cases
underestimated, the FIR-SFR can be overestimated especially at lower
SFR where the AGN contamination would represent a significant fraction of the IR emission.

In any case, the very nature of our objects, and the SED decomposition
technique with which we analyze them, require some special care in the
interpretation of the SFR measurements.
The main problem with type--1 (unobscured) AGN is that the
accretion disk emission strongly contributes in the UV band (see
upper panels of Fig. \ref{fig:sed_decompose}) introducing a degeneracy
in the SED fitting between the UV emission from star formation and
from the central AGN. In practical terms this means that, in 
some cases the SED fitting can give two significantly different 
solutions with a similar  $\chi^2$: an unobscured AGN, with a
prominent Big Blue Bump (BBB) dominating in
the UV range with a passive/moderately star forming galaxy or,
conversely, a moderately obscured AGN with a
strongly star-forming galaxy, which substantially contributes to the
observed UV emission. Since it is not possible to disentangle \textit{a priori} this
kind of degeneracy, we will not take into consideration the SFR
measurements of type--1 AGN for the further discussion. 

On the contrary, since in obscured AGN the UV emission is suppressed
by obscuration (see, as an example, the bottom panels of Fig. \ref{fig:sed_decompose}),
the UV range is clean from AGN contamination and a reliable estimate
of the SFR can be derived from the SED.

Figure \ref{fig:SFRhisto} shows the host galaxy SFR distribution in three redshift bins for
1090 obscured type--2 AGN (thick solid line) and for the different
sub-samples with the usual color code. AGN host galaxies span a very
wide range of SFR, with X-ray
selected AGN populating the entire range, while optically selected
Seyfert--2 galaxies (which are mainly at low redshifts) are hosted
predominantly in galaxies with low SFR (but very low masses as 
shown in Fig. \ref{fig:mass_histo}). This can highlight a physical
difference in the hosts of optically selected Seyfert--2 galaxies or
can be, at least partly, driven by the fact that high SFR implies
strong emission line  dilution which causes a certain  fraction of
AGN to be missed by the emission line ratios (BPT)  selection.

\begin{figure}
\includegraphics[width=8cm,clip]{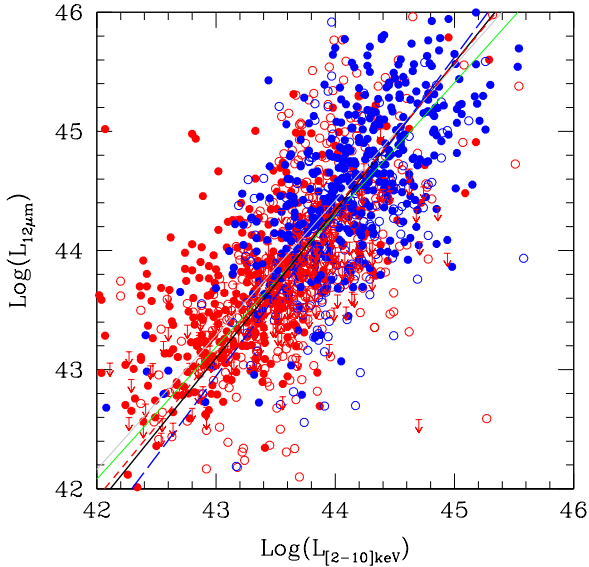}
\caption{X-ray [2-10] keV versus mid-infrared 12$\mu$m luminosities
  for the sample of $\sim$1550 AGN with X-ray detection and measured
  L$_{12\mu{\rm m}}$. Red circles are spectroscopic (filled) and
  photometric (open) type--2 AGN while type--1 AGN are represented
  with filled (spectroscopic) and open (photometric) blue circles.  
The thin solid green diagonal line is the [2-10]keV - 12.3$\mu$m
luminosity relation found by \citet{Gandhi2009} for the sample of 22
well resolved local Seyfert. The thick solid lines are the fitted
correlation considering our entire sample (black) and only the spectroscopic sources (gray), while the long-dashed blue
and the short-dashed red correspond to the type--1 and type--2
sub-sample.}
\label{fig:L12LX}
\end{figure}

\begin{figure}
\includegraphics[width=0.45\textwidth,clip]{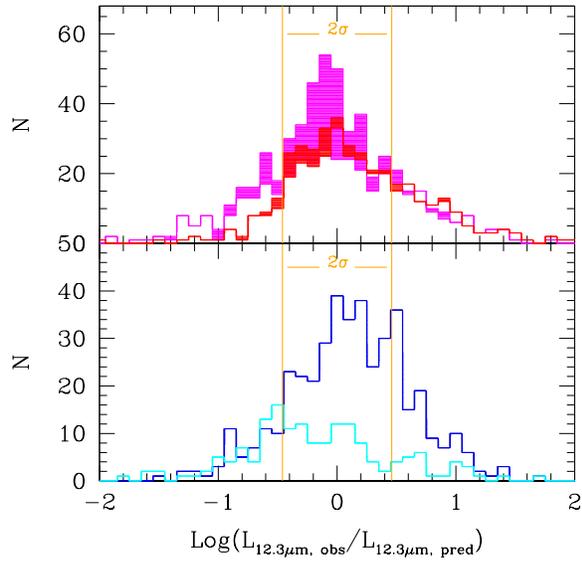}
\caption{Histograms of the ratio between the measured 12.3 $\mu$m
  luminosity and the one predicted by the \citet{Gandhi2009}
  relation on the basis of the observed 2-10 keV X-ray luminosity. 
In the top panel we show type--2 AGN, with the red histogram marking
objects in the spectroscopic sample and the purple one those with
photometric redshifts; filled histograms are for upper limits. 
In the bottom panel we show instead the type--1
AGN, with the dark histogram marking
objects in the spectroscopic sample and the cyan one those with
photometric redshifts. Yellow histograms in both panels show the
outliers in the mixing diagram of Fig.~\ref{fig:HHplot} (see text for
more details).
The two solid orange lines delimit  2$\sigma$ from the Gandhi relation
(mean equal to zero and standard deviation $\sigma$=0.23).}  
\label{fig:L12obspred}
\end{figure}

\subsection{Measuring nuclear L$_{12\mu m}$ and bolometric luminosities}
\label{sec:L12LX}

Before proceeding to a comprehensive discussion of the observable
(and physical) properties of the host galaxies of the AGN selected in
COSMOS, we present some basic ``intrinsic'' properties of the AGN
population itself. This will be also a useful reminder that, in terms
of the physical characteristics of growing black holes and of their
immediate (nuclear) environments, every large survey selects families of
objects occupying specific parts of the generic parameter space
identified by the distance of the source, its accretion luminosity, BH
mass, and level of nuclear obscuration.

To begin with, we note that 
from the best fitting (obscured or un-obscured) AGN component in the SED, 
we can derive an estimate of the
AGN luminosity at (rest-frame) 12$\mu$m which can be used as a reliable estimator of
 the AGN bolometric luminosity.
The intrinsic AGN 12$\mu$m luminosity is calculated by integrating the
AGN component obtained from the best SED fit over a narrow (i.e.,
1$\mu$m) pass-band centered on 12$\mu$m.  
By construction, and given the shape of the BC03 templates used to fit
the galaxy component in every system, the mid-IR band is assumed to
be completely dominated by the nuclear AGN light reprocessed in
the obscuring molecular structure on $\sim 1-10$ pc scale.  
Such a drastic assumption is likely not to be correct for a number of individual objects, 
which may have a significant galaxy contribution (not included in BC03
templates) at 12$\mu$m. However, globally it is not unreasonable for
the studied AGN sample as suggested in
Fig. \ref{fig:L12LX}, where the AGN 12$\mu$m luminosity is plotted
against the intrinsic (de-absorbed)  
X-ray [2-10] keV luminosity\footnote{Calculated on the basis
    of the observed fluxes in the soft (0.5-2 keV) and hard (2-10 kev)
    X-ray energy bands, of the known redshift, and assuming, as X-ray spectral
    model, an intrinsic power-law of slope  $\Gamma=1.9\pm0.2$, absorbed by
    neutral gas.} for the XMM selected sample
of $\sim$1550   
objects with X-ray detections and L$_{12\mu{\rm m}}$ measurements. 

\begin{figure}
\includegraphics[width=0.48\textwidth,clip]{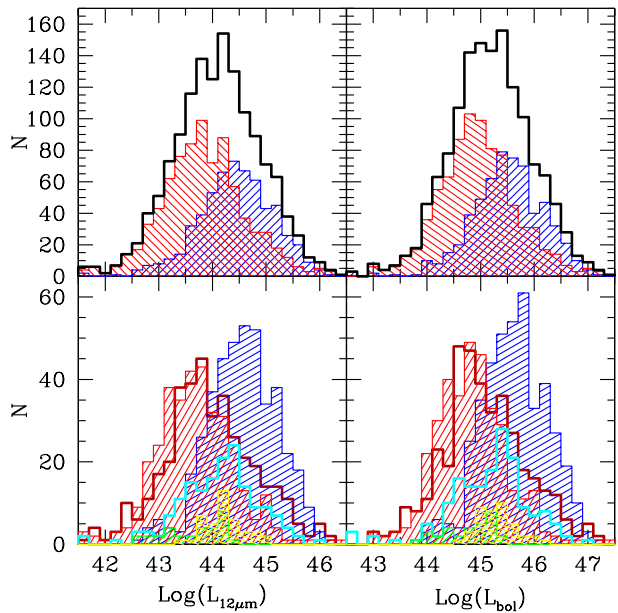}
\caption{\textit{Upper Panels:}  Mid-IR and bolometric luminosity distribution
of the entire sample (thick open histogram) and for the type--1 (shaded
blue) and type--2 (shaded red) AGN sub-samples
separately.
\textit{Bottom Panels:} Mid-IR and bolometric luminosity
  distribution of the different sub-samples included in our AGN
  sample: shaded blue and red  for X-ray spectroscopically confirmed
  type--1 and type--2 AGN, empty cyan and red for X-ray selected
  type--1 and type--2 AGN with photometric redshifts, dashed green and
  yellow for optically selected through emission lines (green: BPT
  diagram and yellow: NeV line) without X-ray detection.}
\label{fig:AGNlum}
\end{figure}

Using a sample of local Seyfert galaxies (both Seyfert-1 and
Seyfert-2) at high spatial resolution, 
\citet{Gandhi2009} found that the uncontaminated nuclear mid-IR
continuum of AGN closely correlates with the X-ray  [2--10] keV AGN
emitted powers over three orders of magnitude in luminosity. To be
specific, the best-fit correlation they obtained considering only the
22 well-resolved sources is shown  as a thin green line in
Fig. \ref{fig:L12LX}
($Log(L_{\rm MIR}/10^{43})=(0.19\pm0.05)+(1.11\pm0.07) Log(L_{\rm X}/10^{43})$).  
Our data follow the Gandhi's relation. 
As a comparison, the fit to our data performed  adopting the ordinary
least-square bisector estimate recommended by \citet{Isobe1990}, which
gives identical weighting to all sources, gives 
$Log(L_{\rm MIR}/10^{43})=(0.10\pm0.03)+(1.24\pm0.03)
Log(L_{\rm X}/10^{43})$ for the entire sample and $Log(L_{\rm MIR}/10^{43})=(0.29\pm0.03)+(1.13\pm0.03)
Log(L_{\rm X}/10^{43})$ if we consider only the spectroscopic sample. The fits are shown respectively as  black and gray solid thick lines in Fig. \ref{fig:L12LX}, together with the fit for the type--1
(blue long-dashed line) and type--2 (red short-dashed) AGN
sub-samples.  
The good agreement between the predicted and the measured 12.3$\mu$m
luminosities is also shown in Fig. \ref{fig:L12obspred} where we show
the histogram of the ratio between the measured 12.3 $\mu$m
luminosity and the one predicted by the \citet{Gandhi2009} relation for type–2 (top panel)
and type–1 AGN (bottom panel).
We notice that, while the distribution of the ratio between the
measured and the predicted MIR luminosity is centered at zero\footnote{in contrast to \citet{Lusso2011} which found a shift of 0.2 towards higher values when considering the ratio between the total (AGN+GAL) SED and the Gandhi predicted values.}, the observed scatter is 
much larger than the one measured by \citet{Gandhi2009} (2$\sigma$ from the Gandhi relation is delimited with orange lines: mean equal to zero and standard deviation $\sigma$=0.23). 
If we consider only the sources with a spectroscopic
redshift (black solid line), we do see a clear asymmetry in
  the distributions, as expected due to the un-accounted 
  contribution of star-formation-related emission in the NIR (see also
  Lusso et al. 2011). Indeed, while
$\sim$61\% of the objects are found within
2$\sigma$ of the distribution, the tail extending to lower values
includes 94 sources (12\%), while for a higher number of sources (211,
27\% of the sample) the observed mid-infrared luminosities are
significantly higher than the predicted ones. This is however
  not the case for the photometric sample. A possible explanation of
  this effect might involve the effect of the uncertainties in the photometric classification in 
  some of the AGN for which we miss spectroscopic information. Indeed,
  we have tested that all the small fraction of outliers in the diagram of
  Fig.~\ref{fig:HHplot} (i.e. type--1 AGN with $\alpha_{\rm OPT}<-1$
  and type--2 AGN with $\alpha_{\rm OPT}>-1$), show a  broader
  distribution in the $L_{12.3\mu{\rm m,obs}}/L_{12.3\mu{\rm m,pred}}$
  ratio with a slight skew towards values less than unity.

Part of the observed differences can also be attributed to
the fact that we 
are extending the Gandhi relation, which was derived for a sample of
local Seyfert galaxies, to a sample spanning a much wider luminosity
and redshift range. However, it is also plausible that  in a fraction
of these sources the SED-fitting procedure overestimates the nuclear
contribution  resulting in a higher  L$_{12.3\mu{\rm m,obs}}$ compared to
the predicted  L$_{12.3\mu{\rm m,pred}}$. 
How much of this scatter is inherently due to the physical conditions
of the AGN and torus clouds, as compared to observational selection
effects, remains an important unresolved issue, which is beyond the
scope of this paper.  However, the comparison points out that on
average the 12.3$\mu$m 
luminosity derived from the SED fitting is a reasonably good measure of the
mid-IR AGN luminosity and that indeed the assumption that the mid-IR
emission is dominated by the AGN emission due to accretion onto the
central black hole rather than SF from the host galaxy is plausible
for most of the sources.

One clear advantage of adopting the rest-frame MIR luminosity,
L$_{12\mu{\rm m}}$, as a universal AGN power estimator for our sample 
is that, contrary to the
[2-10] keV luminosity, this quantity was measured homogeneously for all
the sources, including the objects with no X-ray emission. Hence it
can be used to compare the AGN power in the different sub-samples and
to derive in a homogeneous way the bolometric luminosity for the
entire sample\footnote{However, for a more detailed study of the
  bolometric luminosities of XMM-COSMOS AGN we refer the reader to
  \citet{Lusso2010,Lusso2011}.}. This is shown in
Fig. \ref{fig:AGNlum} where both the 
12$\mu$m and the derived bolometric luminosity using the
\citet{Hopkins2007} conversions are shown for the different
sub-samples (bottom panels) and for the full sample divided in
obscured and unobscured AGN (upper panels). 
Bolometric luminosities range from 10$^{43}$ to 10$^{47}$ erg
s$^{-1}$. 

Optically selected Seyfert--2 galaxies (green histogram) are among the
faintest sources in our sample extending down to L$_{\rm bol}$=10$^{43}$
erg s$^{-1}$.  
Since their 12$\mu$m luminosities are of the same order of the X-ray selected obscured
sources, the lack of X-ray detection can be due to the fact that these
sources lie in the tail of the Gandhi relation
(Fig. \ref{fig:L12obspred}). Alternatively, some of these sources can
be Compton-thick AGN, as shown by \citet{Gilli2010} for the \NeV-based
sample, and, due to 
the high N$_{\rm H}$ absorption, they are not detected in the X-ray band.

\begin{figure*}
\centering
\includegraphics[width=0.9\textwidth]{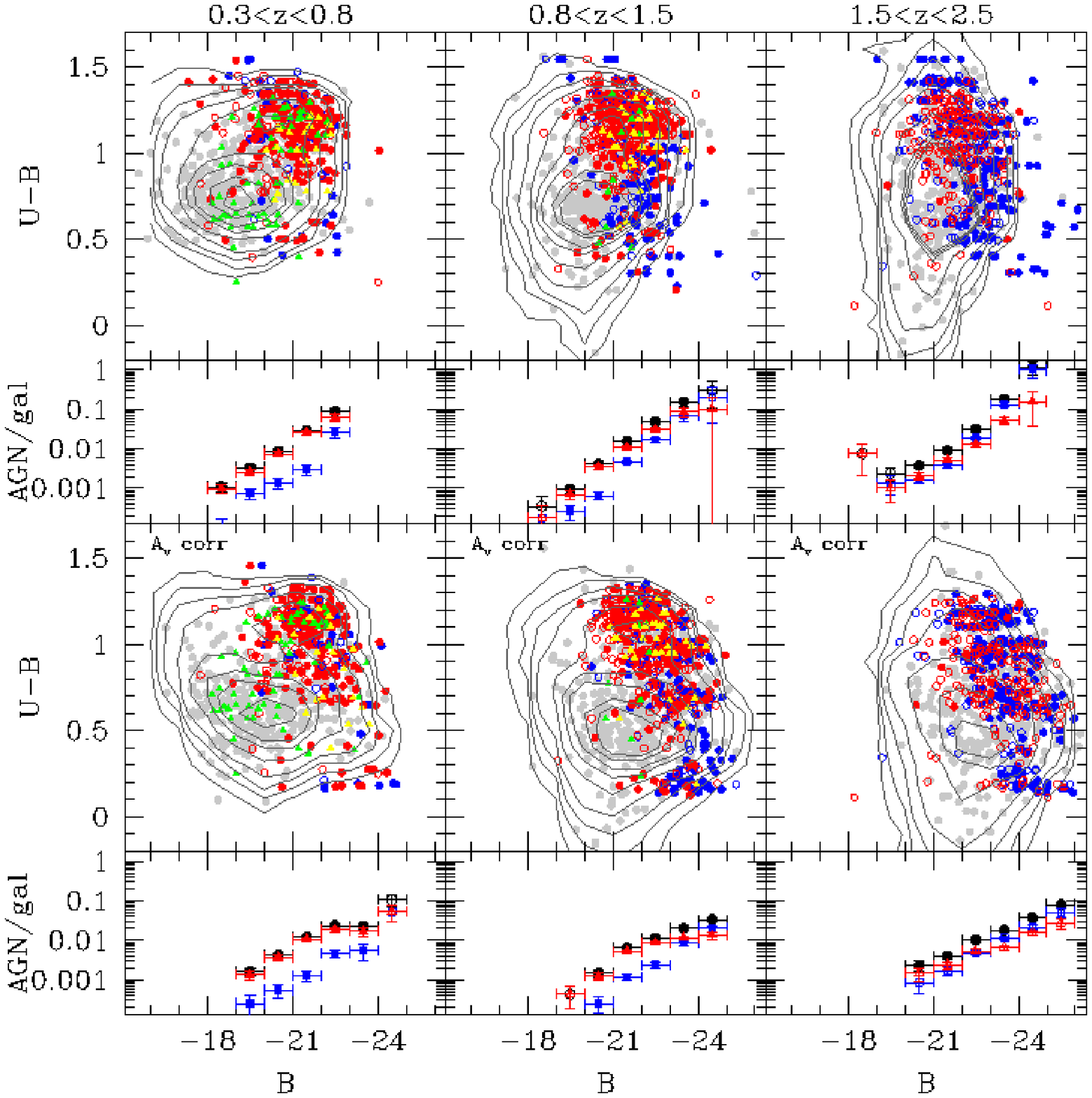}
\caption{\textit{Upper Panels:} U$-$B vs M$_{\rm B}$
  color-magnitude-diagram AGN host
  galaxies of unobscured (blue circles) and obscured (red circle)
  X-ray AGN and optically selected (green and yellow triangles) AGN in
  three redshift bins. Open and filled circles correspond to
  photometric and spectroscopic redshift. Grey contours show the whole
  galaxy sample drawn from \citet{Ilbert2010} while grey points
  correspond to the mass-matched parent sample described in
  Sec. \ref{sec:parent}.  
\textit{Lower Panels:} Same
  as above but now M$_{\rm B}$ and U$-$B are de-reddened using the
  E(B-V) derived from the SED fit. The bottom part of each panel shows
  the fraction of galaxies with an AGN as a function of B magnitude,
  considering all AGN (black circles), only type--1 (blue squares) and
  only type--2 (red triangles) AGN. For these fractions, we consider only X-ray selected AGN and the whole
  galaxy sample.}  
\label{fig:UB_MB}
\end{figure*}

\begin{figure*}
\centering
\includegraphics[width=0.8\textwidth,clip]{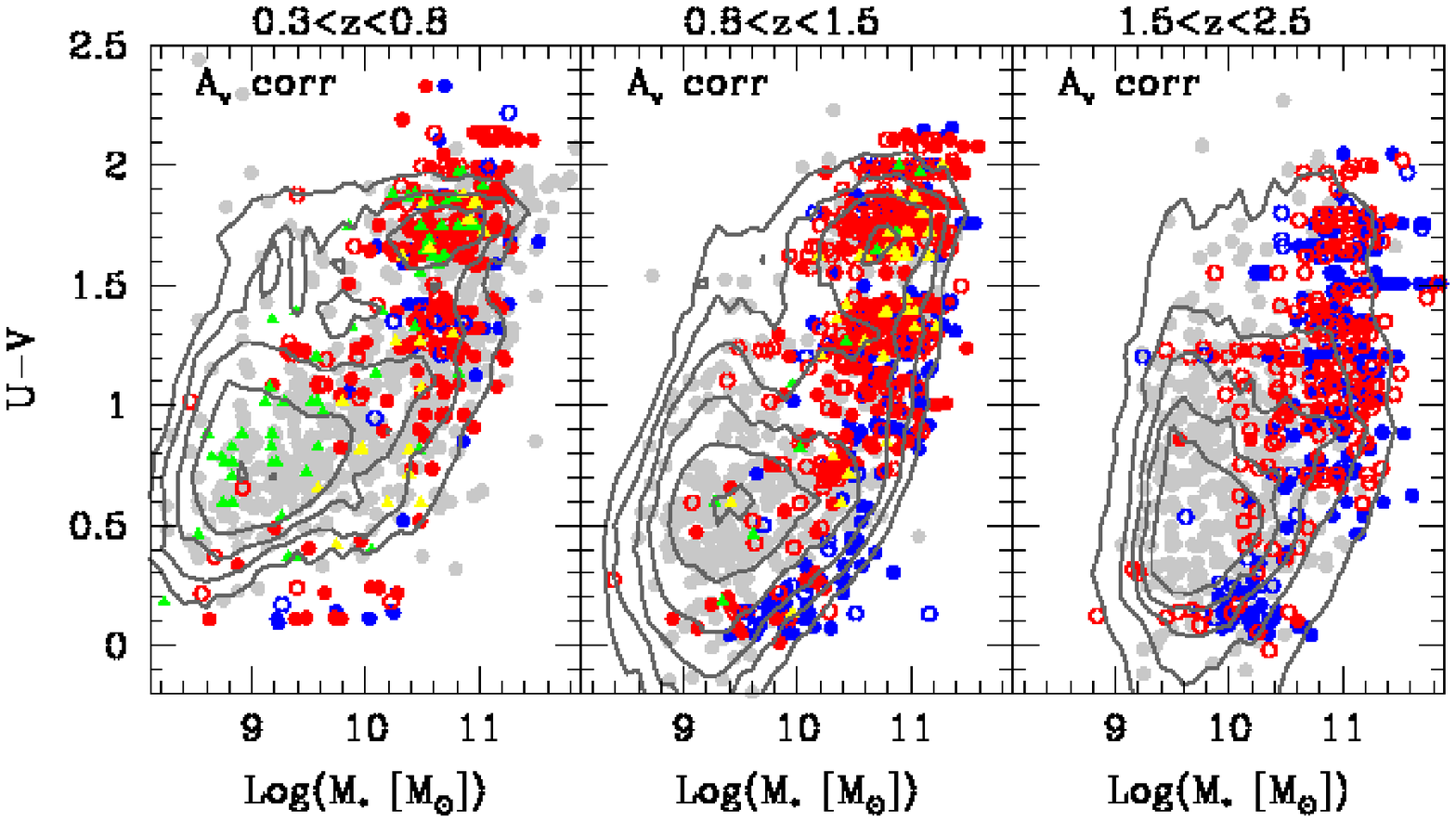}
\caption{Extinction corrected color (U$-$V) vs mass diagram of
  AGN host galaxies of unobscured (blue
  circles) and obscured (red circle) X-ray AGN and optically selected
  (green and yellow triangles) AGN in three redshift bins. Open and
  filled circles correspond to photometric and spectroscopic
  redshift. Grey contours show the whole galaxy sample drawn from
  \citet{Ilbert2010} while grey points correspond to the
  i-band-apparent-magnitude-matched parent sample described in
  Sec. \ref{sec:parent}.}  
\label{fig:UB_M}
\end{figure*}

\section{The physical properties of COSMOS AGN host galaxies}
\label{sec:GALprop}

\subsection{Rest frame colors}
\label{sec:rf_colors}

The unique strength of the COSMOS AGN sample at hand, lies indeed with
the possibility, by means of our SED fitting technique, of reliably removing the
AGN component from the overall SED of any AGN host. In this section,
we thus discuss rest-frame observables (absolute magnitudes and
colors) of the AGN host galaxy
component, which are, by construction, little contaminated by the AGN
colors.  
We also investigate the properties of the AGN host galaxies compared
to normal galaxies.

Figure \ref{fig:UB_MB} shows the rest frame  U$-$B
colors versus the B-band absolute magnitude (the color-magnitude
diagram, CMD) for our sample of
AGN hosts divided in three different redshift bins ([0.3$<$z$\leq$0.8],
[0.8$<$z$\leq$1.5] and [1.5$<$z$\leq$2.5]). The gray contours show for comparison 
the whole galaxy sample drawn from \citet{Ilbert2010} in the same
redshift intervals, while grey points correspond to the mass-matched
parent sample described in Sec. \ref{sec:parent}.  
We have chosen to show both the `measured' values of the U$-$B colors
(upper row) and the 
extinction corrected ones, where the value of the extinction is that
obtained for the galaxy component from the best fit SED decomposition
(see section~\ref{sec:extinction}). This is an important test,
as galaxies with red colors can be either dusty star-forming galaxies 
or galaxies with old stellar populations, and any interpretation of the location
of AGN hosts in the CMD may be strongly affected by the treatment of
dust extinction in these systems \citep[see
e.g.][]{Cardamone2010}. Our rich and comprehensive multi-wavelength
dataset is very well suited to attempt a robust determination of the
large scale extinction properties of the galaxies in the various samples.
 
First of all, a note of caution is in place, as, at any given
host galaxy magnitude, the colors of the type--1 hosts are more and
more unreliable the higher the nuclear AGN luminosity. In this respect, very blue colors of bright
type--1 AGN (see, in particular, the clusters of sources at
U-B$\simeq$0.1-0.3 in all extinction corrected CMD) should not be
too trustworthy, as they most likely indicate objects for which the SED
decomposition, and thus the AGN blue light subtraction, may give
uncertain and possibly unreliable results. We will come back to this
point when discussing the inferred star formation rate properties of
the AGN hosts.

We now move to the discussion of the CMD of the general AGN hosts
population. Underneath  each of the diagrams,
Fig. \ref{fig:UB_MB} shows the fraction of
all AGN (black), type--1 AGN (blue) and type--2 AGN (red) relatively to
the whole galaxy population in the corresponding redshift 
bin as a function of B-band magnitude. For consistency, in the
computation of  these fractions we restrict the analysis to the X-ray
detected sample, the completeness of which can be properly assessed in
a relatively simple way (see e.g. Sec.\ref{sec:AGNfr_mass} below), 
noticing however that the only effect of the pure
optical sources is to  slightly flatten the slope at lower redshifts.  
In agreement with previous studies \citep{Nandra2007,Brusa2009}, the
AGN are almost exclusively hosted in bright (M$_{\rm
  B}<-$20) galaxies and the fraction of galaxies hosting AGN
increases going to higher luminosities. 

No significant difference is found in the host's colors of unobscured
and obscured AGN, in contrast with what expected from the most popular
models of AGN fueling and feedback \citep[e.g.][]{Hopkins2008} but in
agreement with the standard unified model \citep{Antonucci1993}.   
As a matter of fact, the relative fraction of type--1 and type--2 AGN
changes only slightly 
with B-band luminosity in each redshift interval, but the overall
fractions of obscured and 
unobscured AGN are markedly different in the three redshift bins. This
is an effect introduced by the well known anti-correlation between
obscured AGN fraction and luminosity \citep[see e.g.][]{Hasinger2008}
in combination with the almost constant (X-ray) flux limit of the
XMM-COSMOS survey, which implies that high redshift sources are more
luminous than low redshift ones. A comprehensive discussion of the
incidence of obscured and un-obscured AGN in the COSMOS field as a
function of luminosity, redshift and host galaxy properties will be
presented in a separate work (Merloni et al. in prep.).

We find no evidence of color bi-modality for
AGN hosts, which populates mainly the red and intermediate part of the bimodal color
magnitude distribution observed for normal galaxies.
Interestingly, at 0.3$<z<$0.8 the low luminosity, optically selected
(from BPT diagram) Seyfert--2 galaxies without X-ray counterpart
(green triangles) 
seem to behave in a different way, being mainly hosted in 
galaxies populating the blue and the intermediate region.  
The effects of dust extinction correction on the 
AGN distribution in the color-magnitude diagram are clearly visible in the
lower panels of Fig.~\ref{fig:UB_MB}, where
some objects populating the red sequence moved to the
intermediate \ub\ region and some that were in the intermediate region lie
now in the blue region \citep[see also][]{Cardamone2010}.  
However, in contrast to what was found by \citet{Cardamone2010}, no clear
bi-modality is seen in any of the three panels even after dust
correction is applied. More importantly, it is clear from
Fig.~\ref{fig:UB_MB} that dust extinction correction affects more
severely the overall color distribution of the parent galaxy sample.
This can be seen in the fact that the AGN fraction in the B-band most luminous
galaxies in the sample decreases once dust correction is included.

Indeed, at any redshift, AGN hosts are found mainly in red galaxies, albeit
with a substantial tail towards intermediate and bluer colors (see also Fig. \ref{fig:UB_M}).
In many previous studies, AGN were described as populating
preferentially an intermediate
region, the so called ``green valley'' \citep{Silverman2009}, with the possible
interpretation that they live in quenched galaxies going through the
transition phase 
between active and passive star-formation phase.  
However, most of these works were focused on small subsample of the
AGN population and, more important,  the AGN component was not
subtracted in the computation of the host colors leading to a possible blue
contamination from the AGN which in fact would push red hosts into the
green valley.  Moreover, the net effect of the dust correction applied
systematically, and in a uniform way, to both the parent and
the AGN host galaxy samples is to {\it decrease} the fraction of AGN
in the most star-forming objects. 

\begin{figure*}
\centering
\includegraphics[width=0.9\textwidth]{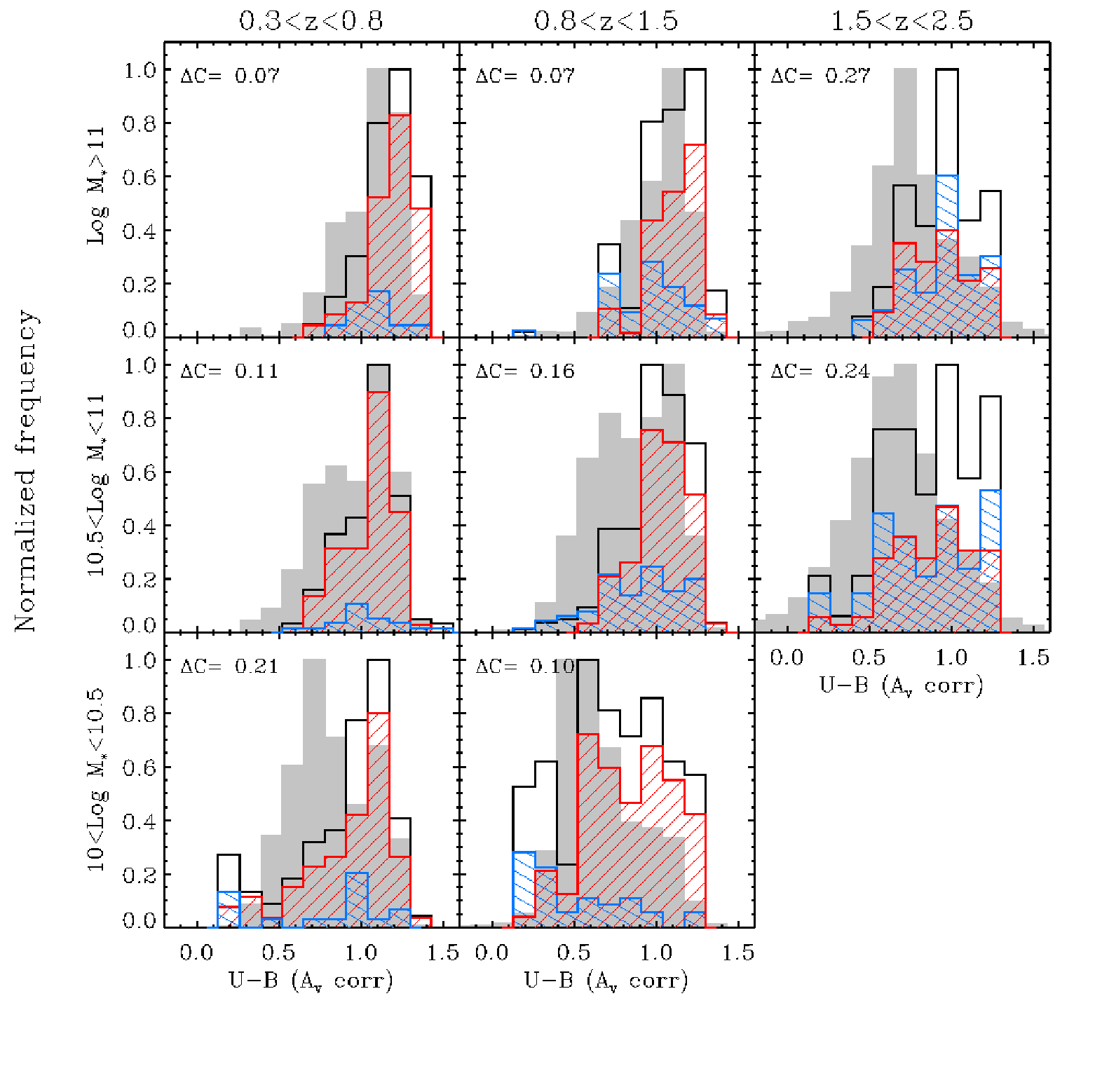}
\caption{Rest frame extinction-corrected  U$-$B color (normalized) distribution of
  AGN host galaxies in three redshift (from left to right) and mass
  (from bottom to top) bins. The black solid line
  corresponds to the total sample which is further divided into unobscured (blue
  shaded histogram) and obscured (red shaded histogram) AGN while the 
filled gray histogram corresponds to the galaxy
sample. For each redshift interval we have considered only masses
above which the parent sample is complete in stellar mass for both passive and star
forming galaxies \citep{Ilbert2010}. In each panel the median shift in
color $\Delta C\equiv \langle(U-B)_{\rm AGN}\rangle - \langle(U-B)_{\rm gal}\rangle$
is also reported.} 
\label{fig:UB_histo}
\end{figure*}

In general, the often contrasting interpretations of many 
previous studies of the CMD of AGN hosts from deep
multi-wavelength surveys
\citep{Nandra2007,Silverman2009,Xue2010,Cardamone2010,Rosario2011}
demonstrates that studying their position within optical
color-magnitude diagrams is 
hardly a robust way to unveil the
star-forming nature of AGN hosts. This is because the interpretation
of optical colors in terms of star-formation properties of a galaxy
can suffer from various, degenerate, biases, especially when dealing
with distant AGN hosts.
To counter such degeneracies, at least partly, in section~\ref{sec:AGNfr_sfr} we
will discuss in more detail how AGN hosts are distributed in the
physical parameter space of stellar mass and star formation rate.

\subsection{Host galaxy color-mass diagrams}

Figure \ref{fig:UB_M} shows the position of our AGN sources in the
color-mass diagram, where the colors (U-V) are corrected for extinction.
Black contours show the whole galaxy sample
drawn from \citet{Ilbert2010} while grey points correspond to the
i-band-apparent-magnitude-matched parent sample described in
Sec.~\ref{sec:parent}.  Comparing with the whole galaxy population, we
find that,  at any 
redshift, AGN are preferentially detected in massive galaxies, in agreement
with previous studies 
\citep{Brusa2009,Silverman2009,Xue2010,Mullaney2012}.  The trend is
still present if we use as comparison sample the
i-band-matched sample. In the next section we will investigate in
greater detail the evolution of the AGN fraction as a function of
redshift, X-ray luminosity, stellar mass of the host, and of the ratio
of these two latter quantities.
Here we note, however, that except for few unobscured AGN in the
intermediate z bin
showing extreme (and probably unreliable) blue colors, obscured and unobscured AGN overlap
very well in the color-mass plane.  

Since massive hosts tend to be redder, the trend with mass and the one
with colors seen in the previous section should go in parallel. In
many previous studies it has been pointed out how, removing the mass
dependency i.e. using a mass-matched control sample, the trend with
color disappears
\citep{Silverman2009,Xue2010,Mainieri2011,Rosario2011}.   
However, this is not the case in our study. Fig \ref{fig:UB_histo}
shows the extinction corrected U$-$B color histogram for AGN hosts in
the usual three redshift bins, for three different ranges of stellar
masses, all above the completeness threshold of the Ilbert et
al. (2010) parent galaxy sample, divided into unobscured (shaded blue) and 
obscured (shaded red) AGN and compared to the mass-matched galaxy
sample. Galaxies hosting an AGN tend to be redder  than normal
galaxies of comparable stellar mass (the median color shift ranges
between $\sim 0.1$ and $\sim 0.25$). 

Following our discussion in the previous section, 
we can speculate about two main reasons for the discrepancy between
our color distributions of AGN hosts in mass-matched samples and
previous studies. First of all, our SED decomposition technique allow
a more reliable correction for AGN light ``pollution'' that would
typically move the color of an AGN host towards the blue, star-forming
sequence. Second, thanks to the unique multi-wavelength coverage of
the COSMOS field, we are able to extinction-correct, in a systematic
way, all galaxy colors, both in the parent and in the AGN host
samples. We find that moving dusty star-forming galaxies back to the blue cloud
with such a correction effectively decreases the incidence of AGN
among star-forming galaxy, while, conversely, the relative fraction of
AGN hosts in truly passive galaxies is enhanced. We caution
  however that, being 
these color shifts relatively small, systematic effects due to a
non-optimal AGN-galaxy decomposition (AGN over-subtraction, see
section~\ref{sec:results_sed_fit}) or to the specific choice of
stellar population templates, could still be held responsible for
some of this observed trend.

\begin{figure*}
\includegraphics[width=\textwidth]{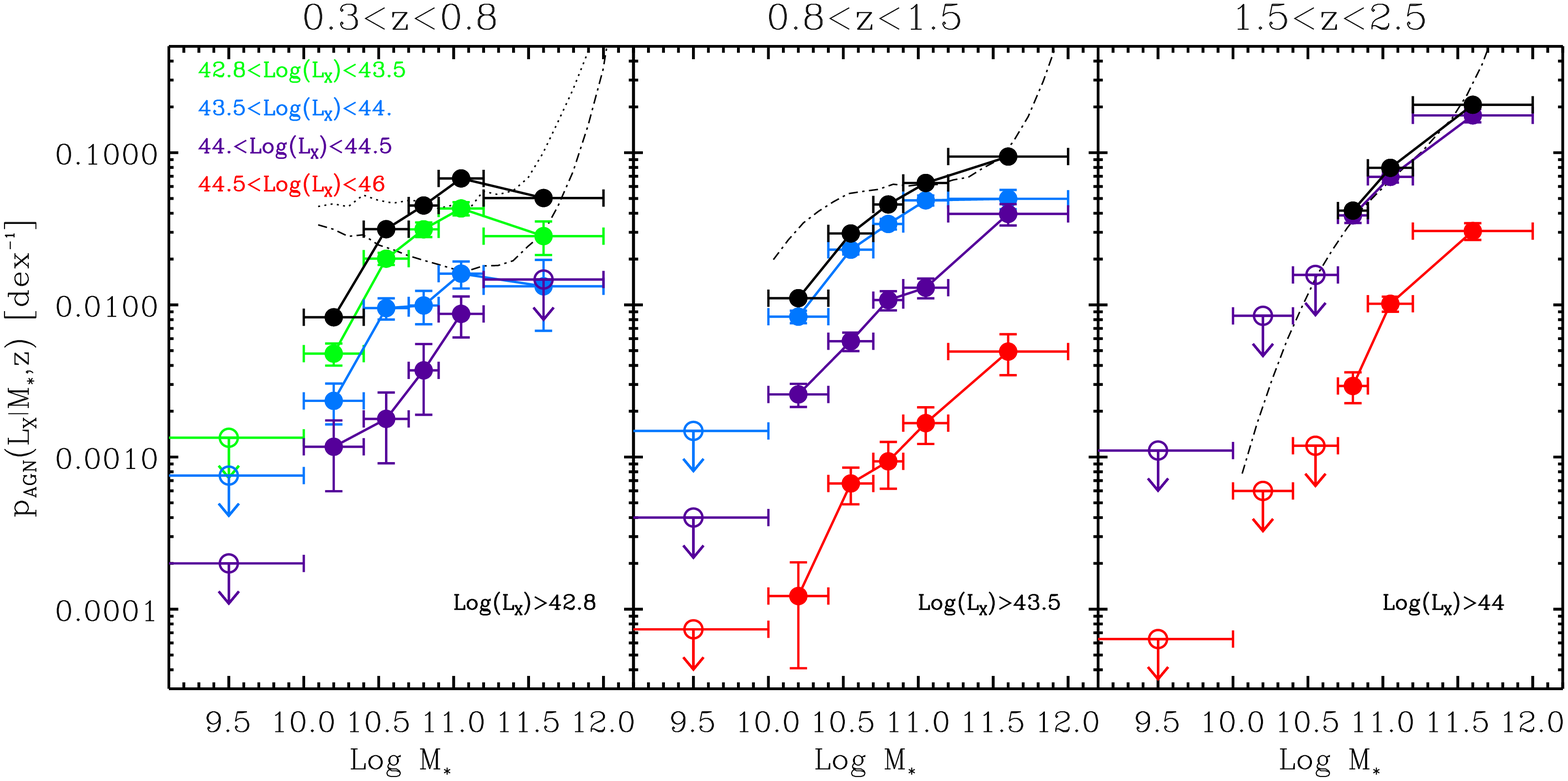}
\caption{Probability of a galaxy to host an AGN of a given luminosity 
  as a function of stellar
  mass in three redshift bins. Different colors show different X-ray
  luminosity bins: green for log(L$_{[2-10keV]}$)=[42.8-43.5]; blue for
  log(L$_{[2-10keV]}$)=[43.5-44]; purple for
  log(L$_{[2-10keV]}$)=[44-44.5] and red for
  log(L$_{[2-10keV]}$)=[44.5-46]; the black points refer to the total
  fraction computed by summing all objects above the given luminosity
  thresholds. Upper limits correspond to the
  mass bins where the host galaxy parent sample is not complete, as
  determined by the completeness limits for quiescent galaxies in COSMOS
  reported in \citet{Ilbert2010}. The thin dot-dashed line
is the prediction for the total AGN fraction, computed in the same
redshift bins, and above the same luminosity thresholds, from the ``AGN
occupation'' model of \citet{Fiore2012} while the dotted line in the first redshift bin is a modified version of it (see text for more details).}
\label{fig:AGNfrac_M_Lbin}
\end{figure*}

\subsection{The incidence of AGN as a function of the host galaxies'
  stellar masses} 
\label{sec:AGNfr_mass}

Let us now move to a quantitative assessment of the fraction of AGN as
a function of host stellar mass. In order to do so, we need to take carefully
into account the selection function of our survey. For this reason, in
this section, we will consider only the X-ray selected AGN, taking
advantage from the well defined exposure map of the XMM-COSMOS survey \citep{Cappelluti2009}.

So far, one of the clearest pieces of evidence emerging from a
variety of comparative studies of AGN and host galaxies in large
multi-wavelength surveys is the fact that the likelihood of finding an
AGN is a very strong function of the galaxy stellar mass, with AGN
fractions (or, equivalently, duty cycles) increasing with increasing
host galaxy stellar mass. This is a particularly strong effect for
radio-selected AGN \citep{Best2005,Smolcic2009}, but has been confirmed also for X-ray
selected AGN \citep{Brusa2009,Xue2010,Mullaney2012}.

\subsubsection{AGN fraction as a function of X-ray luminosity}
First of all, we compute the probability to find an AGN within a
logarithmic bin of a given X-ray
(2-10 keV, absorption corrected) luminosity, $L_{\rm X}$, in different bins of stellar mass by taking the
ratio of the number of AGN with measured host's stellar mass in that
range divided by the number of galaxies (of all morphological, color
and star-formation activity types) having stellar mass in the same
range in the parent sample of \citet{Ilbert2010}:
\begin{equation}
p_{\rm AGN}(L_{\rm X}|M_{*},z)=\frac{dN_{\rm AGN}(z)}{dlogM_{*}
  dlogL_{\rm X}}\frac{dlogM_{*}}{dN_{\rm gal}(z)}
\end{equation}

In practice, the sensitivity limits of the XMM-Newton
observations vary spatially across the field \citep[see Fig. 5 of][]{Cappelluti2009}, hence to derive the fraction of galaxies
hosting an AGN as a function of the stellar mass, we followed
the technique discussed in Sec. 3.1 of \citealt{Lehmer2007}
\citep[see also][]{Mainieri2011}. The $p_{\rm AGN}(L_{\rm X}|M_{*},z)$
for each stellar mass and luminosity bin is determined by the
following equation:

\begin{equation}
p_{\rm AGN}(L_{\rm X}|M_{*},z)=\frac{dN_{\rm AGN}(L_{\rm X}|M_{*},z)}{d\bar N_{\rm gal}(M_{*},z)dLogL_{\rm X}}
\end{equation}
where $dN_{\rm AGN}(L_{\rm X}|M_{*},z)$ is the observed number of AGN in each
bin of X-ray luminosity, stellar mass and redshift and we have defined:
\begin{equation}
d\bar N_{\rm gal}(M_{*},z)=\frac{1}{dN_{\rm AGN}(L_{\rm
    X}|M_{*},z)}\sum_{i=1}^{dN_{\rm AGN}}N_i(L_{\rm X,i}|M_{*},z).
\end{equation}

$N_i(L_{\rm X}|M_{*},z)$ is the number of galaxies that could have
been detected as X-ray sources at their position (given the known
exposure map of the XMM-COSMOS survey) if they had the luminosity
$L_{\rm X,i}$ of the $i$th AGN in the given bin.

Figure \ref{fig:AGNfrac_M_Lbin} shows the probability of galaxies hosting
AGN, $p_{\rm AGN}$, for different stellar mass bins in
 three redshift intervals. For lower mass points, completeness
has been treated conservatively, i.e. we have assumed that the parent
galaxy sample is complete only above the mass completeness limit for
quiescent galaxies as function of redshift, as reported in
\citet{Ilbert2010} (this translates into mass limits completeness of
about $M_{*}=10^{9.7},10^{10.1},10^{10.7}$ for the three redshift
bins, respectively). In those cases, our AGN fractions are considered
as upper limits.

The good statistics of our large, homogeneous and complete
sample, allows for an accurate study of AGN incidence, which accounts
for both redshift and intrinsic AGN luminosity dependence \citep[see
][]{Aird2012}. In each panel, AGN with different intrinsic X-ray luminosity
are plotted with different colors. 
 While the global trend with masses
remains qualitatively the same in each luminosity bin, the
normalization strongly decreases with luminosity, i.e. for a galaxy of
a given mass the probability to host an AGN is higher as the AGN X-ray
luminosity decreases and the same happens
at any redshift. This is an obvious consequence of the shape of the
AGN luminosity function: more luminous AGN are rare, while less
luminous ones are much more common.

Very interestingly, as
already shown by \citet{Aird2012}, the  $p_{\rm AGN}(L_{\rm X}|M_{*},z)$ as a function of
 galaxy stellar mass do show very similar profiles at {\it all}
sampled luminosities. AGN, of any intrinsic luminosity, 
 are more common the more massive the host galaxy is. 
This is by no means a trivial fact: one could have easily expected
that more luminous AGN were to be found in more massive galaxies and
less luminous AGN in lower mass galaxies. This, in fact, is the
prediction of any model in which black hole growth in AGN phases
occurs over relatively narrow Eddington ratio ranges, and the very
different picture we observe in Fig.~\ref{fig:AGNfrac_M_Lbin} (as well as in
Fig.~\ref{fig:AGNfrac_M_Eddbin}, see below) is a strong indication
that AGN of all masses should be characterized by a very broad
distribution of Eddington ratios \citep[see e.g.][]{Merloni2008}. We
quantify this statement by studying the AGN fractions as a function of
the specific accretion rate in section~\ref{sec:agn_frac_ledd} below.

Figure~\ref{fig:AGNfrac_M_Lbin} also shows (black points) the total
AGN fraction above a given luminosity thresholds in each redshift interval
(i.e. computed by summing up the $p_{\rm AGN}(L_{\rm X}|M_{*},z)$ in
different luminosity bins). As a comparison, the thin dot-dashed line
is the prediction for the total AGN fraction, computed in the same
redshift bins and above the same luminosity thresholds, from the ``AGN
occupation'' model of \citet{Fiore2012}. We refer the reader to that
work for more details, but we note here that in \citet{Fiore2012} the AGN
fraction is computed by deriving an active SMBH mass function from the
observed AGN luminosity function, assuming a given shape for the
Eddington ratio distribution, and then comparing it to the observed
galaxy mass function. In the two highest redshift bins, such a model
reproduces our data remarkably well. However, severe discrepancies appear in
our lowest redshift range. 
As we will show in the next section, it
appears that the true Eddington ratio distribution of our AGN is
better described by a simple power-law and the fact that the shape assumed in   
\citet{Fiore2012} is a log-normal can explain the discrepancy. 
In fact, while in the highest redshift bins,
where our COSMOS AGN sample probes just a limited range of Eddington ratios,
the observed distribution is hardly distinguishable from a log-normal distribution with
appropriate parameters, in the lowest redshift range, the log-normal 
shape assumed by \citet{Fiore2012} under-predicts the number of low
luminosity AGN hosted in massive galaxies
($10^{10.6}<M_{*}<10^{11.6}$). Assuming a log-normal distribution with a peak at lower Eddington ratio would already improve the prediction in the first redshift bin as shown by the dotted line in Fig. \ref{fig:AGNfrac_M_Lbin} which corresponds to a distribution with $\lambda_{\rm peak}$=0.03.

%---FIGURE
\begin{figure*}
\includegraphics[width=\textwidth]{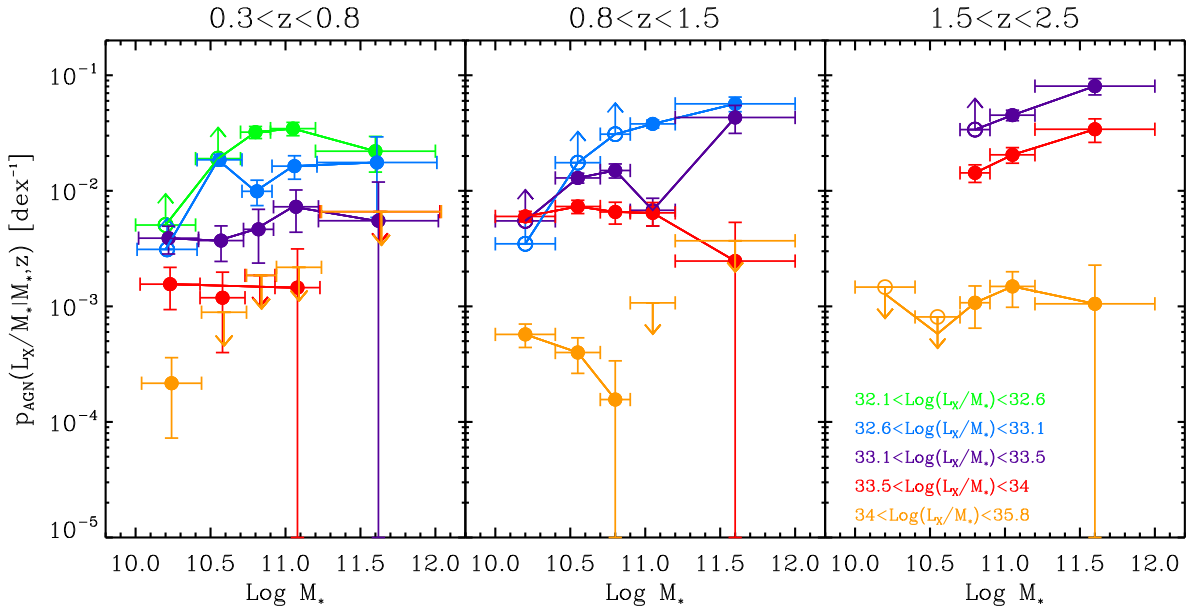}
\caption{Probability of a  galaxies to host an AGN per decade
  specific accretion rate ($L_{\rm X}/M_{*}$) as a function of stellar
  mass in three redshift bins. Different colors show different
  $L_{\rm X}/M_*$ intervals, as explained in the third panel. Empty
  symbols with arrows show upper and lower limits due to the combined effect of
completeness in both $L_{\rm X}$ and $M_{*}$. Upper limits without
empty circles mark the bins where no AGN was found, and were placed
using the \citet{Gehrels1986} formulae for Poisson statistics.} 
\label{fig:AGNfrac_M_Eddbin} 
\end{figure*}

\subsubsection{AGN fraction as a function of specific accretion rate}
\label{sec:agn_frac_ledd}
Indeed, by combining the available information (host galaxy stellar mass
and AGN luminosity) we can study the probability of a galaxy to host
an AGN as a function of mass in bins of ``specific accretion
rate'', i.e., the ratio L$_{\rm X}$/M$_*$ which gives an
approximate estimate of the rate at which the SMBH grows.
To the extent that a proportionality between the
black hole mass and the (total) host galaxy stellar mass can be
assumed \citep[and this is far from being clear, see
e.g.][]{Merloni2010}, this ratio 
could give a rough measure of the black hole Eddington ratio, defined
as the ratio between the AGN
X-ray luminosity and the Eddington luminosity. To be specific, the
Eddington ratio can be expressed as:
\begin{equation}
\lambda_{\rm Edd} = \frac{{\cal A}
\cdot  k_{\rm bol}}{1.3 \times 10^{38}} \times \frac{L_{\rm X}}{M_{*}}\,  %\propto \frac{L_{\rm X}}{M_{\rm BH}} 
\end{equation}
where $k_{\rm bol}$ is the 2-10 keV bolometric correction, and the factor ${\cal A}$ is a constant if
the SMBH mass can be related to the host galaxy mass through scaling
relations \citep[with ${\cal A}\approx 500-1000$;][]{Haring2004,Magorrian1998}.  
Thus, for a mean
bolometric correction\footnote{recent works
  \citep{Lusso2010,Lusso2011} 
have shown that the 2-10 keV bolometric correction is likely an
increasing function of the Eddington ratio itself, a further
complication that we here neglect.} of $k_{\rm bol}=25$ and a constant host
stellar to black hole mass ratio of ${\cal A}=500$, a ratio of
L$_{\rm X}$/M$_*=10^{34}$ would approximately correspond to the Eddington
limit. 

Figure~\ref{fig:AGNfrac_M_Eddbin} shows how the probability $p_{\rm
  AGN}(L_{\rm X}/M_{*}|M_{*},z)$
of a galaxy to host an AGN of a
given specific accretion rate scales with the stellar mass of the host.
Here, of course, the issue of completeness
is far more complicated. Depending on the
$L_{\rm X}/M_*$ ratio interval considered, both the flux limit of the X-ray
survey and the stellar mass completeness limit of the parent galaxy
sample should all be taken into account. 
Figure~\ref{fig:AGNfrac_M_Eddbin} shows that, at any given  specific
accretion rate (a proxy, with all the caveats discussed above, of the
Eddington ratio), the trend with
mass almost disappears: the probability for a galaxy to host an AGN of
a given  $L_{\rm  X}/M_*$ is almost {\it independent} of its mass but
scales only with the  ratio itself. 

\begin{figure*}
\includegraphics[width=\textwidth]{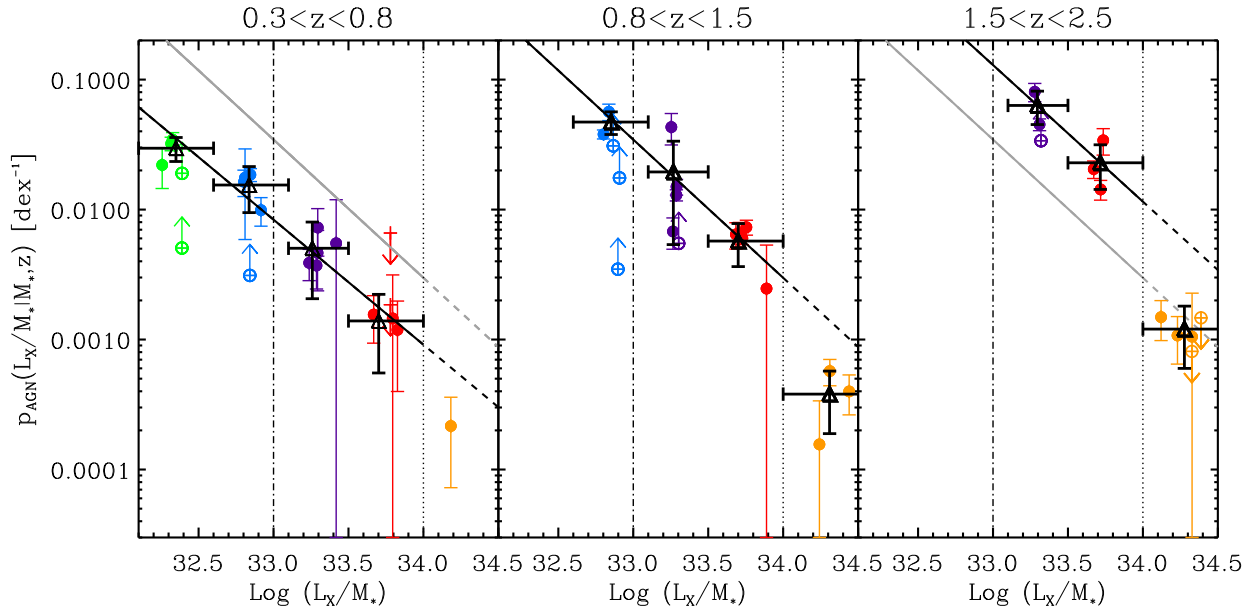}
\caption{The probability of a galaxy to host an AGN of a given specific
  accretion rate (per decade
  $L_{\rm X}/M_{*}$ ratio) is plotted as a function of 
  $L_{\rm X}/M_*$ for three redshift intervals. Colored points, have
  the same coding of Fig.~\ref{fig:AGNfrac_M_Eddbin} and plot the
  individual measurements, while black thick triangles are the mean
  values. The thick black line in each panel is a power-law fit to all
  the mean points obeying Log($L_{\rm X}/M_{*})<34$. For ease of
  comparison, in each panel we also show the best fit to the central
  redshift interval.
The vertical dotted (dot-dashed)
lines mark the approximate location of objects at the Eddington limit
(or ten per cent of it) calculated  assuming a constant
bolometric correction $k_{\rm bol}=25$ and a constant host stellar
to black hole mass ratio of ${\cal A}=500$.}
\label{fig:fAGN_Eddbin} 
\end{figure*}

In Figure~\ref{fig:fAGN_Eddbin}, we then plot $p_{\rm
  AGN}(L_{\rm X}/M_{*}|M_{*},z)$ as a function of the specific
accretion rate itself. As mentioned above, there is little dispersion
among the points corresponding to different mass bins for the same
$L_{\rm X}/M_{*}$ interval. We can 
{\it assume} that the AGN fraction at any given value of
the  $L_{\rm X}/M_*$ ratio is independent of host galaxy mass,
and measure such a constant ``universal'' mean probability for every specific
accretion rate and redshift bin. 
The thick black triangles in Fig.~\ref{fig:fAGN_Eddbin} show
the mean values of $p_{\rm AGN}(L_{\rm X}/M_{*}|M_{*},z)$. Indeed,
all the data seem consistent with the idea
that, at every redshift, there is a {\it universal} distribution 
function that describes the probability of a galaxy of any mass (above
our completeness limit of $\sim 10^{10} M_{\sun}$) to host an AGN of a given
specific accretion rate
$L_{\rm X}/M_{*}$, independent of host stellar mass \citep{Aird2012}.

%---------------FIGURE

 As we have anticipated, to the first
order, the data are consistent with a power-law like distribution
 and a clear break at values of $L_{\rm X}/M_*$
consistent with the Eddington limit. In
Table~\ref{tab:frac_edd_slopes} we report the values of the best fit
power-law distribution for each of the three redshift bins, in the
form\footnote{We note here that the fact that the {\it observed}
  distribution of specific accretion rates (or, for what matters,
  Eddington ratios) is typically found to be log-normal \citep[see
  e.g.][]{Kollmeier2006,Netzer2009,Trakhtenbrot2011,Lusso2012}, rather than a
  power-law, is due to the combined effects of a flux limited survey,
  that depletes the number of low specific accretion rates at
  progressively lower masses, and of the fact that very massive
  galaxies, where slowly accreting objects can be detected are
  exponentially rarer as the stellar mass increases. A thorough
  discussion of these selection effect on the determination of the
  acretion rate distribution function of AGN can be found in
  \citet{Schulze2010} and \citet{Aird2012}.}:
\begin{equation}
 p_{\rm AGN}(L_{\rm X}/M_{*}|M_{*},z)=K\left(\frac{L_{\rm X}}{M_{*}}\right)^{-\gamma}
\end{equation}

As compared to the values found in \citet{Aird2012}, we find a somewhat
steeper slope for the power-law index ($\approx$1 rather than $\approx$0.7). We
speculate that this might have to do with the fact that, differently
from \citet{Aird2012}, we do not include AGN from the deepest {\it
  Chandra} fields, that probe the fainter, shallower end of the X-ray
luminosity function. Hint of such a flattening can
be seen in the decline of the $p_{\rm AGN}$ for the most massive
galaxies at low X-ray luminosities in
Fig~\ref{fig:AGNfrac_M_Lbin}. These are very low specific accretion rate
objects for which we are not complete in the COSMOS sample (they are shown
as lower limits in Figures~~\ref{fig:AGNfrac_M_Eddbin}
and~\ref{fig:fAGN_Eddbin}) and therefore cannot properly assess the
robustness of any change in slope at low $L_{\rm X}/M_{*}$. 
The possibility that the $p_{\rm AGN}(L_{\rm
  X}/M_{*}|M_{*},z)$ might break to a shallower slope at low values of
the specific accretion rate could be tested, indirectly, by comparing
the observed X-ray luminosity function of the AGN with a convolution
of the observed galaxy mass function and the probability distribution 
$p_{\rm AGN}(L_{\rm  X}/M_{*}|M_{*},z)$. We defer this investigation
to a future work.

\begin{table}
\centering
 \caption{The best fit slope and normalization for the universal probability
   distribution function $p_{\rm AGN}(L_{\rm X}/M_{*},z)$ as a
   function of redshift for the entire X-ray
 selected sample, as shown in Fig.~\ref{fig:fAGN_Eddbin}.}  
\label{tab:frac_edd_slopes}
\begin{tabular}{lcc}
  \hline\hline
& \multicolumn{2}{c}{all X-ray} \\
 redshift range & Log $K$ & $\gamma$ \\
\hline
$0.3 < z < 0.8$ & 29.50 & 0.96 \\
$0.8 < z < 1.5$ & 33.41 & 1.06 \\
$1.5 < z < 2.5$ & 33.46 & 1.04 \\ 
\hline
\hline
\end{tabular}
\end{table}
Nevertheless, we see very clearly in Fig.~\ref{fig:fAGN_Eddbin} that 
the $p_{\rm AGN}(L_{\rm X}/M_{*}|M_{*},z)$
{\it normalization} increases with redshift over the entire redshift
range probed by our data. If we compute the value of the $p_{\rm
  AGN}(L_{\rm X}/M_{*}|M_{*},z)$ at Log$(L_{\rm X}/M_*)$=33.2, i.e. at
a specific accretion rate approximately corresponding to $\approx
10\%$ of the Eddington rate, and where
our data best constrain the distribution at all redshifts, we find
that it evolves with redshift as
fast as the specific star formation rate of the overall population of
galaxies \citep[both star-forming and 
quiescent, see][]{Karim2011} $\propto (1+z)^{4}$.

This very important fact, already pointed out, over a more
limited redshift range, by \citet{Aird2012}, implies that the
rapidity with which a black hole grows is not influenced by the host
galaxy mass, while, on the other hand, is a strongly evolving function
of redshift, with  a redshift dependence that follows the overall evolution
of the specific star formation rate of the galaxy population.
This also means that the trend with the hosts' mass visible in
Fig.~\ref{fig:AGNfrac_M_Lbin} and widely confirmed by many authors
before us, is a trend induced by the fact that in more massive
galaxies we are able to detect AGN down to a lower intrinsic Eddington
ratio, while in less massive galaxies only rapidly accreting black
holes are bright enough to be observable, and their total number is
therefore much smaller.

%---FIGURE
\begin{figure*}
\centering
\includegraphics[width=0.8\textwidth,clip]{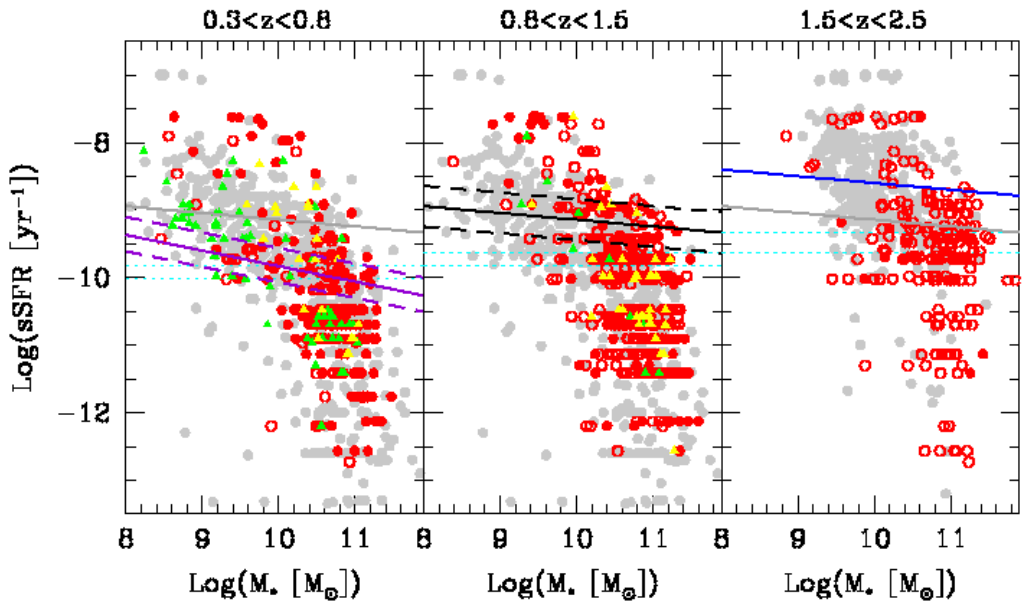}
\caption{Specific star-formation rate versus stellar masses  in three
  redshift bins for the host galaxies of type--2 AGN compared to normal
  galaxies from the i-band-apparent-magnitude-matched sample (gray points). 
  The solid lines are the correlations found
  for blue star-forming galaxies: the purple one in the lowest
  redshift bin corresponds to z$\sim$0
  \citep{Brinchmann2004}, the black one in the intermediate redshift
  bin to z$\sim$1 \citep{Elbaz2007} and
  the blue one in the highest redshift bin to z$\sim$2
  \citep{Daddi2007}. In each panel the $z\sim 1$ curve is reported in
  grey for reference.
 The cyan dotted horizontal lines correspond to the age of the
Universe $t_H$(z) at the minimum and maximum redshift considered in
each panel and indicate roughly the division between galaxies which
are actively forming stars and quiescent.} 
\label{fig:SSFR_M_plot}
\end{figure*}

\subsection{The incidence of AGN as a function of the star-forming
  properties of their hosts} 
\label{sec:AGNfr_sfr}

Studies in the local Universe \citep{Brinchmann2004, Salim2005} have
pointed out the existence of two main galaxy populations: star-forming
galaxies, which are currently forming stars at a rate that was found to
be proportional to their mass, and quiescent galaxies, which are 
massive galaxies in which the star formation has been
quenched\footnote{``Quenching'', regardless of its physical cause, is
  the process that cause the cessation of star formation in some
  star-forming galaxies, which leads to the emergence of the so-called
  “red-sequence” of passive galaxies \citep{Strateva2001}.}.
The specific star formation rate (sSFR) can be 
defined as the ratio, sSFR=SFR/$M_*$, of (total) star formation rate to
stellar mass of a galaxy. The inverse of the sSFR
defines a timescale for the formation of the stellar population of a
galaxy, $t_{\rm SF}$ = sSFR$^{-1}$, which gives the time it would take a
galaxy to double its stellar mass if forming stars at the observed
 rate.

In the sSFR vs mass plane, star-forming galaxies can be divided into a
star-forming sequence \citep[``main sequence''][]{Noeske2007} 
characterized by an almost constant (or slightly decreasing) sSFR as a
function of stellar mass, and a  high
mass region with progressively decreasing sSFR,  the  ``dying
sequence'', occupied by quiescent galaxies. In evolutionary terms, 
such a distribution is observed, almost unchanged, 
at any redshift, with the star-forming galaxies sequence 
maintaining the same slope but increasing normalization as one moves towards
higher redshift \citep{Noeske2007,Elbaz2007,Daddi2007,Pannella2009}.
There are also more extreme objects, i.e. Starburst galaxies
and Ultra Luminous Infrared Galaxies
(ULIRGs), which exhibit highly elevated star formation rates
\citep{Sanders1996} in galaxies 
within the same range of stellar mass of normal galaxies. 
Local ULIRGs are believed to be associated with rare major
merger events \citep{Sanders1988,Sanders1996} and consequently
distinct star formation processes, but recent studies with the {\it
  Herschel} telescope demonstrated that they represent a minor
contribution to the total star formation rate density in the
crucial redshift range $1.4<z<2.5$ \citep{Rodighiero2011}. 
Recent works \citep{Wuyts2011} have shown that a detailed morphological analysis  of
galaxies as a function of their position in the sSFR-stellar mass
plane can provide a unique insight on the physical processes that shape
and grow galaxies, and how the various modes of star formation can be
associated to them.

\begin{figure*}
\includegraphics[width=0.9\textwidth]{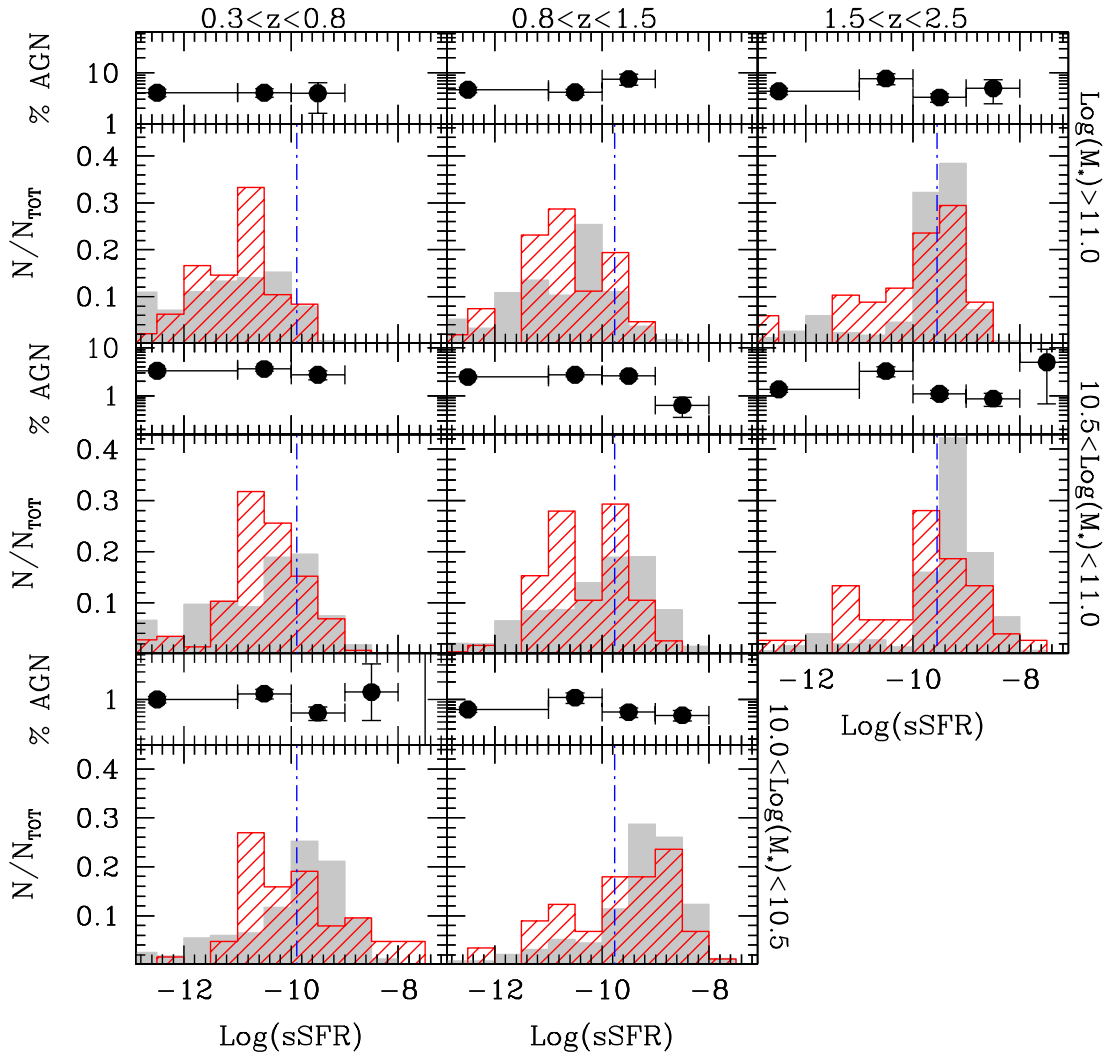}
\caption{Specific star formation rate distributions divided in three
 redshift bins and three mass ranges for type--2 AGN host galaxies
 (red histogram) and normal galaxies (gray histogram). Dot-dashed
  vertical blue lines are used to distinguish quiescent from star-forming galaxies
  based on the ratio between the mass doubling time $t_{\rm SF}$ and
  the age of the Universe in any redshift bin. 
  The upper panel of each histogram shows the
 observed AGN fractions.} 
\label{fig:sSFR_massbin}
\end{figure*}

\begin{figure*}
\includegraphics[width=0.9\textwidth]{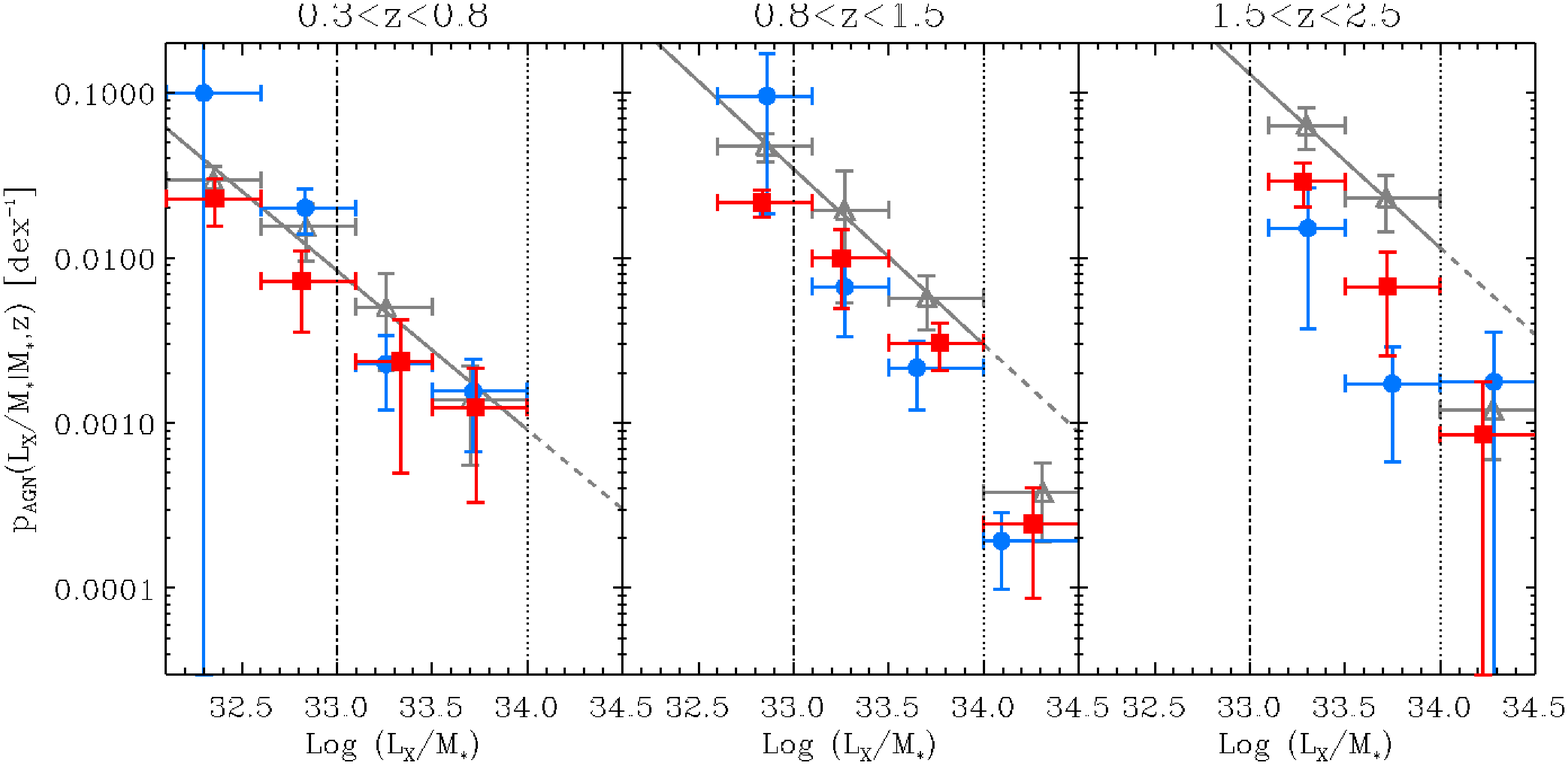}
\caption{The average probability of a galaxy to host an AGN of a given specific
  accretion rate (per decade
  $L_{\rm X}/M_{*}$ ratio) is plotted as a function of 
  $L_{\rm X}/M_*$ for three redshift intervals. In each panel blue
  filled circles are for star-forming galaxies (i.e. those with
  $t_{\rm SF}<t_{\rm H}$) and red squares are for quiescent ones (those with
  $t_{\rm SF}>t_{\rm H}$). Solid lines of corresponding colors show
  the power-law fit to all points obeying Log($L_{\rm
    X}/M_{*})<34$. The grey symbols and lines are the probability
  distribution functions for the overall galaxy population (for all
  AGN types), as from Fig.~\ref{fig:fAGN_Eddbin}. The vertical dotted (dot-dashed)
lines mark the approximate location of objects at the Eddington limit
(or ten per cent of it) calculated  assuming a constant
bolometric correction $k_{\rm bol}=25$ and a constant host stellar
to black hole mass ratio of ${\cal A}=500$.} 
\label{fig:fraction_ledd_ssfr}
\end{figure*}

In light of the above results, it is obvious that a systematic study
of the distribution of AGN host galaxies in the sSFR-stellar mass
plane can reveal key pieces of evidence regarding the long
sought-after association between SMBH growth and galaxy formation. 
Figure \ref{fig:SSFR_M_plot} shows the position of our obscured
(type--2) COSMOS AGN host
galaxies in the sSFR vs Mass plane compared
to the overall non-AGN galaxy population (gray dots). We have
purposely avoided to include type--1 objects in this diagram, as
optical/UV-based SFR measures are too strongly affected by the
unobscured accretion disc light in this class of AGN.

The purple, black and blue solid lines show the position of the main
sequence at $z\sim 0$ \citep{Brinchmann2004},  $z \sim 1$
\citep{Elbaz2007} and $z \sim 2$ \citep{Daddi2007}, while 
the cyan dotted lines correspond to the age of the
Universe $t_{\rm H}(z)$ at the minimum and maximum redshift considered in
each panel. Galaxies of known redshift and sSFR can in fact 
be divided into ``quiescent'' and ``star-forming''  on the
basis of their stellar mass doubling timescale, $t_{\rm SF}$: 
if $t_{\rm SF} < t_{\rm H}(z)$, where $t_{\rm
  H}(z)$ is the age of the 
Universe at the redshift of the source, the galaxy is defined as star-forming; 
 if $t_{\rm SF} > t_{\rm H}(z)$ the galaxy is defined as quiescent.
The i-band-apparent-magnitude matched parent galaxy population from the S-COSMOS survey
\citep{Ilbert2010} does indeed show a broad distinction between a star forming sequence and a dying sequence at the
highest masses, which clearly evolves upwards (in sSFR) with increasing redshift.

Obscured AGN hosts, to a first order, appear to populate the high-mass end of
the distribution quite uniformly. 
%In direct confirmation of our
%suggestion in section~\ref{sec:AGNfr_mass}, these AGN hosts appear to follow
%the global cosmological evolution of the galaxy population in the
%sSFR-stellar mass plane, with no strong preference for either
%star-forming or quiescent galaxies. 
We estimate that the fraction of AGN hosted in starburst galaxies (defined as in \citealt{Rodighiero2011}  to be above 4 times the main sequence) is below 10\% while the fraction of AGN hosted in main sequence galaxies range from 27\% to 37\% in the different redshift bins. The rest of the AGN  (58\%-66\%) are found in quiescent galaxies. 
Considering only the most luminous sources as done by \citet{Mainieri2011}, we found that at z$\sim$1 about 62\% of Type-2 QSOs were actively forming stars (i.e. hosted in starburst and main sequence galaxies).
As comparison, \citet{Mullaney2012}, using the FIR-{\it Herschel} data in the GOODS field and excluding the galaxies in which a strong contribution from the AGN to FIR wavelengths was evident, estimated for moderate luminous AGN the following percentages:  $<$10 \% of AGN are hosted in strongly starbursting galaxies; 79\%  in main-sequence galaxies, and 15\% in quiescent galaxies. The discrepancy between the two findings could be attributed to the differences (especially at low SFR) found between the SFR derived from SED fitting and the FIR ones as showed in Appendix \ref{sec:sfr}. 
To test this in detail, we fit  below and above 20 M$_{\odot}$/yr$^{-1}$ the SFR$_{\rm SED}$-SFR$_{\rm IR}$ relation found (see left panel of Fig. \ref{fig:SFRsed_SFRIR}) and we assign to each of our SFR values, the corresponding SFR$_{\rm IR}$. We now find that the fraction of AGN hosted in strongly starbursting galaxies is 9\%, while 67\% are hosted  in main-sequence galaxies, and 24\% in quiescent galaxies, in much better agreement with the {\it Herschel}-based results.

Figure~\ref{fig:sSFR_massbin} shows, as histograms, the sSFR
distributions of AGN hosts and normal galaxies in
bins of stellar mass and redshift. Once again, the AGN do appear to follow
the evolution of the specific star formation rate with stellar mass
and redshift, with just a marginal increase in the AGN fraction 
towards lower values of sSFR in almost all bins, indicating a slightly
higher incidence of (obscured) AGN in non-star-forming galaxies.  

According to the models that assume that major mergers are the main
AGN fueling mechanism \citep[e.g.][]{Hopkins2008}, one could have expected an
enhancement of the SFR activity in obscured AGN, which is however not
visible. From our analysis, there is no clear evidence that the COSMOS
obscured AGN hosts are preferentially found among the extreme, most
rapidly star-forming galaxies.

To analyze this in further detail we divide the host galaxies in two
 classes depending on the ratio between $t_{\rm SF}$ and $t_{\rm H}$. 
According to this division of the total sample of obscured AGN discussed here,
the fraction of quiescent hosts (always computed for stellar masses
above the completeness limit of the COSMOS-IRAC survey, as discussed
in \citealt{Ilbert2010}) are 75\%, 65\% and 61\% in the
redshift bins 0.3--0.8, 0.8--1.5 and 1.5--2.5, respectively. In total, 
about two thirds of the obscured AGN in COSMOS that we find in massive
galaxies are hosted by objects with stellar
mass doubling time longer than the age of the Universe. These
fractions are higher than those of the 
galaxy sample, where the fraction of quiescent hosts are 51\%, 41\% and 32\% in the
redshift bins 0.3--0.8, 0.8--1.5 and 1.5--2.5, respectively.
Such a predominance of quiescent (or, better, not strongly star
forming hosts) is indeed consistent with
the predominantly red colors exhibited by the AGN hosts that we discussed in
section~\ref{sec:rf_colors}. 

In section~\ref{sec:agn_frac_ledd} we have computed the probability
of a galaxy (above the stellar mass completeness threshold of the
COSMOS-IRAC survey) to host an AGN of a given specific accretion rate,
irrespective of the star-formation rate properties of the
galaxies. For the sub-sample of AGN which are obscured, for which a
reliable and robust determination of sSFR is possible, we have also
computed separately the probability of star-forming and quiescent
galaxies to host a (type--2) AGN as a function of specific accretion
rate $L_{\rm X}/M_*$. The results are shown in
figure~\ref{fig:fraction_ledd_ssfr}, where we also report, for the sake of
comparison the overall $p_{\rm AGN}$ of
Fig.~\ref{fig:fAGN_Eddbin}.
Taken at face value, the plot shows that there is little significant
evidence for a difference in the probability of hosting a type--2 AGN
in quiescent and star-forming galaxies, as the data points agree
within the errors (that here are representative of the dispersion
of $p_{\rm AGN}(L_{\rm X}/M_{*}|M_{*},z)$ among the different stellar
mass bins). The slight increase in the probability of hosting AGN in
passive galaxies is hinted at in the highest redshift bin, while
in the lowest one, star-forming galaxies appear to have a higher
probability of hosting AGN with low specific accretion rate, but the
number statistics is far too low to make any firmer statement. 
Even the size of our COSMOS sample is not large enough, once we
subdivide the sources into spectral class, X-ray luminosity, stellar
mass and star formation properties of the hosts.

\section{Discussion}
\label{sec:discussion}

Consistently with earlier studies, X-ray selected AGN are found to
reside mainly in massive galaxies, while optically selected Seyfert--2 galaxies, which
are also the low luminosity ones, seem to prefer galaxies with smaller
masses.  
The total stellar masses of the AGN host galaxies in our sample 
range from 10$^{10}$M$_{\sun}$ to 10$^{11.5}$M$_{\sun}$ with a
peak at 10$^{10.9}$M$_{\sun}$. No difference is found between X-ray
obscured and unobscured AGN which on average show the same mass
distribution.  

In accordance with their high stellar masses, AGN hosts are more likely
to be found in galaxies with red colors.
Many authors have pointed out that the trend with color was just
a mass driven effect and that removing the mass dependency would make
it disappear \citep{Silverman2009,Xue2010,Mainieri2011,Rosario2011}.  
Indeed, several previous studies
\citep[e.g.][]{Silverman2008GV} claimed that AGN were found mainly in
a region of intermediate colors, the so called `green valley'. This was 
interpreted as evidence of a close association between the migration
of galaxies from the `blue cloud' to the `red sequence' and the
feedback from active nuclei. However, in most of these works, the AGN
component was not subtracted in the computation of the host colors,
and the dust extinction correction could not be properly assessed, 
leading to a possible blue contamination from the AGN and a higher
incidence of AGN in the blue cloud and green valley.  These two effects could explain our findings. 
In fact, in disagreement with the above studies, we find that compared to normal galaxies of the
same mass, galaxies hosting an AGN are on average (slightly) redder and form stars less rapidly.  
 It is only by properly taking into account both the dust
extinction properties and the contamination from the nuclear light of
the entire galaxy population that we can assess the true incidence of AGN as a
function of star formation in their hosts. 
However, we cannot exclude that there is a difference in 
color between low luminosity AGN and intermediate-high luminosity ones and 
that this differences are, at least partially, responsible for the discrepancy between our result 
and previous studies based on deep fields \citep[e.g.][]{Silverman2008GV, Aird2012}.  

In agreement with the standard unified model \citep{Antonucci1993} and
contrarily to what (maybe naively) expected from the most popular
models of AGN 
fueling and feedback \citep{Hopkins2008}, and with a number of caveats regarding the
robustness of color determination in type--1 AGN, 
we do not find any significant difference between the
host colors of obscured and unobscured AGN. 
We note here that recent works by \citet{Pierce2010} and \citet{Rosario2011} 
found possible evidence of a weak anti-correlation between the color of
the AGN hosts and the hydrogen column density of X-ray obscuring gas
$N_{\rm H}$, in the sense that redder AGN seem to have on average
higher values of $N_{\rm H}$. As our classification into obscured and
un-obscured AGN is a complex one, and mainly based on optical
spectroscopy, it is not straightforward to compare our results with
those of  \citet{Pierce2010} and \citet{Rosario2011}, but a common
wisdom is emerging that (at least mildly) obscured AGN are not to be
found among the most active star-forming galaxies.

When looking in detail at the relationships between AGN luminosity and
host stellar masses, in agreement with \citet{Aird2012}, we find that
the probability to host an AGN {\it of a given
specific accretion rate $L_{\rm X}/M_*$} (a proxy for the SMBH Eddington ratio) 
is practically independent of the  galaxy stellar mass, but
depends strongly on redshift and on the specific accretion rate
itself, i.e. there is a strongly 
 decreasing probability to host an AGN with increasing $L_{\rm X}/M_*$
 ratio.

This also means that the higher
AGN fraction is observed in massive red galaxies simply because it is
easier to detect the more prevalent, low Eddington ratio AGN in
more massive galaxies, which are more likely to host the more massive
SMBH.

The specific accretion rate distributions we can indirectly infer
from the observed probability of a galaxy to host a AGN is better
described by a power-law function, the slope and
normalization of which appear to
increase with redshift. 
We note here that \citet{Kauffmann2009},  studying the present-day growth rates
of black holes in SDSS galaxies using the [OIII] luminosity as a
tracer of AGN accretion activity and stellar masses as tracer of the BH masses, 
pointed out the existence of two
trends in the Eddington ratio distribution as a function of the
stellar population (a `power-law' and a `log normal'). These observed
trends have been interpreted by them as the indication of two distinct
accretion ``modes'' for black holes: the log-normal distribution
identifies a mode in which the supply of gas is ample (``feast''), 
while the power-law distribution corresponds to a  mode where
accretion is limited by the available supply of gas (``famine'').
Another interpretation has been proposed by \citet{Hopkins2008}, who
identify the origin of the bi-modality in the evolution of triggering
rates (mergers) in combination with a luminosity-dependent AGN
lifetime.  
Our result shows that at higher redshift the picture might be
different, as we do not find clear sign of a log-normal component in
the specific accretion rate distribution, even when dividing the population of
type--2 AGN hosts into actively star-forming and quiescent. This
difference could be indicative of an interesting evolutionary effect to be
further investigated, although we cannot exclude possible selection effect due
to the fact that the work from \citet{Kauffmann2009} is based on a
pure optically selected sample for which the AGN luminosity is computed based on the highly uncertain narrow line [OIII]-to-bolometric luminosity conversion. 
%Also, our specific accretion rate distributions are based on a proxy for which incompleteness severely affect our data in many of the available redshift and stellar mass bins. 
Nonetheless, the apparent 
universality of the observed probability distribution $p_{\rm
  AGN}(L_{\rm X}/M_{*}|z)$ suggests that, at least in the redshift
range populated by our COSMOS sample, the
same physical processes might be responsible for triggering and fueling AGN
activity in galaxies of different masses.  

We also find that 
the overall probability to host an AGN of any given $L_{\rm X}/M_{*}$ 
(as determined by the
distribution normalization) increases rapidly
with redshift as (1+z)$^{4}$. Such an evolution can
be due to either an increase in the number of sources accreting
(vertical shift) or a shift of the characteristic Eddington ratio
distribution towards higher values (horizontal shift). The second
possibility would also explain the ``AGN downsizing'' \citep[the
comoving number density of high luminosity objects peaks at higher
redshifts than low-luminosity
AGN.][]{LaFranca2005,Hasinger2005,Bongiorno2007} as a consequence of
the different Eddington ratio distribution at different redshift.  

The fact that the probability of a galaxy to host an AGN of a given
specific accretion rate evolves with redshift in the same way as
the  overall evolution of the specific star formation rate of the galaxy population
\citep{Karim2011} strongly suggests that AGN activity and star formation
are globally correlated, at least in a statistical sense. 
From the physical point of view, the drop with time in the
global star formation rate density reflects a drop in the availability
of cold gas in galaxies to fuel star formation and potentially
AGN. In that respect, AGN seem to follow the evolution of the galaxy
population as a whole, a possibility reinforced by our detailed study
of the position of AGN hosts in the sSFR-stellar mass diagram
(section~\ref{sec:AGNfr_sfr}).

\section{Summary and Conclusion}
\label{sec:summary}

In this paper we studied a sample of $\sim$ 1700 obscured and
unobscured AGN, selected in the COSMOS field using X-ray and optical
(spectroscopic) criteria. Of these, 602 are unobscured type--1 AGN (430 with
spectra) and 1100 obscured type--2 AGN (549 with spectra) with $L_{\rm
  bol}$ ranging from 10$^{43}$ to 10$^{47}$ erg s$^{-1}$.  
Thanks to the wide COSMOS multi-wavelength coverage we were able to
study in detail the source SEDs and to fit the observed fluxes with a
two component model based on a combination of AGN and host-galaxy
templates, accounting for dust extinction in both components independently.  
This method allowed us to decompose the entire spectral energy
distribution into a nuclear AGN and a host galaxy component for almost
all the sources and to derive robust measurements of both the AGN and
the host galaxy properties (i.e. rest-frame colors, masses and SFRs). 
A test considering 1-component (galaxy) versus 2-components
(galaxy+AGN) SED fitting technique, discussed in Appendix
\ref{sec:1component}, revealed that while considering only the galaxy
component can be used to roughly estimate the blue colors of obscured
AGN hosts, caution has to be used when deriving red magnitudes
(e.g. M$_K$) and physical parameters, e.g. stellar mass and SFRs. On
the contrary, for unobscured AGN hosts no study should be performed
without properly taking into account the AGN component; even with this
caution, we find that for unobscured AGN while total stella masses of
their host can be determined quite reliably, start-formation rate are
too affected by nuclear light contamination to be deemed robust.
 
Our main findings can be summarized as follows: 

\begin{itemize}
\item COSMOS AGN host galaxy masses range from $10^{10}M_{\sun}$ to
  $10^{11.5}M_{\sun}$ with a peak at $\approx 10^{10.9}M_{\sun}$. No significant
  difference is found in the stellar mass distributions of X-ray
  obscured and unobscured AGN hosts. Only optically selected
  Seyfert--2 galaxies without X-ray counterpart populate the tail in the distribution at
  lower masses down to 10$^{8.5}M_{\sun}$. 

\item  Host galaxies of obscured AGN span a very wide range of SFRs, with X-ray
  selected AGN populating the entire range, while optically selected Seyfert--2 galaxies 
   being hosted mainly in galaxies with low SFRs (and very low
  masses). As extensively discussed in Appendix
\ref{sec:sfr}, we find that above $\sim 20 M_{\sun}$ yr$^{-1}$ the SFRs derived from the
  SED and the one derived from the {\it Herschel} FIR observations are in good agreement.  
At low SFR ($< 20 M_{\sun}$ yr$^{-1}$), however, the disagreement
becomes evident, with our SED fitting method indicating substantially
lower SFR than the FIR one.

We interpret this discrepancy as the possible combination of two effects: (i)
contamination from star-formation-related emission
 in the NIR not properly taken into account in
the SED fitting and (ii) AGN contamination in the FIR bands not taken
into account in the FIR SFR estimates.
We demonstrated that at least for the XMM-COSMOS sample, which is dominated by high luminosity AGN, the driven effect is the AGN contamination on the 60$\mu m$ luminosities (see Appendix \ref{sec:sfr} for e detailed description), which, on the contrary, is low or even negligible for moderate and low luminosity AGN samples.

\item In contrast to what was found by \citet{Cardamone2010}, no clear
bi-modality is seen in the color magnitude diagram of the COSMOS AGN hosts.
At any redshift, the hosts of the AGN in our sample are mainly
massive, redder galaxies.  
In many previous studies on AGN samples of much smaller size, 
AGN were found to have a higher incidence in star forming
galaxies of blue colors \citep{Silverman2009,Xue2010}, once mass matched
samples of active and non-active galaxies are considered. 
Even when accounting properly for the mass-color degeneracy, and when
looking at mass-matched sample, we 
do not confirm these findings with our much improved statistics. Even
at fixed stellar masses, well above the completeness limit of the
COSMOS sample, we find a slight trend for AGN hosts to be redder (by
$\sim 0.1$--$0.2$ magnitudes) than the non active galaxies. 
We argue that the main reason for such a discrepancy is the proper
subtraction of the AGN light from our SED decomposition technique,
complemented by a uniform and systematic treatment of dust extinction
in both parent and AGN host galaxy sample. We caution,
  however, that degeneracies in color-magnitude diagrams may easily
  arise due to alternative choices of spectral templates and residual
  issues with the AGN-host decomposition technique.

\item For a galaxy of any given mass above the stellar mass
  completeness limit of the COSMOS survey, 
the probability to host an AGN of a given X-ray luminosity
  increases with stellar mass, and this holds true for any
  value of the X-ray luminosity we are sensitive to. 
  In fact, against the naive expectation that more luminous AGN should be
  found in more massive galaxies and less luminous ones in lower mass
  galaxies, we find that AGN {\it of any intrinsic luminosity} are more
  common the more massive the galaxy is.   

\item We have used the observable ``specific accretion rate'', i.e. the
  ratio between X-ray luminosity and
  host galaxy stellar mass as a proxy for the Eddington ratio of the
  growing black hole. By doing so we find that 
  the probability for a galaxy to host an AGN {\it of any given
    Eddington ratio} (or specific accretion rate) is roughly independent of
  the host galaxy stellar mass, but strongly decreases with increasing
  $L_{\rm X}/M_{*}$, and we also found a clear evidence of a
    break in the specific accretion rate distribution at values
    consistent with the Eddington limit.
  These results imply that the higher incidence of AGN
  observed in massive galaxies by essentially all multi-wavelength
  surveys is just a consequence of the fact that low specific accretion
  rate objects are more common than high specific accretion rate ones,
  and, at the same time, low  $L_{\rm X}/M_{*}$ objects in low mass
  galaxies drop out of flux-limited AGN samples. This effect, closely
  associated to the broad Eddington ratio distribution of AGN was 
  anticipated in Merloni \& Heinz (2008) and has been confirmed
  observationally by \citet{Aird2012}.

\item We found that the probability for a
  galaxy to host an AGN {\it of any given
    specific accretion rate} strongly increases with redshift, approximately as
  (1+z)$^{4}$ for SMBH accreting at about 10\% of the Eddington
  limit, where most of the SMBH mass assembly is supposed to take
  place. Such a very strong evolutionary trend follows closely the overall evolution of the
  specific star formation rate of the galaxy population
  \citep{Karim2011}. This is yet another strong indication that AGN activity and star
  formation are globally correlated.  

\item To study in detail the star formation properties of the COSMOS
  AGN hosts, we have looked at the type--2 (obscured) AGN only, in
  order to avoid any bias/contamination from the inaccurate
  subtraction of the dominant blue nuclear light in type--1 objects.
  Type--2 AGN hosts have on average the same, or slightly lower SFRs than
  non active  galaxies of the same mass and at the same redshift. 
We find that the fraction of AGN hosted in strongly starbursting galaxies is very low ($<$10\%) while is higher for main sequence galaxies (27\% to 37\%) and quiescent ones (58\%-66\%). 
Compared to {\it Herschel}-based results \citep[e.g.][]{Mullaney2012}, we find a much higher fraction of quiescent host. We tested that the discrepancy between the two results is explained by the differences in the SFR (especially below 20 M$_{\odot}$ yr$^{-1}$) derived from SED fitting and FIR which have been attributed to the combination of the two opposite effects explained above.

The
  probabilities of star-forming or quiescent galaxies to host a
  type--2 AGN also decline as a power-law as a function of specific
  accretion rate and there are tantalizing hints of difference in the
  two populations and in their redshift dependence. However, more
  detailed investigations of the incidence of AGN in galaxies as a
  function of redshift, stellar mass, star-formation rate and nuclear
  luminosity are hampered by the limited statistics of the
  sample. Wide area surveys probing volumes at $z>1$ comparable to that
  explored by SDSS in the local Universe will be needed to properly
  address the fundamental questions on the physical nature of the
  AGN-galaxy co-evolution.

\end{itemize}

The most striking piece of evidence that emerges from our study is the
fact that AGN hosts do follow quite closely the overall evolution of the stellar
build-up of the parent galaxy population and populate both
   star-forming and more passive galaxies almost equally. 

We argue that, on the basis of the evidence presented in this work,
it is hard to find strong indication that AGN play a direct
role in shaping the global properties of their host galaxies, or their
evolution. The same, probably stochastic, process of AGN activation
and triggering  
seems to be in place in galaxies of all
masses, with a slight predominance in red ones with specific star
formation rates below the nominal threshold that distinguish actively
star-forming from quiescent galaxies. Because of the robustness of
this picture across the wide redshift range probed, we believe that
whatever physical process is 
responsible for triggering and fueling AGN activity, it must be the same
between $z\sim$2.5 and  $z\sim$0.3, but must decrease in frequency or
shift towards lower accretion rates. 
Although AGN activity and star
formation appear to have a common triggering mechanism, we do not find
strong evidence of any 'smoking gun' signaling powerful AGN
influence on star forming galaxies SFR. The extent
to which this lack of evidence can be meaningfully used to constrain
and/or rule out theoretical models of AGN feedback, remains to be investigated.

\section*{Acknowledgments}

We thank the referee Dr. James Mullaney for the constructive comments which helped improving the manuscript and for the interesting discussion. 
We also thank R. Hyckox and K. Schawinski for the useful discussions.  AB work 
is supported by the INAF-Fellowship Program.
The HST COSMOS Treasury program was supported through NASA grant
HST-GO-09822.  This work is mainly based on observations obtained with
XMM-Newton, an ESA Science Mission with instruments and contributions
directly funded by ESA Member States and the USA (NASA), and with the
European Southern Observatory under Large Program 175.A-0839,
Chile. In Germany, the XMM-Newton project is supported by the
Bundesministerium f{\"u}r Wirtschaft und Technologie/Deutsches Zentrum
f{\"u}r Luft und Raumfahrt (BMWI/DLR, FKZ 50 OX 0001), the Max-Planck
Society, and the Heidenhain-Stiftung. 
In Italy, the XMM-COSMOS project is supported by ASI-INAF grants
I/009/10/0 and ASI/COFIS/WP3110,I/026/07/0.  
Part of this work was supported by INAF PRIN 2010 `From the dawn of galaxy formation to the peak of the mass assembly'. 
%We also used data collected at the Subaru Telescope, which is
%operated by the National Astronomical Observatory of Japan; Kitt Peak
%National Observatory, Cerro Tololo Inter-American Observatory, and the
%National Optical Astronomy Observatory, which are operated by the
%Association of Universities for Research in Astronomy, Inc. (AURA)
%under cooperative agreement with the National Science Foundation; the
%National Radio Astronomy Observatory which is a facility of the
%National Science Foundation operated under cooperative agreement by
%Associated Universities, Inc; and the Canada-France-Hawaii
%Telescope with MegaPrime/MegaCam operated as a joint project by the
%CFHT Corporation, CEA/DAPNIA, the National Research Council of Canada,
%the Canadian Astronomy Data Centre, the Centre National de la
%Recherche Scientifique de France, TERAPIX and the University of
%Hawaii.  
We gratefully acknowledge the contribution of the entire COSMOS
collaboration; more information on the COSMOS survey is available at
http://www.astro.caltech.edu/cosmos.

%\bibliographystyle{mn2e_mod}

%\bibliography{angi}

\appendix

 \section{2-components versus 1-component SED fitting}
 \label{sec:1component}

\begin{figure*}
\centering
\includegraphics[width=5.5cm,clip]{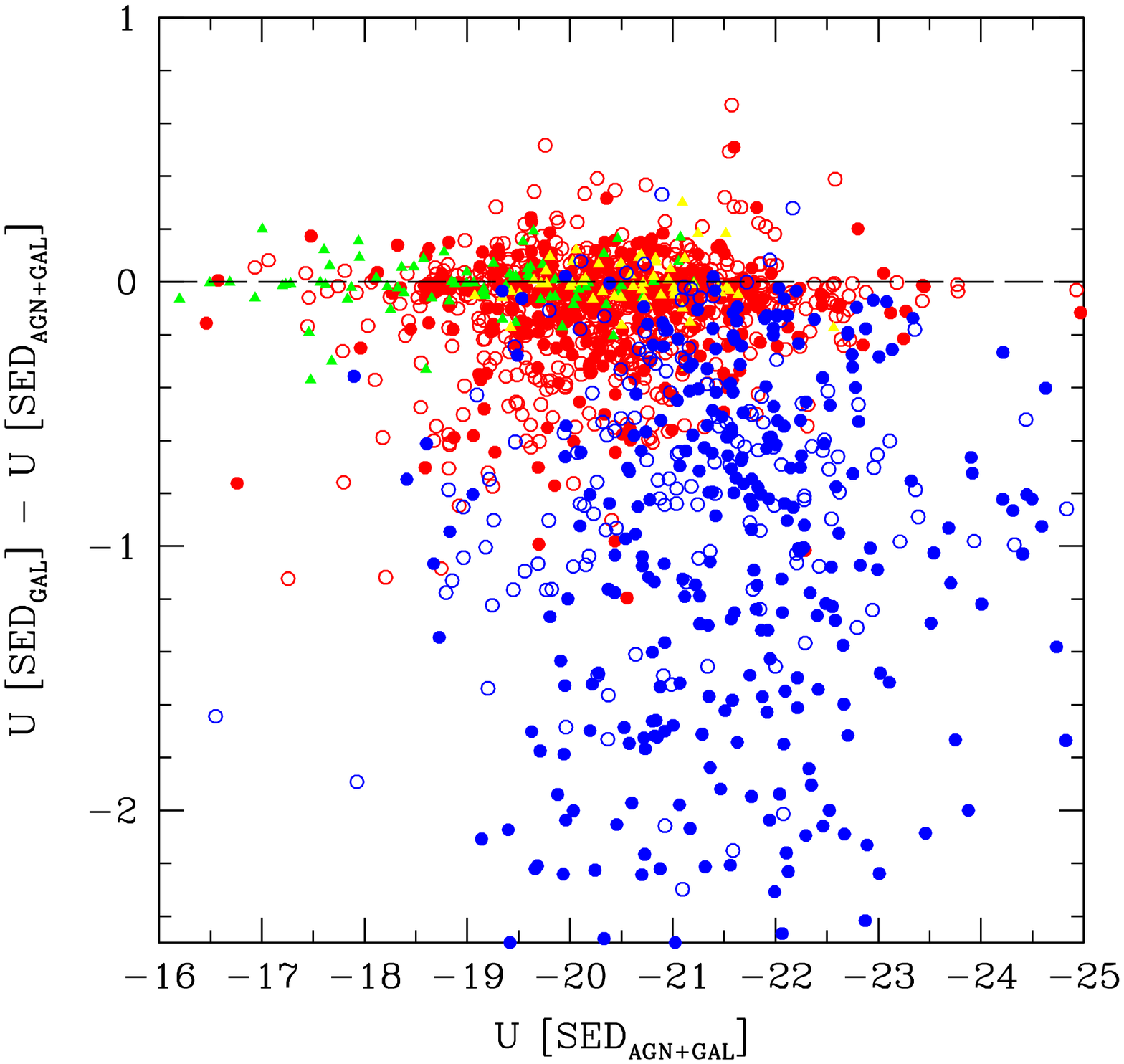}
\includegraphics[width=5.5cm,clip]{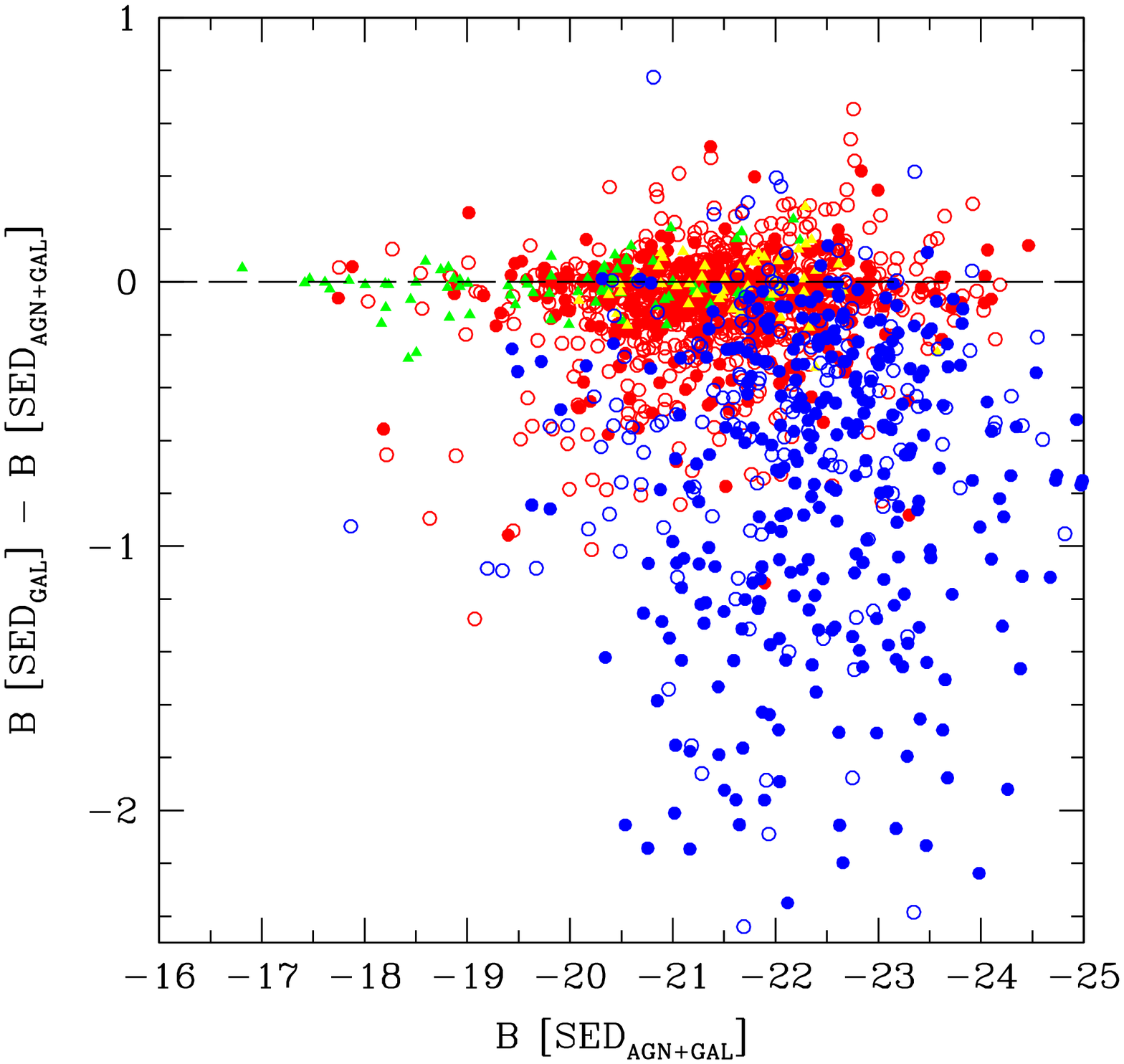}
\includegraphics[width=5.5cm,clip]{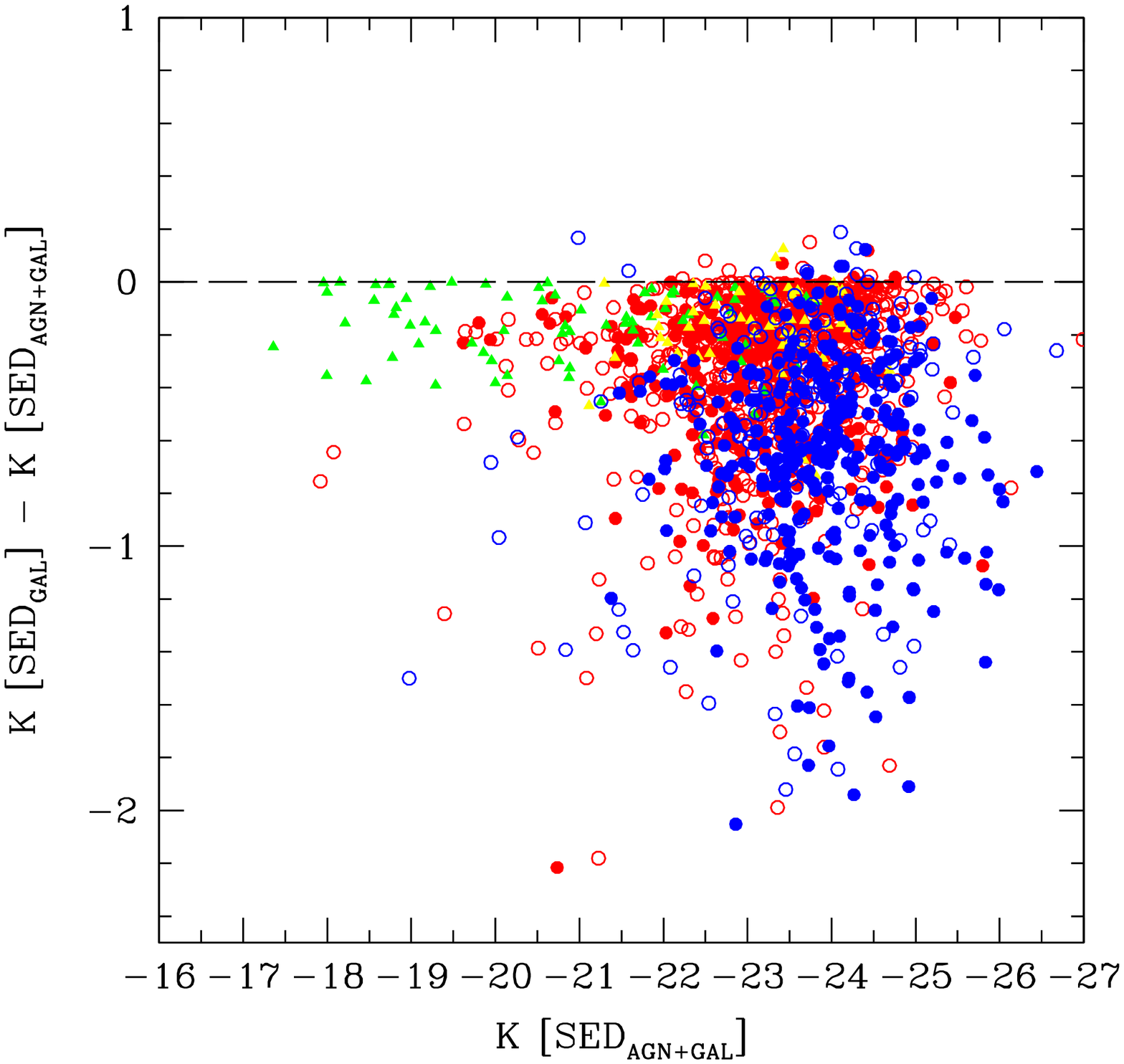}
\caption{Comparison between the rest frame U-, B- and K-band
  magnitudes of our AGN host sample obtained using the 2-component SED
  fitting presented in this work versus the one obtained using a
  standard SED fitting considering only the galaxy component.} 
\label{fig:magn_compare}  
\end{figure*}

In this section we will discuss the differences between the measured
galaxy parameters that one obtains when using the SED fitting with two components (AGN and galaxy)
compared to what would be obtained ignoring the AGN component, as
in other studies \citep[e.g.][]{Silverman2009,Rosario2011}.  
The color code for the symbols in the figures is the same as used throughout paper.

\subsection{Rest-frame magnitudes and galaxy colors:}

Rest frame magnitudes in any given band are computed integrating the rest-frame galaxy
template found as best fit solution in the corresponding
filter. They are thus related to the ``type'' of galaxy chosen as best
fit (which can introduce a difference up to 4 magnitudes) and, more
important, to the normalization of the template, which is directly
related to the total flux emitted by the source. 
In case of obscured AGN, the total emission in the U- and B-bands
($\lambda_{\rm eff}\sim 3800$\AA\ - 4500\AA) is hardly influenced by
the nuclear emission which is, at these wavelengths, partially or
entirely absorbed (see lower panels Fig. \ref{fig:sed_decompose}). For
this reason the shape of the chosen galaxy at these wavelengths and
the normalization of the template should not strongly change when
considering (or not) the AGN component. In order to test the effect of
using one or two components fitting, we re-run our code using only galaxy
templates. As shown in Fig. \ref{fig:magn_compare}, the U and B rest
frame host galaxy magnitudes of obscured AGN with (e.g. U[SED$_{\rm
  AGN+GAL}$]) and without (e.g. U[SED$_{\rm GAL}$]) the AGN component
lie on the one-to-one relation with a roughly symmetric scatter of
about 0.4 magnitudes (red, yellow and green symbols). On the contrary,
the correlation is much worse for unobscured AGN (blue circles) whose
U- and B-band rest frame magnitudes with and without AGN can be up to
more than 2 magnitudes different. Moreover, the spread is not
symmetric and not using the AGN component always leads to a much
brighter galaxy. This is due to the fact that unobscured AGN emission
is strong in the U and B bands (see upper panels of
Fig. \ref{fig:sed_decompose}), i.e. not taking into account the AGN
component corresponds to additional attributed flux to the galaxy
hence resulting in a overestimation of the galaxy luminosity.  
The conclusion is different if we look at the comparison in the K-band
($\lambda_{\rm eff}\sim2.2\times 10^4$\AA). Contrarily to the U- and
B-band, at these wavelengths the contribution of AGN emission can be
relevant even in the obscured sources (see e.g. the bottom right panel
of Fig. \ref{fig:sed_decompose}). For this reason,  the K-band
magnitude of both obscured and unobscured AGN hosts will be
overestimated (i.e. up to 2 magnitudes brighter) if we do not take
into account the AGN component. 

The objects whose host luminosities are less affected by the presence
of an AGN component are the optically selected type--2 AGN 
without X-ray emission (triangles). For these objects the obscuration
must be very high in order to suppress the AGN emission up to the
K-band wavelengths. As suggested is Sec. \ref{sec:L12LX}, these
sources are likely Compton-thick AGN and hence a high level of
obscuration is expected.

\begin{figure}[h]
\centering
\includegraphics[width=6cm,clip]{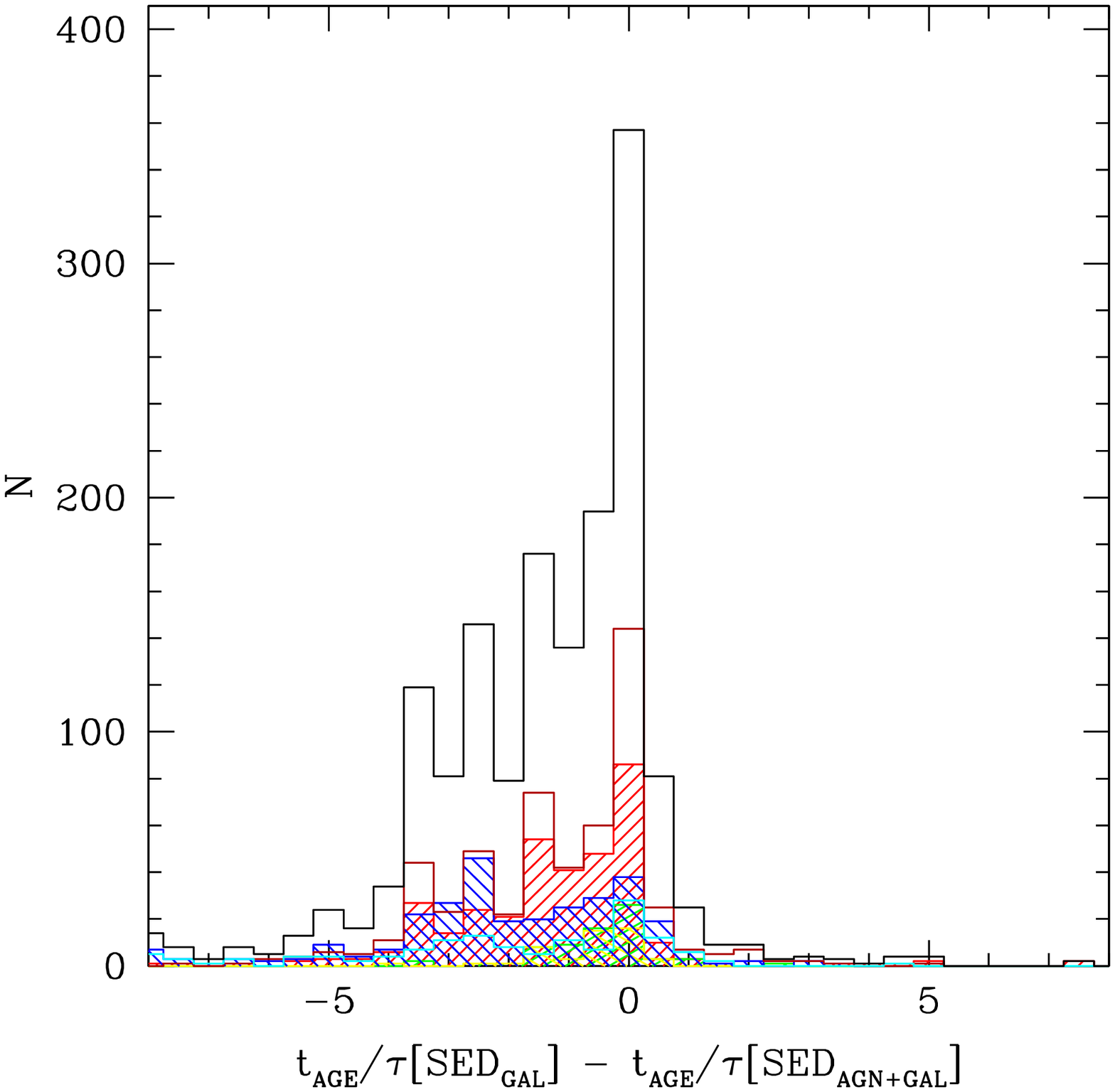}
\caption{Comparison between the SED fitting parameters
  t$_{age}$/$\tau$ obtained using the 2-component SED fitting
  [GAL+AGN] presented in this work versus the one obtained using a
  standard SED fitting considering only the galaxy component.} 
\label{fig:TAU}
\end{figure}

\begin{figure*}
\includegraphics[width=8cm,clip]{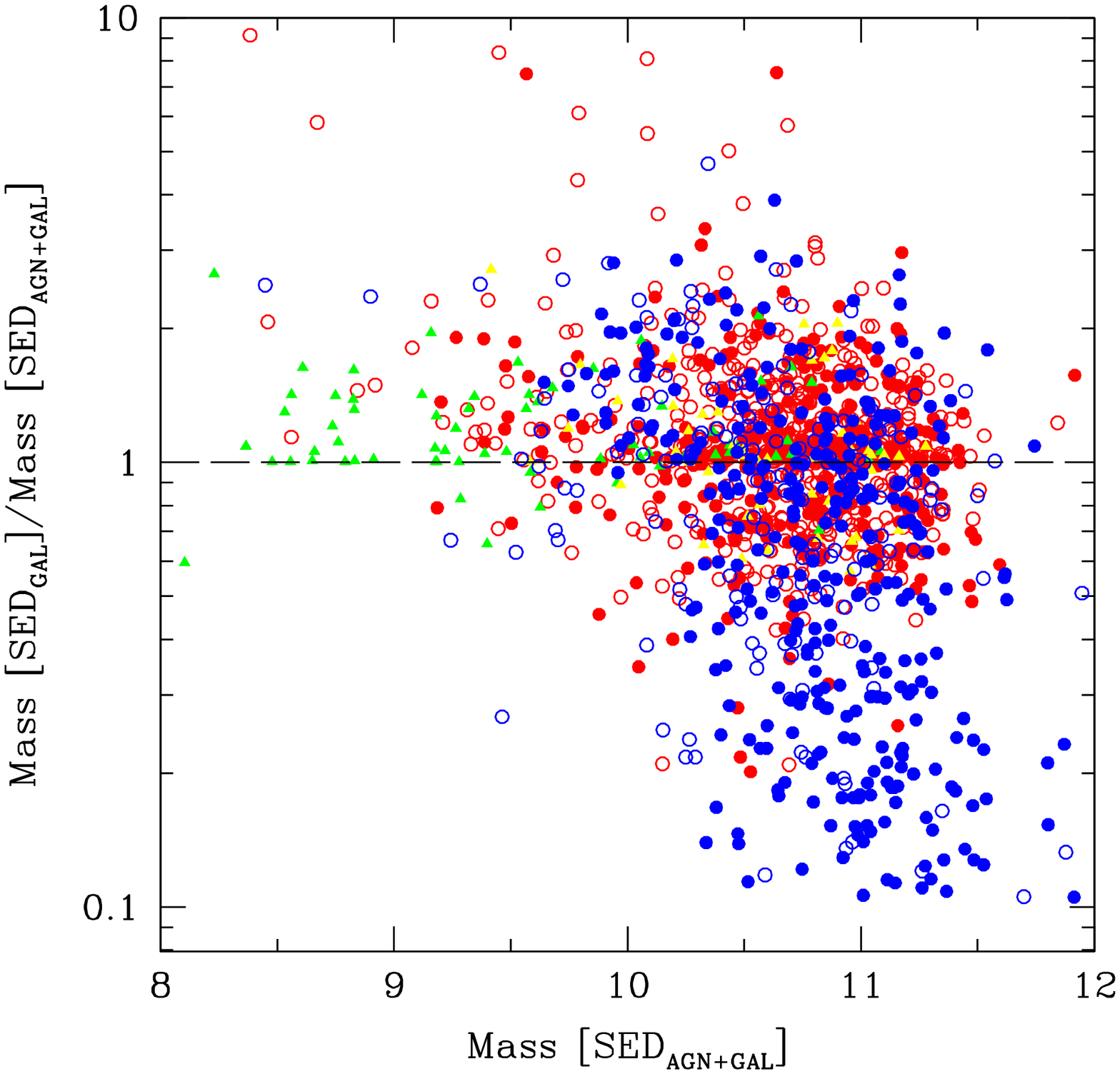}
\includegraphics[width=8cm,clip]{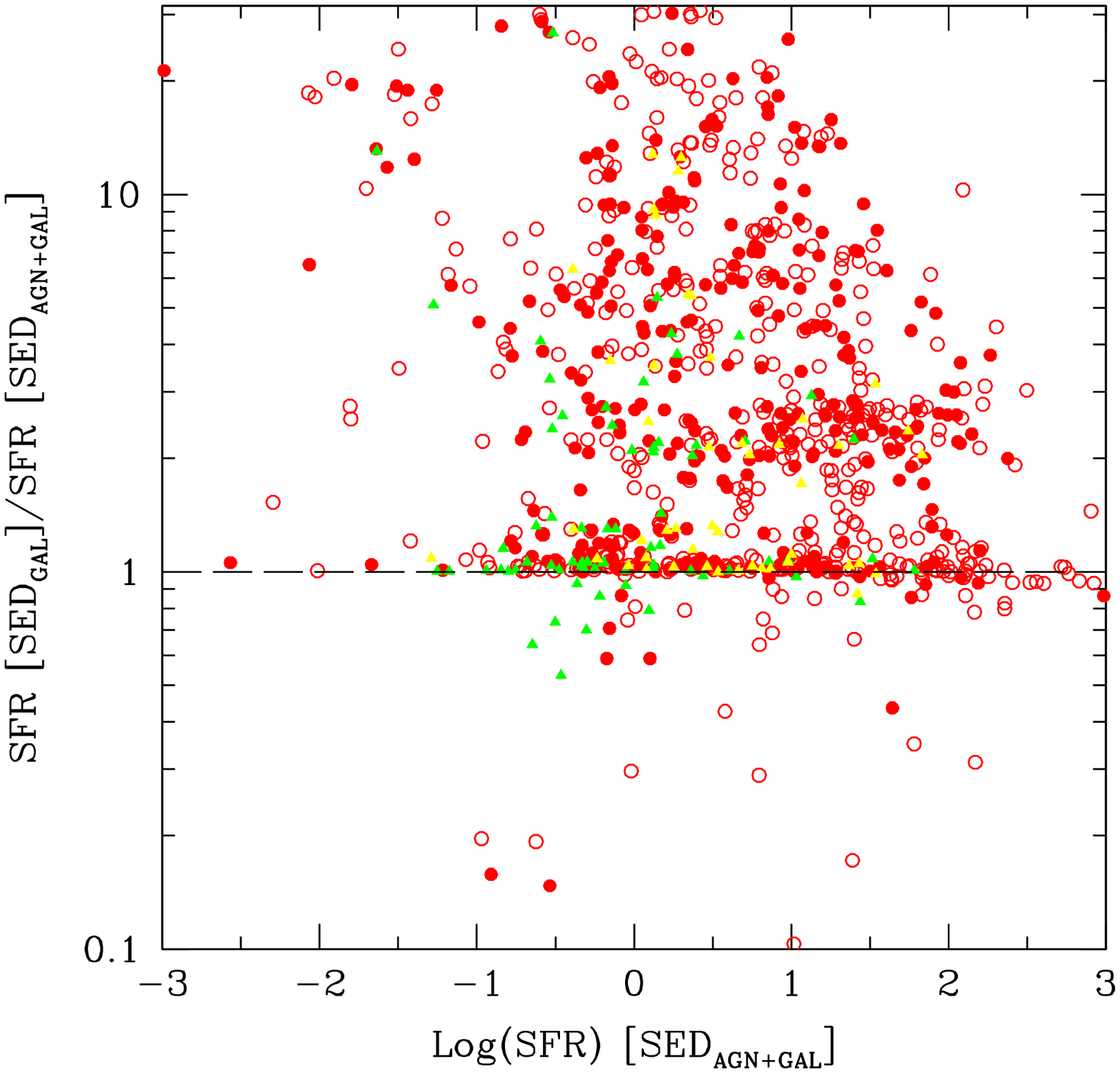}
\caption{Comparison between the stellar masses and SFR of our AGN
  hosts obtained using the 2-component SED fitting presented in this
  work versus the one obtained using a standard SED fitting
  considering only the galaxy component.} 
\label{fig:MSFR_compare}
\end{figure*}

\subsection{Host galaxy stellar mass and SFR}

Apart from the rest-frame luminosity in various bands, 
the other two important parameters we derive from the SED fitting are
the stellar masses and the star formation rates. Besides the dependence
on the template normalization, these physical parameters are linked
also to the principal parameters used in the SED fitting, i.e. the
e-folding time ($\tau$), the age since the beginning of star formation
(t$_{age}$), and to some degree the extinction (E(B-V)). All 
SFHs considered in this work have an exponentially
decreasing shape with $SFR\propto$ e$^{-t_{age}/\tau}$ (but one,
which is constant). For such an exponentially decreasing SFH,
a large value of t$_{age}$/$\tau$ implies a quasi passive evolution
with a negligible ongoing SF and that the majority of the stars 
in the galaxy are old \citep{Grazian2007}.

Fig. \ref{fig:TAU} shows the difference in the $t_{age}/\tau$ ratio
when considering or not the AGN component in the SED fitting. In
almost all cases, not considering the AGN component results in lower
values of $t_{age}/\tau$, which directly translates into galaxies
with higher star formation rates, as visible in the right panel of figure
\ref{fig:MSFR_compare}. The right panel of this figure shows, for obscured type--2 AGN in
the COSMOS sample, the
ratio between the  SFR obtained without the AGN component and once the
AGN component is taken into account. For about 40\% of the objects,
the two values agree within a factor of 2, while for the rest
of the objects the SFR estimated without AGN component is up to more
than 20 times higher than the one obtained including the AGN.

Finally, the measured stellar masses show a non-negligible spread in both
directions, for both obscured and unobscured AGN. As showed in the
left panel of Fig. \ref{fig:MSFR_compare}, the masses computed not
including the AGN component range from 6 times higher (2 times if we
consider only spectroscopic redshifts) to 10 times smaller (i.e. in
unobscured AGN).

Summarizing, for obscured AGN a standard galaxy SED fitting technique 
that does not include a nuclear component 
results in small differences in rest frame U and B absolute magnitudes
($\sim$0.4 mag brighter) and larger ones for the K-band magnitudes
($\sim$0.8 mag brighter). The mass and SFR estimates are instead much
more affected (up to a factor of 6 in mass, and SFR up to 20 times
higher).   
This is not the case for optically selected type--2 AGN with no X-ray
counterpart for which, due to the extremely high obscuration, the
differences in magnitudes, mass and SFR are relatively small when
considering (or not) the AGN component. 
For unobscured AGN on the contrary, all derived quantities (absolute
magnitudes, stellar mass and SFR) are strongly offset from the values
obtained when taking into account the AGN component.
 
In conclusion, while the SED fitting technique considering only the
galaxy component can be used to roughly estimate the blue colors of
obscured AGN hosts, caution has to be used when deriving red magnitudes
(e.g. M$_K$) and physical parameters, e.g. stellar mass and SFRs. On
the contrary, for  unobscured AGN hosts no study can be performed
without properly taking into account the AGN component.

\section{Comparison between different SFR indicators}
\label{sec:sfr}

Evidences of star formation in galaxies can be seen in both the UV and IR bands.
The UV emission is a direct measure of the starlight produced by a
mixture of young (O, B and A type) stars. Using the UV emission to
measure the SFR suffers from the limitation that part of the emitted
radiation is absorbed, which is not taken into account.  
On the other hand, when using only the IR emission to constrain the
SFR, the assumption is that a dominant fraction of the total bolometric
stellar luminosity of the star-forming population is absorbed and
re-radiated as thermal infrared dust emission.  
The most robust way to compute the SFR is thus the sum of both the
unobscured UV star formation with the obscured one re-emitted in the
IR by dust  (SFR$_{\rm UV+IR}$).

Thanks to the advent of the \textit{Herschel} Space Observatory, deep
far-infrared observations of the COSMOS field are now available. 
Here, we use the deep COSMOS PACS 100 and 160$\,\mu$m observations of
the PACS Evolutionary Probe
\citep[PEP\footnote{http://www.mpe.mpg.de/ir/Research/PEP};][]{Lutz2011}
guaranteed time key program, reaching a $3\sigma$ limit of
$\sim$$\,4$~mJy and $\sim$$\,10$~mJy at 100 and 160~$\mu$m,
respectively \citep[see][]{Berta2011}. 
Using these observations we infer the obscured SFR of our host
galaxies and test the accuracy of SFR derived from SED fitting. 

\begin{figure*}
\includegraphics[width=8cm,clip]{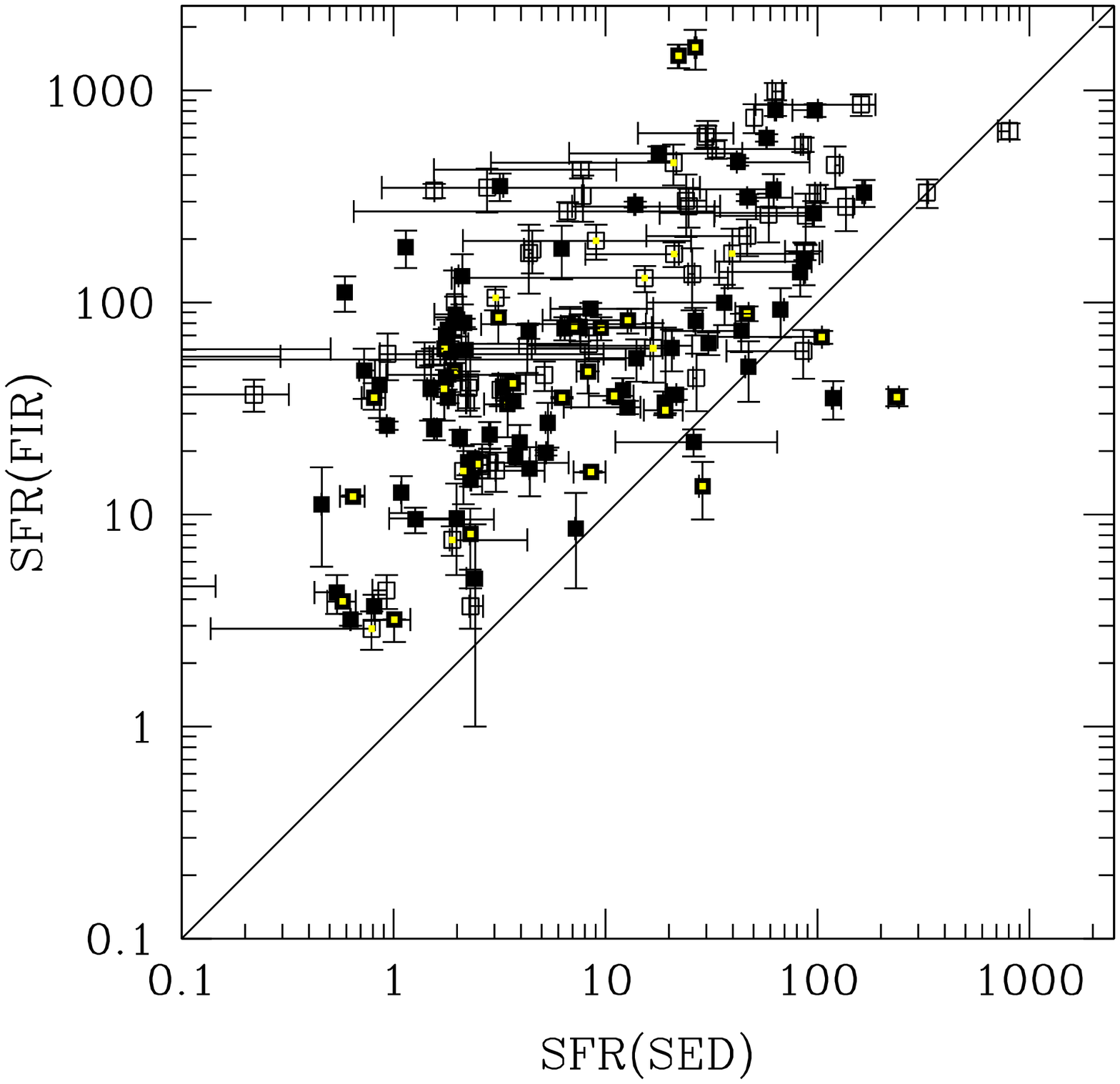}
\includegraphics[width=8cm,clip]{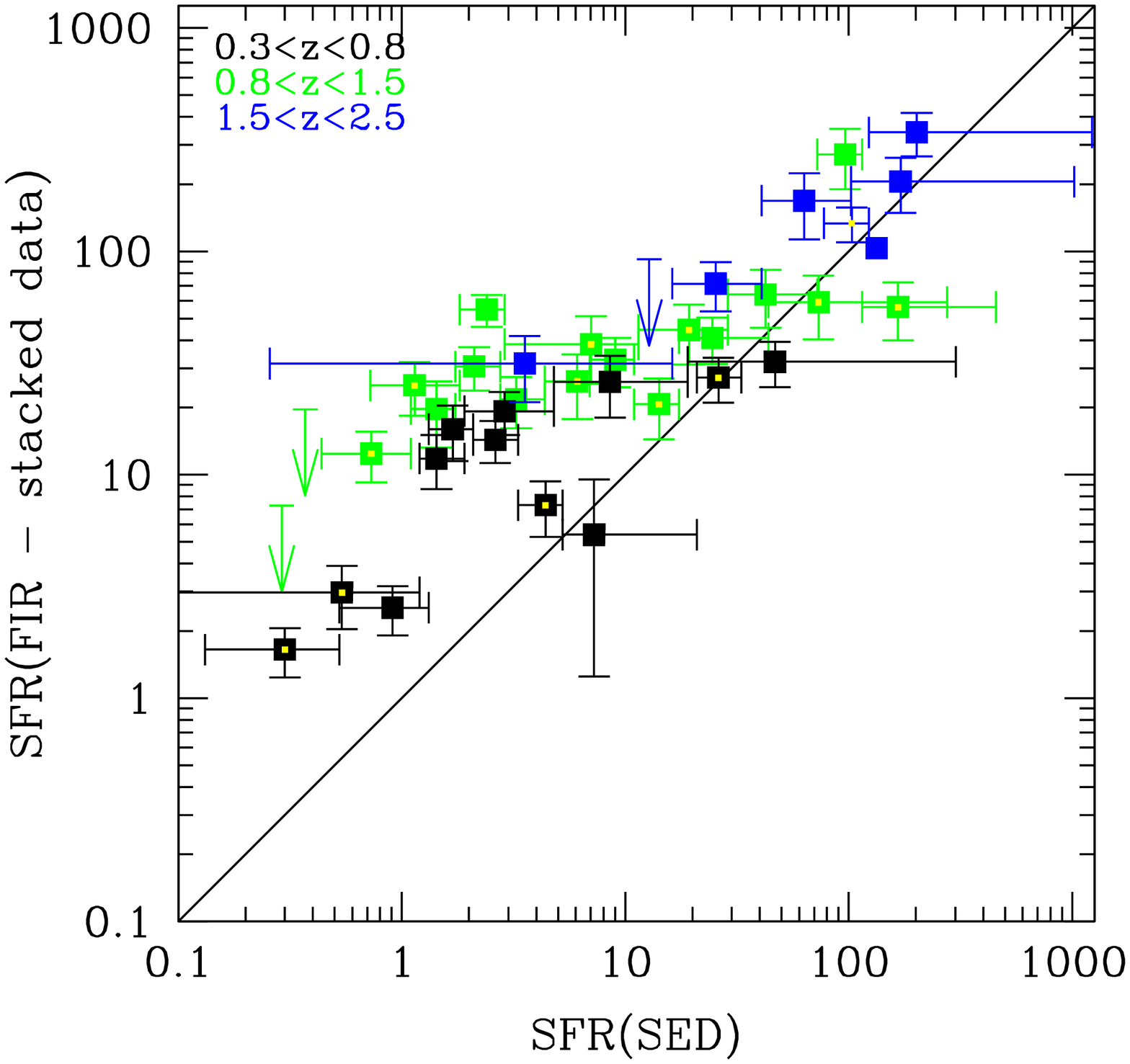}
\caption{
\textit{Left:} Comparison between the SFR derived from the SED fitting
and the one derived from the FIR band for the sources detected in at least one FIR band. Open and filled squares corresponds to objects with photometric and spectroscopic redshift respectively. 
\textit{Right:} Comparison between the SFR derived from the SED fitting
and the one derived from FIR using stacked data (binning in SFR$_{\rm sed}$ and
redshift). Different colors correspond to different redshift ranges:
0.3$<$z$<$0.8 black, 0.8$<$z$<$1.5 green and 1.5$<$z$<$2.5
blue. } 
\label{fig:SFRsed_SFRIR}
\end{figure*}

Only a small fraction of our sources ($\sim$$\,10\%$) are detected above the $3\sigma$ limits of the PEP catalogues. For these sources we computed the SFR using their infrared luminosities
and assuming the L$_{IR}$ to SFR conversion of \citet{Kennicutt1998}, i.e., $SFR=10^{-10} L_{IR}$ for a Chabrier IMF. The IR luminosity L$_{IR}$ is derived from the monochromatic luminosity at the longest PACS wavelength available converted using the infrared SED templates of \citet{Chary2001} as it has
been proven to provide reliable conversions over a broad range of redshift \citep{Elbaz2011}.
In the left panel of of Fig. \ref{fig:SFRsed_SFRIR} we show the comparison between the SFRs computed using the IR data and the one
computed with the 2-components SED fitting. As visible,
the majority of the sources lie above the 1 to 1 relation with the
SFR computed from the SED being lower than the one derived
from the IR. This is driven by a selection effect due to the fact
that, given a broad distribution of SFR$_{IR}$/SFR$_{SED}$ values, the Herschel detected sources will be the most powerful ones corresponding by definition to the upper envelope of the SFR$_{IR}$/SFR$_{SED}$ distribution.

However, to test the accuracy of the SFR derived from SED fitting, we
have to rely also on a statistical approach using a stacking analysis. 
Images of galaxies in a given redshift and SFR bin are stacked,
allowing us to detect their typical far-infrared flux densities well
below the detection limits of the current PACS observations (the noise
of the stacked images scales roughly with the square root of the
number of the stacked sources). 
We start by separate our galaxies into the usual three redshift bins
(i.e., $0.3<z<0.8$, $0.8<z<1.5$ and $1.5<z<2.5$). 
Within each redshift bin, we then separate our galaxies per SFR bins. 
The size of the SFR bins are optimized in order to obtain the highest
number of bins with clear PACS detections (i.e., $>$$\,3\sigma$). 
The first SFR bin is initialized with a size of 0.1 dex and an upper
range equal to the highest SFR observed in the redshift bin. 
Then, we progressively decrease the value of the lower range of the
bin (using steps of 0.1 dex), until we obtain a $3\sigma$ detection in
at least one PACS passband. 
The lower range of the bin is then taken as the upper range of the
next SFR bin and the same procedure is repeated in order to find the
size of the new bin. 
The process is stopped once the lower range of the current bin reaches
the lowest SFR observed in the redshift bin. 
Photometry of the stacked images are measured using the PSF fitting
analysis presented in \citet{Magnelli2009,Magnelli2011} and using the
relevant PSFs and aperture corrections of the COSMOS field
\citep{Berta2011}.  
Errors are measured using a standard bootstrap analysis which takes
into account both the photometric uncertainties of the stacked images
and the intrinsic dispersion of the underlying galaxy population.

We derive the obscured SFR (i.e., SFR$_{\rm IR}$) of each redshift
and SFR bin using their stacked PACS flux densities and the
\citet{Chary2001} SED library. 
In the case where there is only one PACS detection (at 100$\,\mu$m
\textit{or} 160$\,\mu$m), SFR$_{{\rm IR}}$ is obtained simply using
the scaled \citet{Chary2001} SED library. 
In the case where there are two PACS detections (at 100$\,\mu$m
\textit{and} 160$\,\mu$m), we fit the PACS flux densities with the
Chary \& Elbaz (2001) SED library leaving the normalization of each
template as a free parameter. 
In both cases, SFR$_{{\rm IR}}$ is given by integrating the best
\citet{Chary2001} SED template from 8-to-1000$\,\mu$m ($L_{{\rm IR}}$)
and using the standard SFR-$L_{{\rm IR}}$ relation of
\citet{Kennicutt1998} for a \citet{Chabrier2003} IMF (SFR~$[{\rm
  M}_{\sun}~ {\rm yr}^{-1}] = 1\times 10^{-10} L_{\rm IR}~[{\rm
  L}_{\sun}]$).

\begin{figure}
\includegraphics[width=8cm,clip]{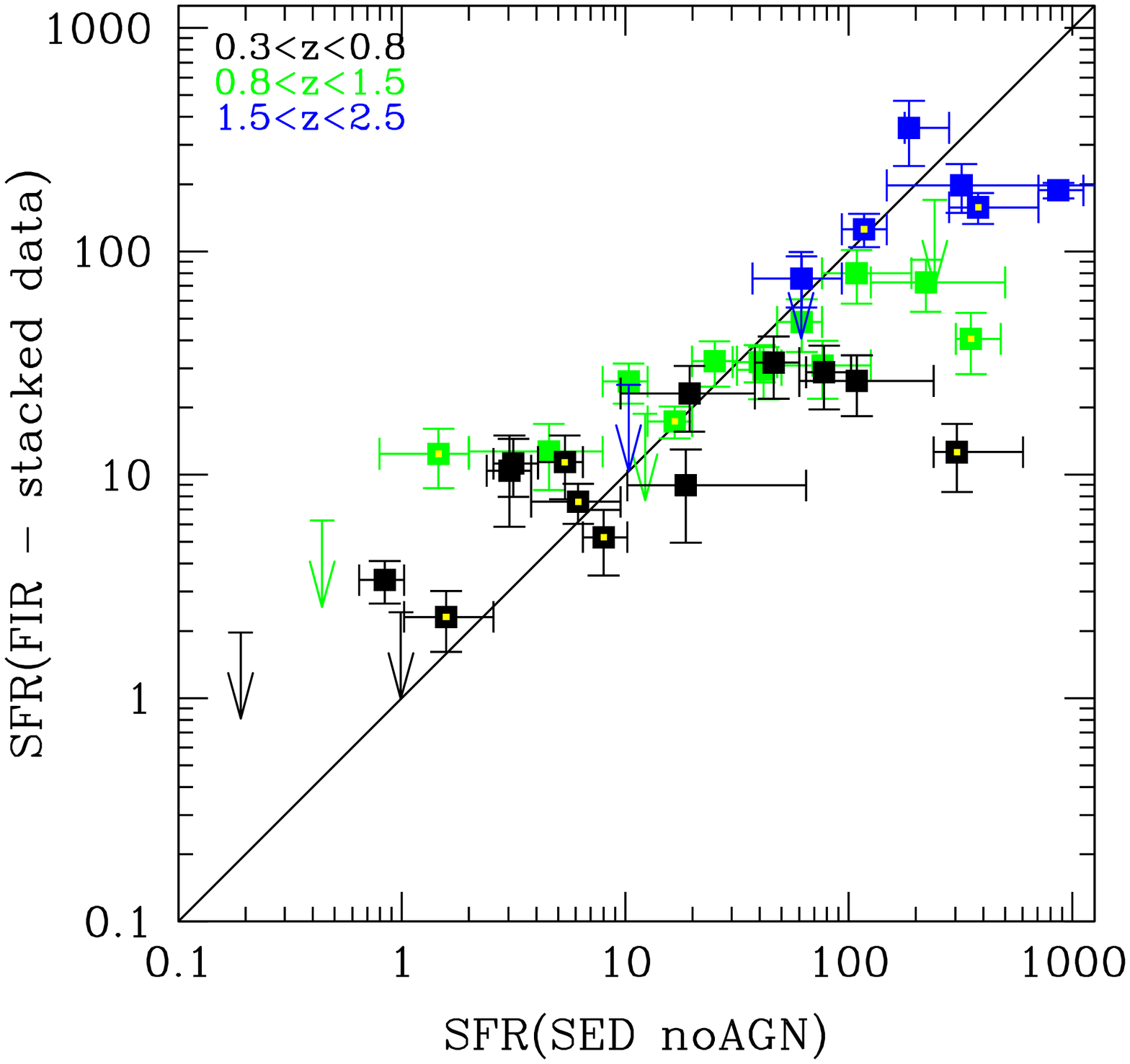}
\caption{Comparison between the SFR derived from the SED fitting without AGN component 
and the one derived from FIR using stacked data (binning in SFR$_{\rm sed}$ and
redshift). Different colors correspond to different redshift ranges:
0.3$<$z$<$0.8 black, 0.8$<$z$<$1.5 green and 1.5$<$z$<$2.5
blue. } 
\label{fig:SFRsed_SFRIR_noAGN}
\end{figure}

\begin{figure}
\includegraphics[width=8cm,clip]{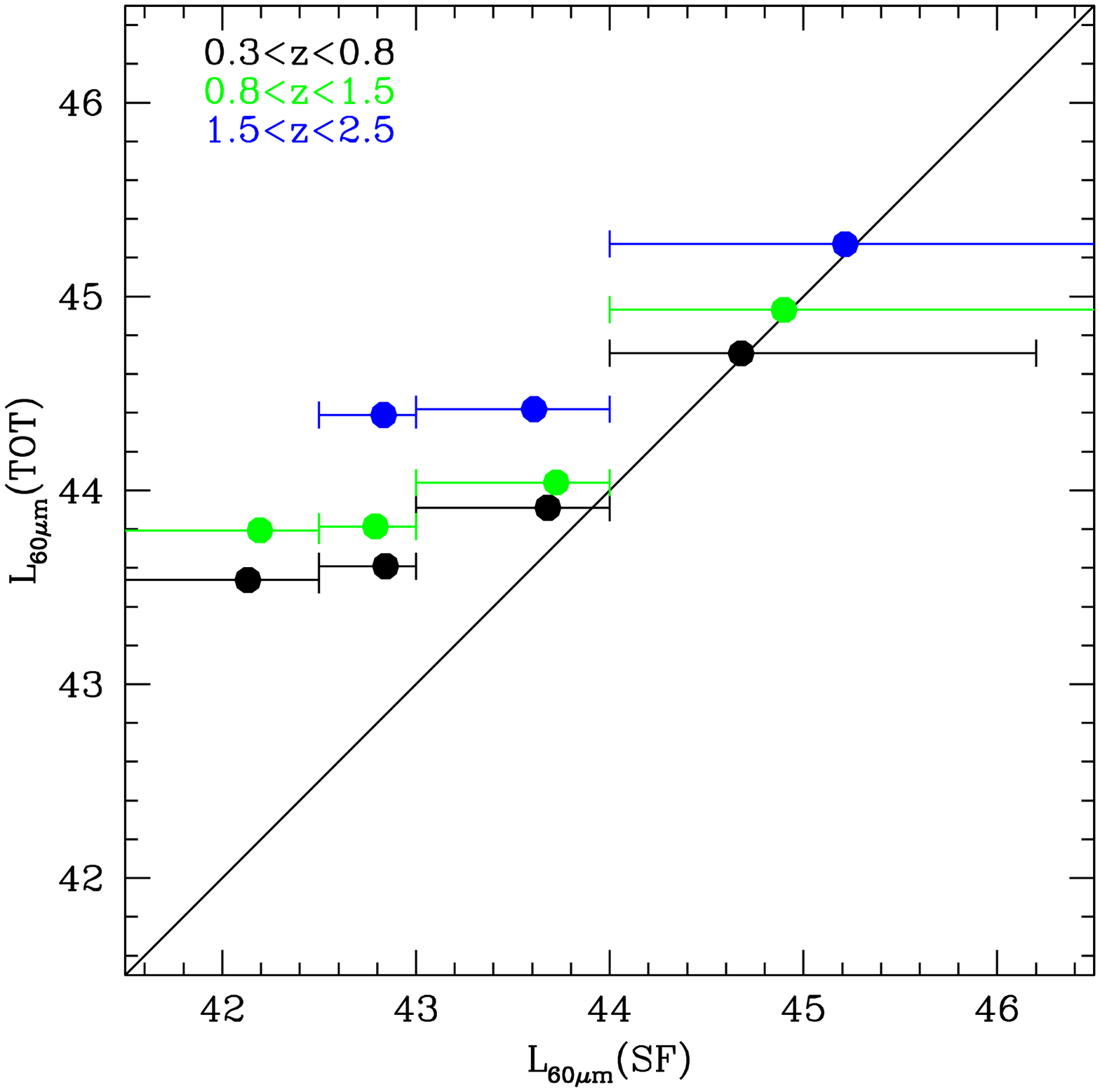}
\caption{Mean total (AGN+SF) L$_{60\mu m}$  versus mean L$_{60\mu m}$ due only to the SF contribution obtained using a mock catalog of galaxies with different levels of SF and AGN contaminations \citep{Rosario2012}, statistically weighed using the observed L$_{bol}$ distribution of our AGN sample in the usual three redshift bins. Error bars show the range in L$_{60\mu m}$(SF) used in the computation of the mean.}
\label{fig:simulation_w}
\end{figure}

In the right panel  of Fig. \ref{fig:SFRsed_SFRIR} we compare for
type--2 AGN the SFRs computed using the stacked IR data and the one
computed with the 2-components SED fitting.  
Above $\sim$20 M$_{\sun}$ yr$^{-1}$ the comparison is good with the
SFR derived from the SED fitting being slightly underestimated, while
at low SFR ($<$20 M$\sun$ yr$^{-1}$) the disagreement becomes
evident. 
This trend seems to be in contrast with previous studies which,
however, were based on samples of normal galaxies. In fact,
\citet{Santini2009} and \citet{Wuyts2011} studying a sample of
galaxies at high redshift (up to z$\sim$3)  found an overall good
correspondence between the SFR$_{\rm UV+IR}$ and SFR$_{\rm SED}$ up to
intermediate SFR, while at the largest SFR  ($>50 M_{\sun}$ yr${-1}$)
the discrepancy was significant and hence somewhat worrisome.   
However, \citet{Wuyts2011}  showed that the deviation from a
one-to-one relation between SFR$_{\rm UV+IR}$ and SFR$_{\rm SED}$ is most
severe for galaxies that have a large SFR$_{\rm IR}$/SFR$_{\rm UV}$ ratio
which is not the case for most of our sources (especially at high
X-ray luminosities) for which the IR emission is dominated by the
emission from hot dust heated by the AGN rather than young stars (see
Sec. \ref{sec:L12LX} and Fig. \ref{fig:L12LX}).    
However, for the sources for which this is not the case, we still
assume that the IR band ($\sim$12$\mu$m at $z\sim$1) is dominated by
the AGN emission and  this assumption can lead to underestimate the
SFR attributing the IR emission due to obscured young stars to the AGN
component. On the other hand, the IR estimation of the SFR assume zero
contamination from the AGN component in the IR and attribute the whole
IR flux to the emission of the young stars. Again, this
assumption, reasonable in most cases, can have the effect of
overestimating the SFR, and the effect would be higher at low SFR
where the contamination from the AGN is a significant fraction of the
IR emission. 
As explained below, we performed two kinds of tests to explore the above hypotheses.
First of all, we compared the SFR obtained from the SED
fitting without including the AGN component with the IR SFR (see Fig. \ref{fig:SFRsed_SFRIR_noAGN}). 
We find that in this case there is a much better agreement
between the two SFR indicators at low SFR. However, at high SFR we no longer have an agreement i.e. the SFR from the SED fitting is higher
than the one derived from FIR band (Fig.  \ref{fig:SFRsed_SFRIR_noAGN}). 
The result of this comparison indicates that the discrepancy seen in the right panel of Fig. \ref{fig:SFRsed_SFRIR}  can be partly due to the fact that the SED fitting method tends to overestimate 
the AGN component thus under-predicting the SFR.
 
On the other hand, to test the AGN contamination to the FIR bands, we built a mock catalog of galaxies with different levels of SF and AGN contaminations\footnote{intrinsic AGN luminosity at 60 $\mu m$ was estimated using the mean SED (normalized to the bolometric luminosity of the AGN) from \citet{Mullaney2011a} and assuming a scatter around this value of 0.3 dex, which corresponds to the expected scatter in the SED shapes from \citet{Rosario2012}.  Since the simulation assumes that the AGN have a uniform distribution in AGN bolometric luminosity, we used the bolometric luminosity function of \citet{Hopkins2007} to derive statistical weights for the AGN sample depending on their L$_{bol}$ and we applied them to get the real AGN distribution.} From this catalog and using the computed weights, we calculated the average L$_{60\mu m}$ due to AGN+SF as a function of the L$_{60\mu m}$ due only to the SF.   We find that, if we include the lowest luminosity tail of the AGN distribution i.e.  AGN with L$_{bol}$ down to 10$^{41}$ erg s$^{-1}$, there is hardly any difference between the L$_{SF}$ and L$_{TOT}$ since the weighed mock is dominated by low luminosity AGN. On the contrary, if we restrict the comparison to L$_{bol}>10^{44}$ erg s$^{-1}$, we find that the 60$\mu m$ luminosity is contaminated by the AGN contribution especially at low L$_{60\mu m}$. Since L$_{60\mu m}$ is commonly used as a proxy for L$_{IR}$ and, therefore, SFR, this plot can be directly compared with Fig. \ref{fig:SFRsed_SFRIR} right panel. As clearly visible, the two plots show a very good agreement in the overall shape.

To perform a direct comparison with the observed trend in Fig. \ref{fig:SFRsed_SFRIR} right panel, we used the same mock catalog as before which this time is weighed using the observed L$_{bol}$ distribution of our sample in the usual three redshift bins.
The result is shown in Fig. \ref{fig:simulation_w} where the mean total (AGN+SF) L$_{60\mu m}$  versus the mean L$_{60\mu m}$ due only to the SF contribution are plotted. As clearly visible in the plot, the shape of the points in this figure and in our Fig. \ref{fig:SFRsed_SFRIR} right panel are in perfect agreement. This demonstrates that the trend observed when comparing SFR-IR and SFR-SED is mostly driven by the effect of AGN contamination on the 60$\mu m$ luminosities. This effect is strong for the XMM-COSMOS sample, which is dominated by high luminosity AGN, but can be much smaller or even negligible when considering lower luminosity samples as demonstrated in the above test.

We conclude that  the discrepancy seen in the right panel of Fig. \ref{fig:SFRsed_SFRIR} is due
to the combination of two effects: (1) the SED fit tendency of overestimating
the AGN component thus under-predicting the SFR
and (2) the FIR-overestimation of the  SFR especially at high AGN luminosities and low SFR, where
the AGN contamination in the IR band is not negligible as demonstrated by the tests performed using the mock catalog.

\section{AGN host galaxy parameters derived from the 2-component SED fitting}
 \label{sec:table}
Here we report as example the first 16 lines of the tables published online.

%__________________________________________________ Table 
\begin{table*}
\centering
 \caption{\textbf{Source identification}\textit{(the full version of
     the table is available online)}. Columns are (1) XMM ID from
   \citep{Cappelluti2009}; (2) zCOSMOS ID \citep{Lilly2009} in case
   the object is in the BPT selected sample \citep{Bongiorno2010}; (3)
   zCOSMOS ID \citep{Lilly2009} in case the object is in the NeV
   selected sample (Mignoli et al. in prep); (4) photometric ID from
   \citet{Ilbert2009}; (5) RA; (6) DEC; (7) I-band apparent magnitude;
   (8) spectroscopic redshift; (9) photometric redshift when the
   spectroscopy is not available \citep{Salvato2009}; (10)
   Classification derived from the spectra: 1=unobscured AGN,
   2=obscured AGN classified from the X-ray \citep{Brusa2010},
   22.2=obscured AGN selected using the BPT diagram
   \citep{Bongiorno2010} and 22.4=obscured AGN selected through the
   \NeV\ line (Mignoli et al. in prep); (11) Classification derived
   from the SED \citep{Salvato2011} in case the spectra are not
   available (1=unobscured AGN, 2=obscured AGN).}   
\label{tab:tab1}
\begin{tabular}{ccccccccccc}
  \hline\hline
 XMM-ID &  zC-ID(BPT)  &  zC-ID(NeV) & PHOT-ID&$ \rm \hfill  \alpha_{J2000} \hfill $&$
  \hfill \rm  \delta_{J2000} \hfill $&$ \rm I_{AB} $& $ \rm z_{spec}$ &$ \rm z_{phot}$& $
\rm Class_{spec} $& $\rm Class_{phot}$\\
\hline
   1   &     -    &	-   &  786683 &  150.10521 &  1.981182  &    19.12 &  0.373 &	-     &     1.0  &  -99.0 \\
   6   &     -    &	-   &  767213 &  150.17979 &  2.110378  &    18.34 &  0.360 &	-     &     1.0  &  -99.0 \\
   19  &     -    &	-   & 1024197 &  149.99362 &  2.258538  &    20.12 &  0.659 &	-     &    2.0  &  -99.0 \\
   26  &     -    &	-   & 1627881 &  150.39963 &  2.687948  &    18.25 &  0.216 &	-     &    2.0  &  -99.0 \\ 
   21  &     -    &	-   & 1425464 &  150.23079 &  2.578183  &    20.12 &  1.403 &	-     &   1.0 &   -99.0 \\
   63  &  826475  &     -   & 1067126 &  149.78189 &  2.139104  &    18.97 &  0.355 &   -     &    2.0  &  -99.0 \\  
  181  &     -    &	-   & 751114  &  150.26744 &  2.055690  &    24.93 &	-   &  1.719  &   -99.0 &    2.0 \\ 
  182  &     -    &	-   & 1215214 &  150.32085 &  2.332960  &    25.48 &	-   &  2.030  &   -99.0 &    2.0 \\ 
  184  &     -    &	-   & 1453559 &  150.22397 &  2.550771  &    22.26 &	-   &  1.111  &   -99.0 &    2.0 \\ 
  185  &     -    &	-   & 808830  &  149.99526 &  2.006661  &    21.90 &	-   &  2.189  &   -99.0 &    1.0 \\ 
   -   &   803488 &	-   &  272411 &  150.65016 &  1.657098  &   20.43  &  0.587 &	-     &     22.2 &  -99.0 \\
   -   &     -    & 803886  &  286168 &  150.52974 &  1.725584  &   20.93  &  0.896 &	-     &     22.4 &  -99.0 \\
   -   &     -    & 804237  &  291946 &  150.42996 &  1.684480  &   22.30  &  1.000 &	-     &     22.4 &  -99.0 \\
   -   &   804862 &    -    &  303288 &  150.27359 &  1.778924  &   22.44  &  0.370 &	-     &     22.2 &  -99.0 \\
   -   &     -    & 804431  &  308347 &  150.38304 &  1.745181  &   22.06  &  0.702 &	-     &     22.4 &  -99.0 \\
   -   &   805059 &    -    &  311801 &  150.23288 &  1.724173  &   22.36  &  0.338 &	-     &     22.2 &  -99.0 \\
  ...  &    ...   &    -    &	...   &     ...    &	   ...    &    ...   &   ...  &  ...	&     ...  &   ... \\
  ...  &    ...   &	-   &	...   &     ...    &	   ...    &    ...   &   ...  &  ...	&     ...  &   ... \\
\hline
\end{tabular}
\end{table*}

\begin{table*}
\centering
 \caption{\textbf{Parameters derived from the SED fitting}\textit{(the
     full version of the table is available online)}. Columns are:
   (1), (2), (3) as above; (4)  extinction E(B-V) of the AGN component
   (5) rest-frame U-band magnitude of the host galaxy not corrected
   for extinction; (6) rest-frame B-band magnitude of the host galaxy
   not corrected for extinction; (7) host galaxy stellar mass; (8)
   flag indicating whether the previous value is a measurement (=0),
   an upper limit (=1) or if there is no measurement (=2); (9) host
   galaxy star formation rate; (10) flag indicating whether the
   previous value is a measurement (=0), an upper limit (=1) or if
   there is no measurement (=2 or =3); (11)  extinction E(B-V) of the
   galaxy component. }   
\label{tab:tab2}
\begin{tabular}{ccccccccccc}
  \hline\hline
 XMM &  zC(BPT)  &  zC(NeV) & E$_{B-V}^{AGN}$&$\rm M_{U_{jkc}}^{gal} $ &$\rm M_{B_{jkc}}^{gal} $ & $ \rm log(M_{*}^{gal}[M_{\sun}])$ & $\rm f_{M_{*}^{gal}}$ & $ \rm log(SFR^{gal}[M_{\sun} yr^{-1}]) $ & $\rm f_{SFR^{gal}}$& E$_{B-V}^{gal}$\\
\hline
   1   &     -    &  	  -   & 0.0 &	 -20.235 &    -20.641 &     9.238   &	  0	&    -99.0   &  3   &  0.3  \\
   6   &     -    &  	  -   & 0.1 &	 -21.655 &    -22.406 &    10.975   &	  0	&    -99.0   &  3   &  0.0  \\
   19  &     -    &  	  -   & 0.3 &	 -21.352 &    -22.254 &    11.158   &	  0	&    1.459   &  0   &  0.2  \\
   26  &     -    &  	  -   & 5.6 &	 -20.108 &    -20.983 &    10.546   &	  0	&    0.364   &  0   &  0.0  \\
   21  &     -    &  	  -   & 0.1 &	 -99.000 &    -99.000 &    10.537   &	  1	&   -99.0    &  3   &  0.0  \\
   63  &  826475  &  	  -   & 8.7 &	 -20.647 &    -21.582 &    10.994   &	  0	&    0.861   &  0   &  0.1  \\
  181  &     -    &  	  -   & 0.6 &	 -20.417 &    -21.728 &    11.051   &	  0	&    1.628   &  0   &  0.5  \\
  182  &     -    &  	  -   & 0.6 &	 -20.746 &    -22.006 &     10.984   &     0	 &   -0.152   &  0   & 0.1 \\
  184  &     -    &  	  -   & 9.0 &	 -21.045 &    -21.952 &     10.866   &     0	 &    1.005   &  0   & 0.1  \\
  185  &     -    &  	  -   & 0.0 &	 -23.113 &    -24.295 &     11.296   &     0	 &   -99.0    &  3   & 0.0 \\
   -   &   803488 &  	  -   & 9.0 &	 -21.056 &    -22.238 &    10.483   &	  0	&   -0.818   &  0   &  0.0\\
   -   &     -    &   803886  & 9.0 &	 -21.468 &    -22.497 &    10.979   &	  0	&    0.997   &  0   &  0.1 \\
   -   &     -    &   804237  & 3.3 &	 -20.700 &    -21.656 &    10.454   &	  0	&    1.536   &  0   &  0.4  \\
   -   &   804862 &  	 -    & 2.1 &	 -18.115 &    -18.743 &    8.7610   &	  0	&   -0.296   &  0   &  0.0  \\
   -   &     -    &   804431  & 0.3 &	 -19.425 &    -20.458 &    10.325   &	  0	&   -0.15    &  0   &  0.0  \\
   -   &   805059 &  	 -    & 0.3 &	 -17.459 &    -18.166 &    8.6090   &	  0	&   -0.622   &  0   &  0.0  \\
  ...  &    ...   &  	 -    & ...  &     ...    &	...    &     ...    &	...	&     ...    &  ...  &  ...\\
  ...  &    ...   &  	  -   & ...  &     ...    &	...    &     ...    &	...	&     ...    &  ...  &  ...\\
\hline
\end{tabular}
\end{table*}

%--------------------------------------------------

\bsp
\label{lastpage}

\end{document}